\documentclass[aps,amssymb,amsmath,notitlepage,superscriptaddress,twocolumn]{revtex4-1}

\usepackage{amsfonts}
\usepackage{graphicx,graphics,epsfig,times,bm,bbm,mathrsfs} %amssymb,amsmath,amsfonts
\usepackage[normalem]{ulem}
\usepackage{subfigure}
\usepackage[pdfstartview=FitH]{hyperref}
\hypersetup{
    colorlinks=true,       % false: boxed links; true: colored links
    linkcolor=cyan,          % color of internal links
    citecolor=magenta,        % color of links to bibliography
    filecolor=magenta,      % color of file links
    urlcolor=cyan,           % color of external links
    runcolor=cyan
}
\usepackage[pdftex]{color}
\usepackage{upgreek}

\newcommand{\beq}{\begin{equation}}
\newcommand{\eeq}{\end{equation}}
\newcommand{\beqnn}{\begin{equation*}}
\newcommand{\eeqnn}{\end{equation*}}
\newcommand{\beann}{\begin{eqnarray*}}
\newcommand{\eeann}{\end{eqnarray*}}

\newcommand{\mc}{\mathcal}

\newcommand{\bes} {\begin{subequations}}
\newcommand{\ees} {\end{subequations}}
\newcommand{\bea} {\begin{eqnarray}}
\newcommand{\eea} {\end{eqnarray}}

\newcommand{\Tr}{\mathrm{Tr}}

\newcommand{\ignore}[1]{}

\newcommand{\ave}[1]{{\langle #1 \rangle}}

\newcommand{\braket}[2]{\langle #1 | #2\rangle}
\newcommand{\ket}[1]{ | #1\rangle}
\newcommand{\bra}[1]{\langle #1 | }
\newcommand{\ketbra}[2]{|#1\rangle\langle #2|}

\def\Tr{\mathop{\mathrm{Tr}}}

\def\md{\mathcal{D}}

\def\al{\alpha}

\begin{document}

%\title{Nested Error Correcting Codes for Quantum Annealing}
%\title{Energy-Boosted Quantum Annealing Correction on Complete Graphs via Nesting\\ \red{other title ideas welcome}}
\title{Nested Quantum Annealing Correction}

\author{Walter Vinci}
\affiliation{Department of Electrical Engineering, University of Southern California, Los Angeles, California 90089, USA}
\affiliation{Department of Physics and Astronomy, University of Southern California, Los Angeles, California 90089, USA}
\affiliation{Center for Quantum Information Science \& Technology, University of Southern California, Los Angeles, California 90089, USA}
\author{Tameem Albash}
\affiliation{Department of Physics and Astronomy, University of Southern California, Los Angeles, California 90089, USA}
\affiliation{Center for Quantum Information Science \& Technology, University of Southern California, Los Angeles, California 90089, USA}
\affiliation{Information Sciences Institute, University of Southern California, Marina del Rey, California 90292, USA}
\author{Daniel A. Lidar}
\affiliation{Department of Electrical Engineering, University of Southern California, Los Angeles, California 90089, USA}
\affiliation{Department of Physics and Astronomy, University of Southern California, Los Angeles, California 90089, USA}
\affiliation{Center for Quantum Information Science \& Technology, University of Southern California, Los Angeles, California 90089, USA}
\affiliation{Department of Chemistry, University of Southern California, Los Angeles, California 90089, USA}

\begin{abstract}
We present a general error-correcting scheme for quantum annealing that allows for the encoding of a logical qubit into an arbitrarily large number of physical qubits.  Given any Ising model optimization problem, the encoding replaces each logical qubit by a complete graph of degree $C$, representing the distance of the error-correcting code.  A subsequent minor-embedding step then implements the encoding on the underlying hardware graph of the quantum annealer. We demonstrate experimentally that the performance of a D-Wave Two quantum annealing device improves as $C$ grows.  We show that the performance improvement can be interpreted as arising from an effective increase in the energy scale of the problem Hamiltonian, or equivalently, an effective reduction in the temperature at which the device operates.  The number $C$ thus allows us to control the amount of protection against thermal and control errors, and in particular, to trade qubits for a lower effective temperature that scales as $C^{-\eta}$, with $\eta \leq 2$. This effective temperature reduction is an important step towards scalable quantum annealing.
\end{abstract}
\maketitle
%
%\section{Introduction}
 Quantum annealing (QA) attempts to exploit quantum fluctuations to solve computational problems faster than it is possible with classical computers \cite{kadowaki_quantum_1998,Brooke1999,brooke_tunable_2001,farhi_quantum_2001,morita:125210,RevModPhys.80.1061,EPJ-ST:2015}. As an approach designed to solve optimization problems, QA is a special case of adiabatic quantum computation (AQC) \cite{farhi_quantum_2000}, a universal model of quantum computing \cite{aharonov_adiabatic_2007,PhysRevLett.99.070502,Gosset:2014rp,Lloyd:2015fk}. In AQC, a system is designed to follow the instantaneous ground state of a time-dependent Hamiltonian whose final ground state encodes the solution to the problem of interest. This results in a certain amount of stability, since the system can thermally relax to the ground state after an error, as well as resilience to errors, since the presence of a finite energy gap suppresses thermal and dynamical excitations \cite{childs_robustness_2001,PhysRevLett.95.250503,TAQC,Lloyd:2008zr,amin_decoherence_2009,Albash:2015nx}.

Despite this inherent robustness to certain forms of noise, AQC requires error-correction to ensure scalability, just like any other form of quantum information processing \cite{Lidar-Brun:book}.  Various error correction proposals for AQC and QA have been made \cite{jordan2006error,PhysRevLett.100.160506,PhysRevA.86.042333,Young:13,Sarovar:2013kx,Young:2013fk,PAL:13,Ganti:13,Bookatz:2014uq,Mizel:2014sp,PAL:14,Vinci:2015jt,Mishra:2015ye,MNAL:15}, but 
an accuracy-threshold theorem for AQC is not yet known, unlike in the circuit model (e.g., \cite{Aliferis:05}). 
A direct AQC simulation of a fault-tolerant quantum circuit leads to many-body (high-weight) operators that are difficult to implement \cite{Young:13,Sarovar:2013kx} or myriad other problems \cite{Lloyd:2015fk}. Nevertheless, a scalable method to reduce the effective temperature would go a long way towards approaching the ideal of closed-system AQC, where only non-adiabatic transitions constitute the source of errors.

Motivated by the availability of commercial QA devices featuring hundreds of qubits \cite{Dwave,Johnson:2010ys,Berkley:2010zr,Harris:2010kx}, we focus on error correction for QA.
There is a consensus that these devices are significantly and adversely affected by decoherence, noise, and control errors \cite{q108,SSSV,Albash:2014if,q-sig2,Crowley:2014qp,Martin-Mayor:2015dq,King:2015zr,vinci2014hearing}, which makes them particularly interesting for the study of tailored, practical error correction techniques. Such techniques, known as quantum annealing correction (QAC) schemes, have already been experimentally shown to significantly improve the performance of quantum annealers \cite{PAL:13,PAL:14,Vinci:2015jt,Mishra:2015ye}, and theoretically analyzed using a mean-field approach \cite{MNAL:15}. However, these QAC schemes are not easily generalizable to arbitrary optimization problems since they induce an encoded graph that is typically of a lower degree than the qubit-connectivity graph of the physical device. Moreover, they typically impose a fixed code distance, which limits their efficacy. 

To overcome these limitations, here we present a family of error-correcting codes for QA, based on a ``nesting'' scheme, that has the following properties: (1) it can handle arbitrary Ising-model optimization problem, (2) it can be implemented on present-day QA hardware, and (3)  it is capable of an effective temperature reduction controlled by the code distance. Our ``nested quantum annealing correction" (NQAC) scheme thus provides a very general and practical tool for error correction in quantum optimization. 

We test NQAC by studying antiferromagnetic complete graphs numerically, as well as on a D-Wave Two (DW2) processor featuring $504$ flux qubits connected by $1427$ tunable composite qubits acting as Ising-interaction couplings, arranged in a non-planar Chimera-graph lattice \cite{Bunyk:2014hb} (complete graphs were also studied for a spin glass model in Ref.~\cite{Venturelli:2014nx}). We demonstrate that our encoding schemes yields a steady improvement for the probability of reaching the ground state as a function of the nesting level, even after minor-embedding the complete graph onto the physical graph of the quantum annealer. We also demonstrate that NQAC outperforms classical repetition code schemes that use the same number of physical qubits.\\  

%%%%%%%%%%%%%%%%%%%%%%%%%%%%%%%%%%%%%%%%%
%\emph{Quantum annealing and encoding the Hamiltonian.}---%
\section{Quantum Annealing and Encoding the Hamiltonian}
\begin{figure}[ht]
\begin{center}
\subfigure[\ Logical graph: 1st level.]{\includegraphics[width=0.23\textwidth]{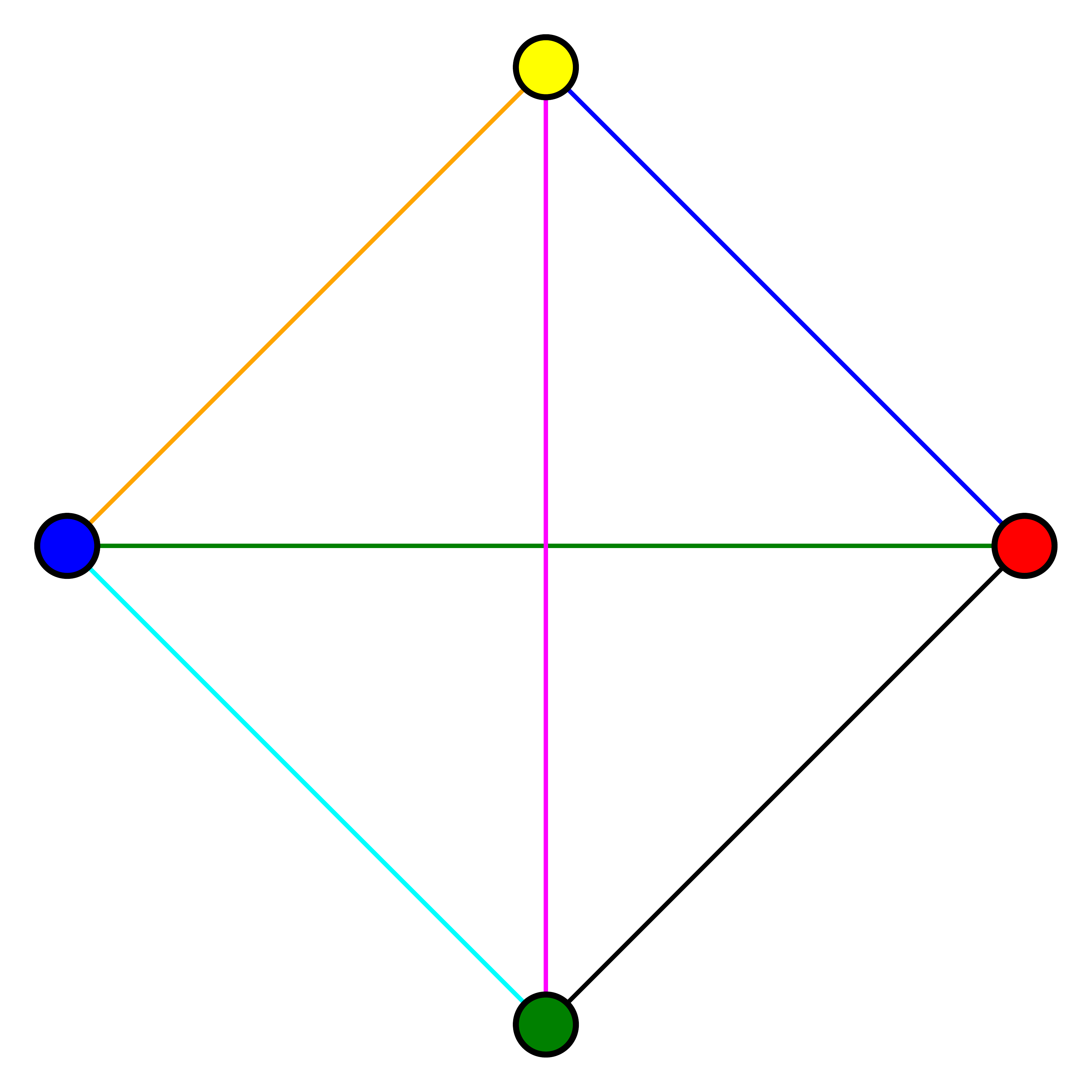}\label{fig:4x4_ideal_logical_plot_1}}
\subfigure[\ 1st level ME.]{\includegraphics[width=0.23\textwidth]{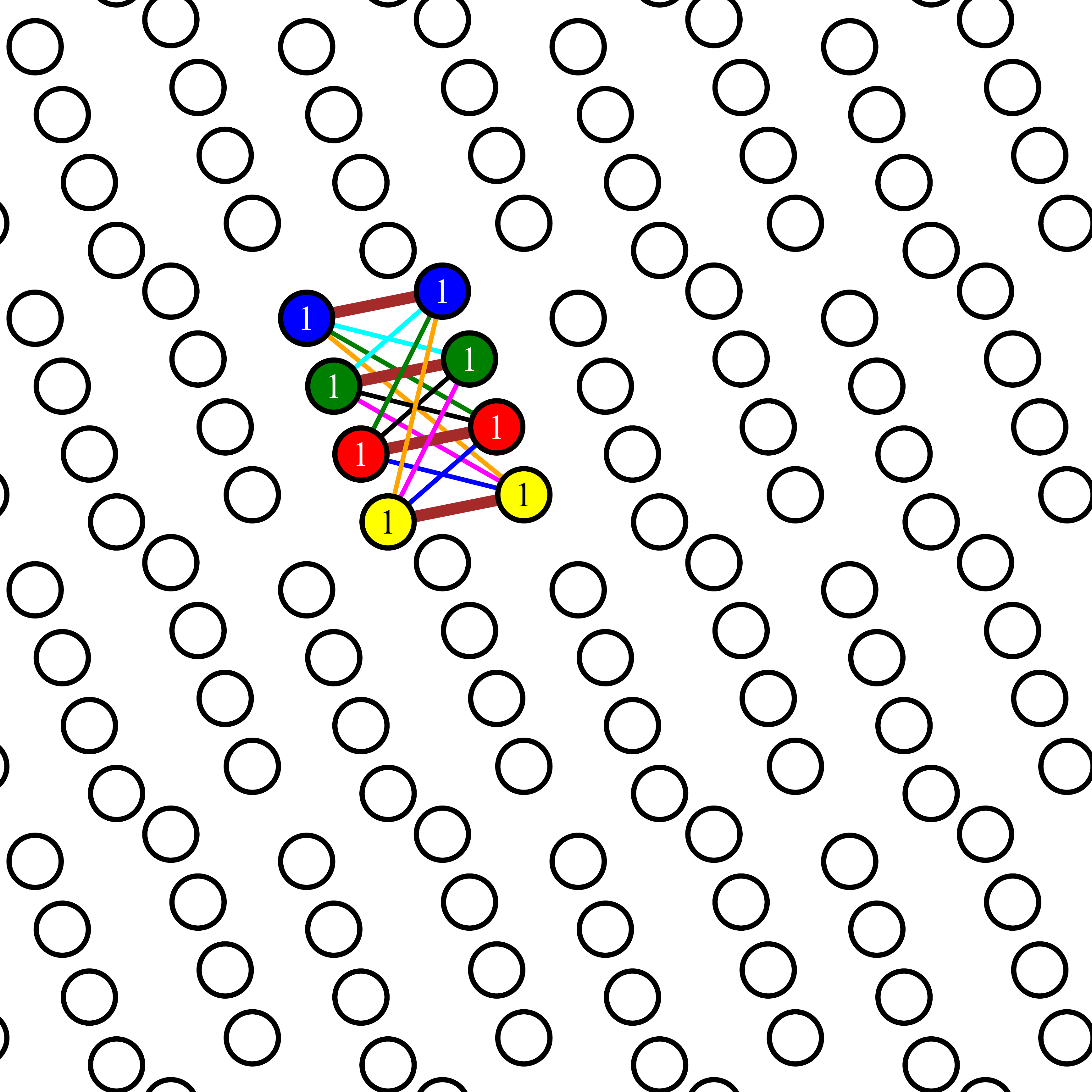}\label{fig:4x4_ideal_physical_1}}
\subfigure[\ Nested graph: 4th level.]{\includegraphics[width=0.23\textwidth]{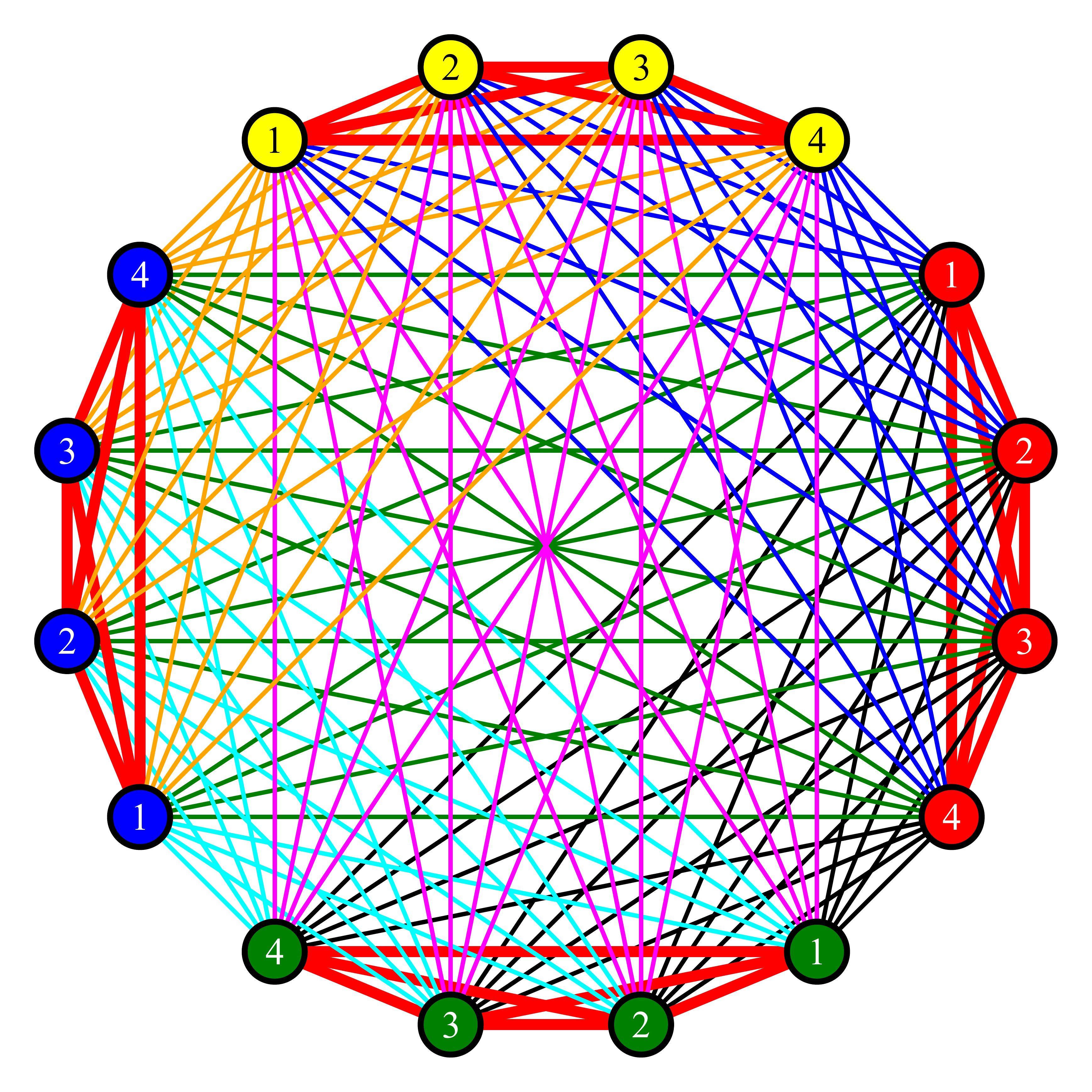}\label{fig:4x4_ideal_logical_plot_4}}
\subfigure[\ 4th level ME.]{\includegraphics[width=0.23\textwidth]{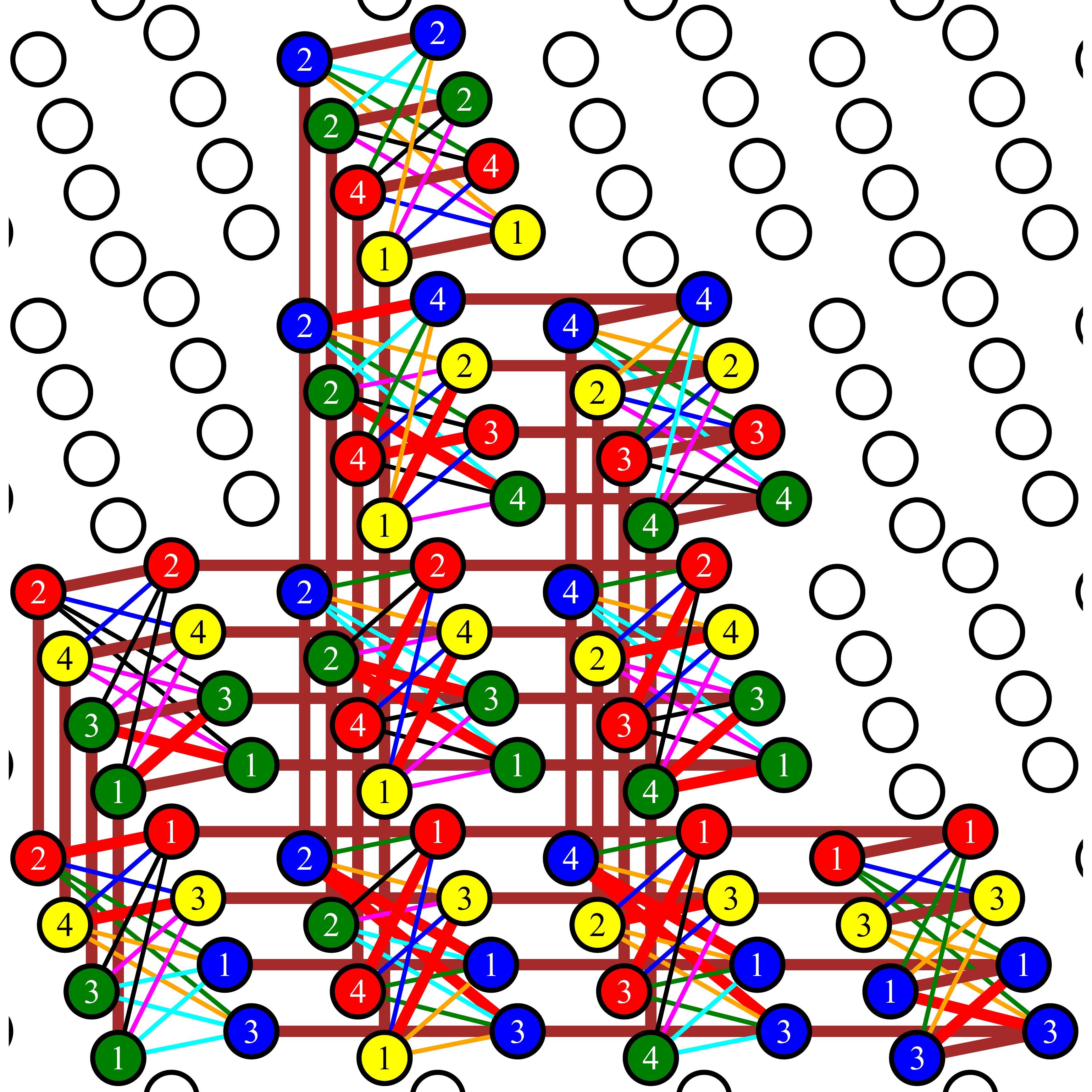}\label{fig:4x4_ideal_physical_4}}
\caption{Illustration of the nesting scheme. In the left column, a  $C$-level nested graph is constructed by embedding a $K_N$ into a $K_{C\times N}$, with $N=4$ and $C=1$ (top) and $C=4$ (bottom). Red, thick 
%(lighter grey, thick) 
couplers are energy penalties defined on the nested graph between the $(i,c)$ nested copies of each logical qubit $i$. 
The right column shows the nested graphs after ME on the DW2 Chimera graph. Brown, thick couplers %(darker grey, thick) 
correspond to the ferromagnetic chains introduced in the process.
} 
\label{fig:log-nesting}
\end{center}
\end{figure}

%%%%%%%%%%%%%%%%%%%%%%%%%%%%%%%%%%%%%%%%%
%
 In QA the system undergoes an evolution governed by the following time-dependent, transverse-field Ising Hamiltonian:
\beq
H(t) = A(t) H_X + B(t)  H_{\mathrm{P}}\ , \qquad t\in[0,t_f] \ ,
\label{eq:adiabatic}
\eeq
with respectively monotonically decreasing and increasing ``annealing schedules" $A(t)$ and $B(t)$. The ``driver Hamiltonian'' $H_X = -\sum_i \sigma_i^x$ is a transverse field whose amplitude controls the tunneling rate. The solution to an optimization problem of interest is encoded in the ground state of the Ising problem Hamiltonian $H_{\mathrm{P}}$, with
\beq 
\label{eq:HP}
H_{\mathrm{P}} = \sum_{i \in \mc{V}} h_i \sigma^z_i + \sum_{(i,j) \in \mc{E}} J_{ij}\sigma^z_i\sigma^z_j\, ,
\eeq
where the sums run over the weighted vertices $\mc{V}$ and edges $\mc{E}$ of a graph $G = (\mc{V},\mc{E})$, and $\sigma_i^{x,z}$ denote the Pauli operators acting on qubit $i$. The D-Wave devices use an array of superconducting flux qubits to physically realize the system described in Eqs.~\eqref{eq:adiabatic} and \eqref{eq:HP} on a fixed ``Chimera" graph (see Fig.~\ref{fig:log-nesting}) with programmable local fields $\{h_i\}$, couplings $\{J_{ij}\}$, and annealing time $t_f$ \cite{Johnson:2010ys,Berkley:2010zr,Harris:2010kx}. 

For closed systems, the adiabatic theorem \cite{Kato:50,Jansen:07} guarantees that if the system is initialized in the ground state of $H(0) = A(0) H_X$, a sufficiently slow evolution relative to the inverse minimum gap of $H(t)$ will take the system with high probability to the ground state of the final Hamiltonian $H(t_f) =B(t_f) H_{\mathrm{P}}$. Dynamical errors then arise due to diabatic transitions, but they can be made arbitrarily small via boundary cancellation methods that control the smoothness of $A(t)$ and $B(t)$, as long as the adiabatic condition is satisfied \cite{lidar:102106,Wiebe:12,Ge:2015wo}. For open systems, specifically a system that is weakly coupled to a thermal environment, the final state is a mixed state $\rho (t_f)$ that is close to the Gibbs state associated with $H(t_f)$ if equilibration is reached throughout the annealing process \cite{Avron:2012tv,Albash:2015nx,Venuti:2015kq}. In the adiabatic limit the open system QA process is thus better viewed as a Gibbs distribution sampler. The main goal of QAC is to suppress the associated thermal errors and restore the ability of QA to act as a ground state solver. In addition QAC should suppress errors due to noise-driven deviations in the specification of $H_{\mathrm{P}}$ \cite{Young:2013fk}.

Error correction is achieved in QAC by mapping the logical Hamiltonian $H(t)$ to an appropriately chosen encoded Hamiltonian $\bar H(t)$:
\beq
\bar H(t) = A(t) H_X + B(t)  \bar H_{\mathrm{P}}\ , \qquad t\in[0,t_f] \ ,
\label{eq:encoded}
\eeq
defined over a set of physical qubits $\bar N$ larger than the number of logical qubits $N = |\mc{V}|$. Note that $\bar H_{\mathrm{P}}$ also includes penalty terms, as explained below. The logical ground state of $H_{\mathrm{P}}$ is extracted from the encoded system's state $\bar \rho(t_f)$ through an appropriate decoding procedure. A successful error correction scheme should recover the logical ground state with a higher probability than a direct implementation of $H_{\mathrm{P}}$, or than a classical repetition code using the same number of physical qubits $\bar N$. Due to practical limitations of current QA devices that prevent the encoding of $H_X$, only $H_{\mathrm{P}}$ is encoded in QAC.  

 %%%%%%
%%%%%%

\begin{figure*}[ht]
\begin{center}
{\includegraphics[width=0.33\textwidth]{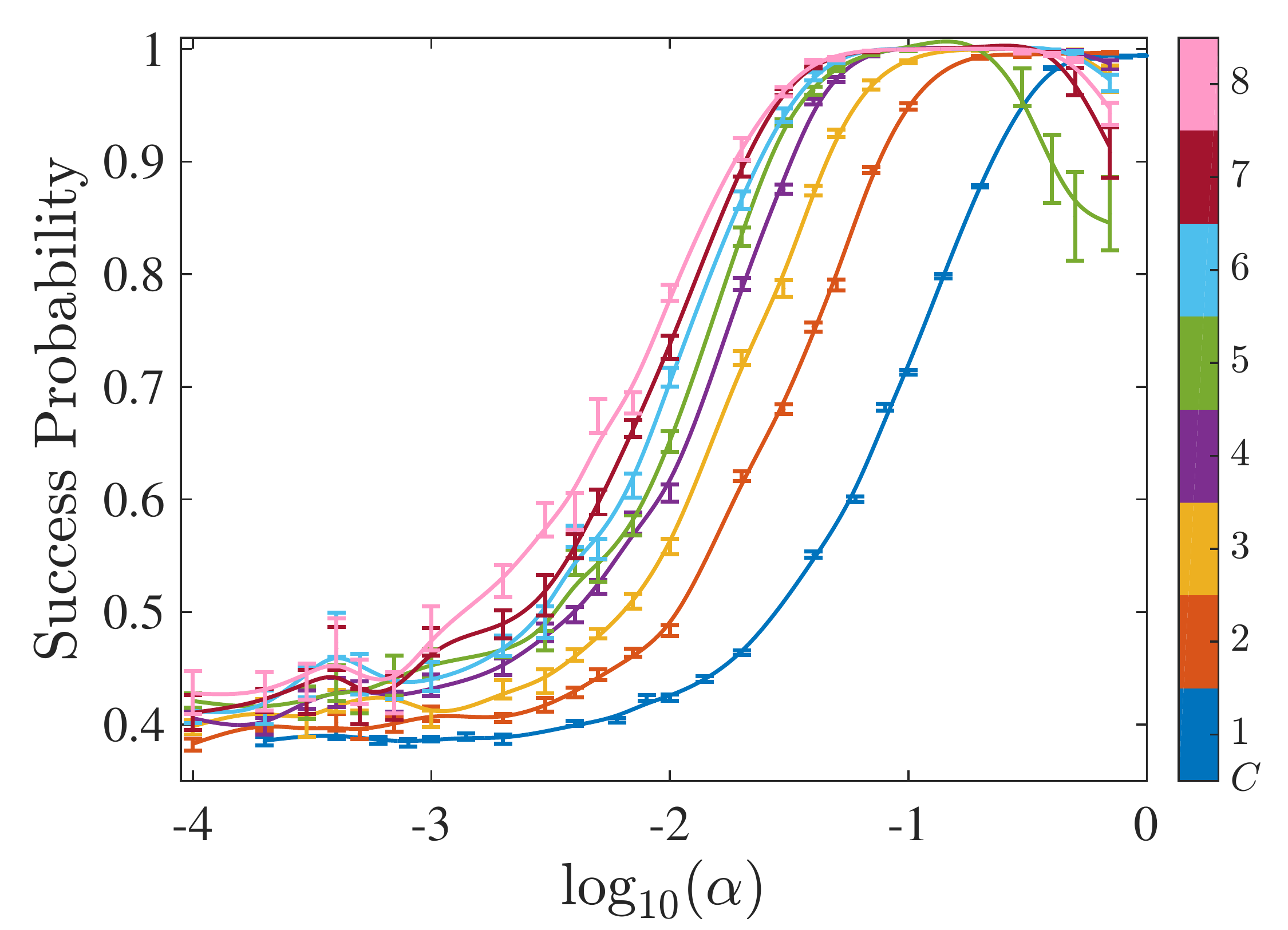}\label{fig:4x4_ideal_per_logical-}}
{\includegraphics[width=0.33\textwidth]{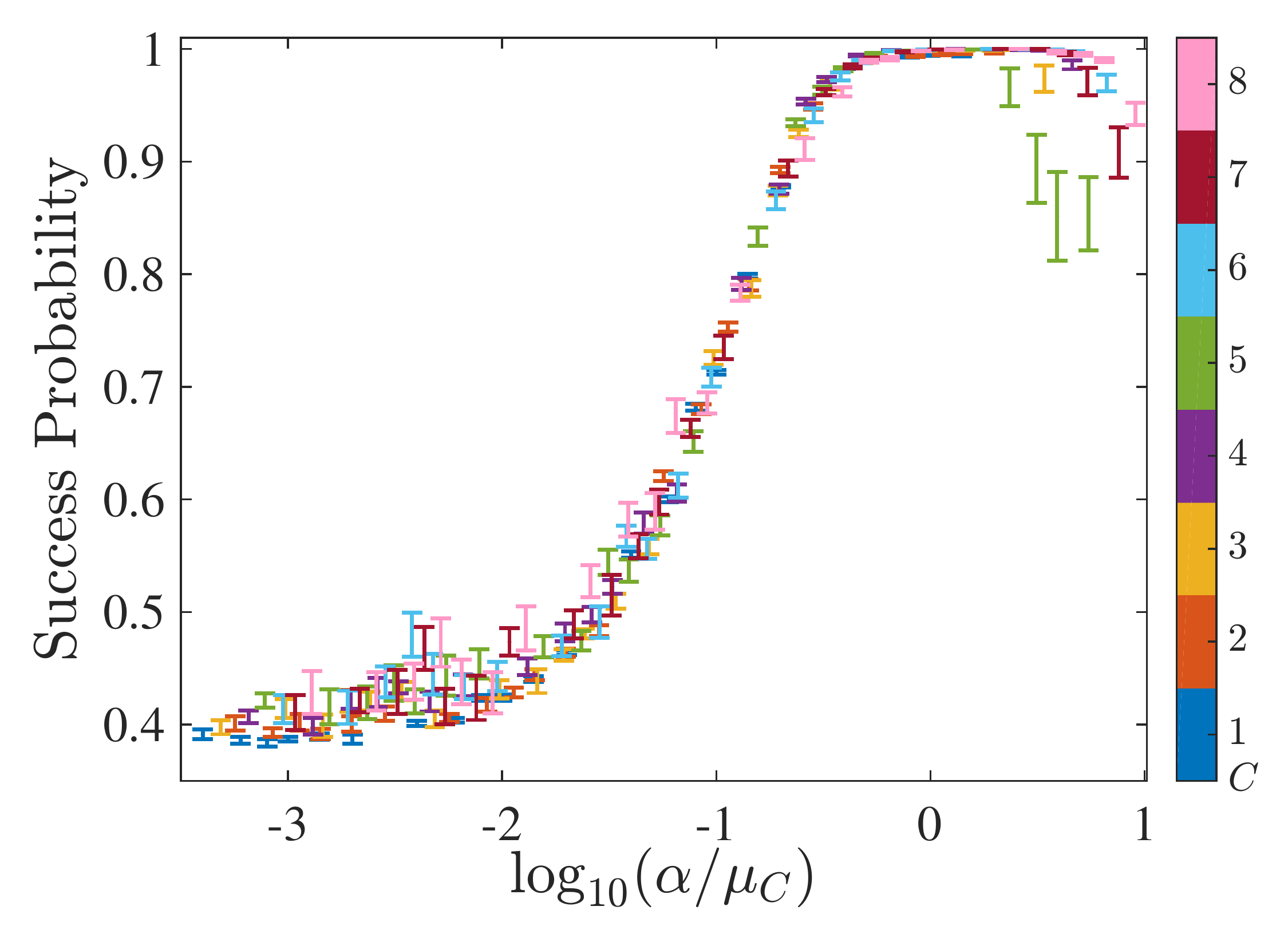}\label{Overlapped_K4}}
{\includegraphics[width=0.33\textwidth]{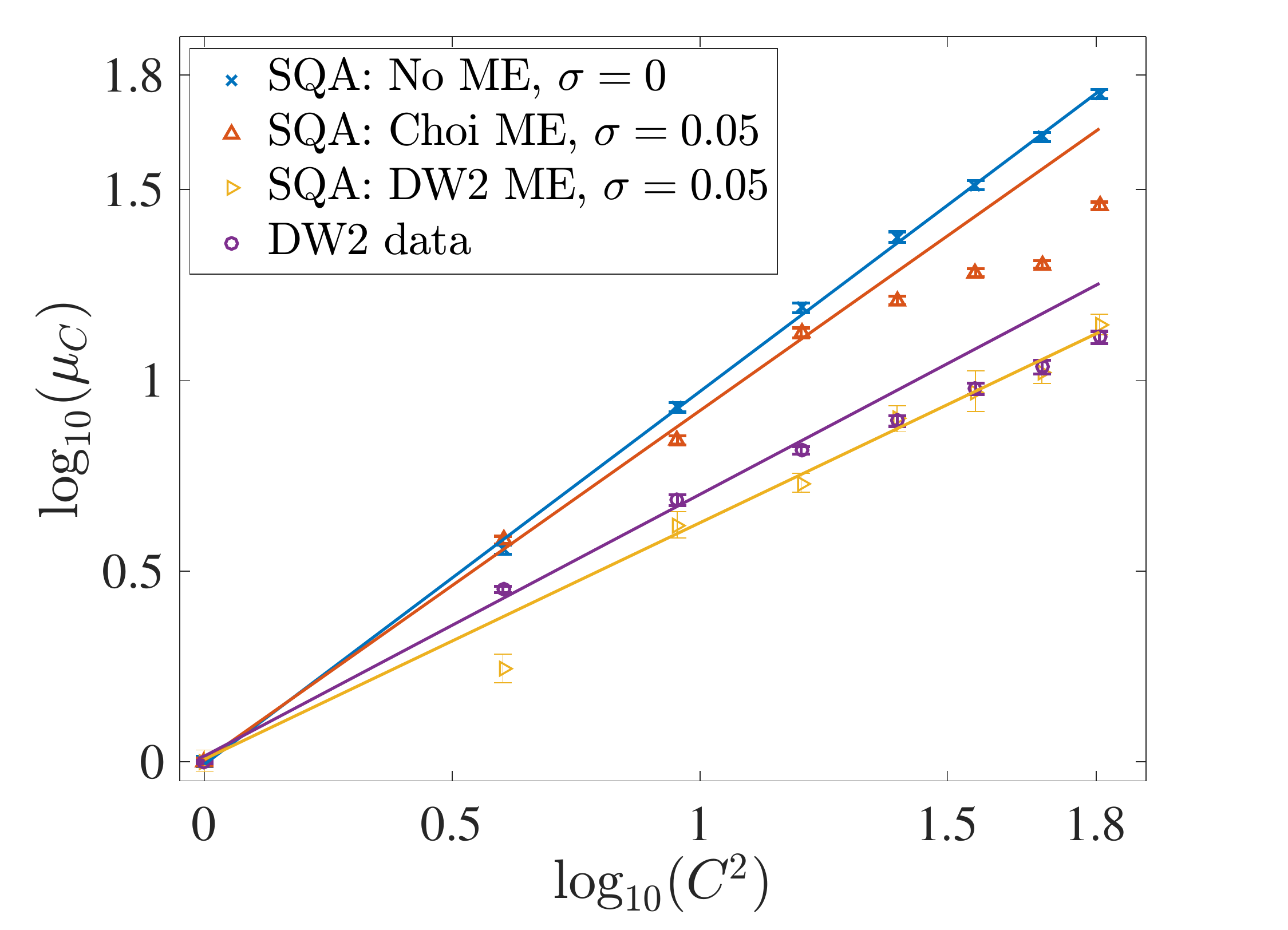}\label{fig:EBoost_SQA}}
\caption{Experimental and numerical results for the antiferromagnetic $K_4$, after encoding, followed by ME and decoding. Left: DW2 success probabilities $P_C(\alpha)$ for eight nesting levels $C$. Increasing $C$ generally increases $P_C(\alpha)$ at fixed $\alpha$. Middle: Rescaled $P_C(\alpha\mu_C)$ data, exhibiting data-collapse. Right: scaling of the energy boost $\mu_C$ \textit{vs} the maximal energy boost $\mu_C^{\max}$, for both the DW2 and SQA. Purple circles: DW2 results. Blue stars: SQA for the case of no ME (i.e., for the problem defined directly over $K_{C\times N}$ and no coupler noise). Red up-triangles: SQA for the Choi ME \cite{Choi2} (for a full Chimera graph), with $\sigma = 0.05$ Gaussian noise on the couplings. Yellow right-triangles: SQA for the DW2 heuristic ME \cite{Cai:2014nx,Boothby2015a} (applied to a Chimera graph with $8$ missing qubits) with $\sigma = 0.05$ Gaussian noise on the couplings. The flattening of $\mu_C$ suggests that the energy boost becomes less effective at large $C$. However, this can be remedied by increasing the number of SQA sweeps (see Appendix \ref{sec:Num_Add}), fixed here at $10^4$. Thus the lines represent best fits to only the first four data points, with slopes %$0.9766$
  $0.98$, %$0.9163$
  $0.91$, %$0.6195$,
  $0.62$ and %$0.6856$
  $0.69$ respectively. In all panels $N_{\mathrm{phys}}\in [8,288]$.}
\label{fig:exp-k4-nesting}
\end{center}
\end{figure*}
%%%%%%

%%%%%%%%%%%%%%%%%%%%%%%%%%%%%%%%%%%%%%%%%
%\emph{The nesting scheme.}---%
%%%%%%%%%%%%%%%%%%%%%%%%%%%%%%%%%%%%%%%%%
%%%%%%
In order to allow for the most general $N$-variable Ising optimization problem, we now define an encoding procedure for problem Hamiltonians $H_{\mathrm{P}}$ supported on a complete graph $K_N$.  The first step of our construction involves a ``nested" Hamiltonian $\tilde H_{\mathrm{P}}$ that is defined by embedding the logical $K_N$ into a larger $K_{C\times N}$. The integer $C$ is the ``nesting level" and controls the amount of hardware resources (qubits, couplers, and local fields) used to represent the logical problem.  $\tilde H_{\mathrm{P}}$ is constructed as follows. Each logical qubit $i$ ($i = 1,\dots,N$) is represented by a $C$-tuple of encoded qubits $(i,c)$, with $c = 1,\dots,C$. The ``nested" couplers $\tilde J_{(i,c),(j,c')}$ and local fields $\tilde h_{(i,c)}$ are then defined as follows: 
\bes
\label{eq:nesting}
\begin{align}
\tilde J_{(i,c),(j,c')} &= J_{ij}\,, \quad  \forall c,c', i\neq j\ ,   \\
\tilde h_{(i,c)} &= C h_{i}\,, \quad  \forall c, i\ , \label{eqt:h} \\
\tilde J_{(i,c),(i,c')} &= -\gamma \,, \quad  \forall c\neq c' \ .
\end{align}
\ees
This construction is illustrated in the left column of Fig.~\ref{fig:log-nesting}. Each logical coupling  $J_{ij}$ has $C^2$ copies $\tilde J_{(i,c),(j,c')}$, thus boosting the energy scale at the encoded level by a factor of $C^2$. Each local field $h_{i}$ has $C$ copies $\tilde h_{(i,c)}$; the factor $C$ in Eq.~\eqref{eqt:h} ensures that the energy boost is equalized with the couplers. For each logical qubit $i$, there are $C(C-1)/2$ ferromagnetic couplings $\tilde J_{(i,c),(i,c')}$ of strength $\gamma>0$ (to be optimized), representing energy penalties that promote agreement among the $C$ encoded qubits, i.e., that bind the $C$-tuple as a single logical qubit $i$.

The second step of our construction is to implement the fully connected problem $\tilde H_{\mathrm{P}}$ on given QA hardware, with a lower-degree qubit connectivity graph. This requires a minor embedding (ME) \cite{Kaminsky-Lloyd,Choi2,klymko_adiabatic_2012,Cai:2014nx,Boothby2015a}.  The procedure involves replacing each qubit in $\tilde H_{\mathrm{P}}$ by a ferromagnetically-coupled chain of qubits, such that all couplings in $\tilde H_{\mathrm{P}}$ are represented by inter-chain couplings.  The intra-chain coupling represents another energy penalty that forces the chain qubits to behave as a single logical qubit.
The physical Hamiltonian obtained after this ME step is the final encoded Hamiltonian $\bar H_P$. 
%If we specialize the ME process to the Chimera graph, the native graph of the DW2 processor, we 
We can minor-embed a $K_{C\times N}$ nested graph representing each qubit $(i,c)$ as a physical chain of  length $L = \lceil C N/4 \rceil+1$ on the Chimera graph \cite{Choi2}. This is illustrated in the right column of Fig.~\ref{fig:log-nesting}. The number of physical qubits necessary for a ME of a $K_{C\times N}$ is $N^{\mathrm{phys}}_{C} =  CNL \sim C^2N^2/4$. 

At the end of a QA run implementing the encoded Hamiltonian $\bar H_{\mathrm{P}}$ and a measurement of the physical qubits, a decoding procedure must be employed to recover the logical state. For the sake of simplicity we only consider majority vote decoding over both the length-$L$ chain of each encoded qubit $(i,c)$ and the $C$ encoded qubits comprising each logical qubit $i$ (decoding over the length-$L$ chain first, then over the $C$ encoded qubits, does not affect performance; seeAppendix~\ref{sec:Exp_Meth}). The encoded and logical qubits can thus be viewed as forming repetition codes with, respectively, distance $L$ and $C$. Other decoding strategies are possible wherein the encoded or logical qubits do not have this simple interpretation; e.g., energy minimization decoding, which tends to outperform majority voting \cite{Vinci:2015jt}. In the unlikely event of a tie, we assign a random value of $+1$ or $-1$ to the logical qubit.\\

%%%%%%
\begin{figure*}[ht]
\begin{center}
{\includegraphics[width=0.33\textwidth]{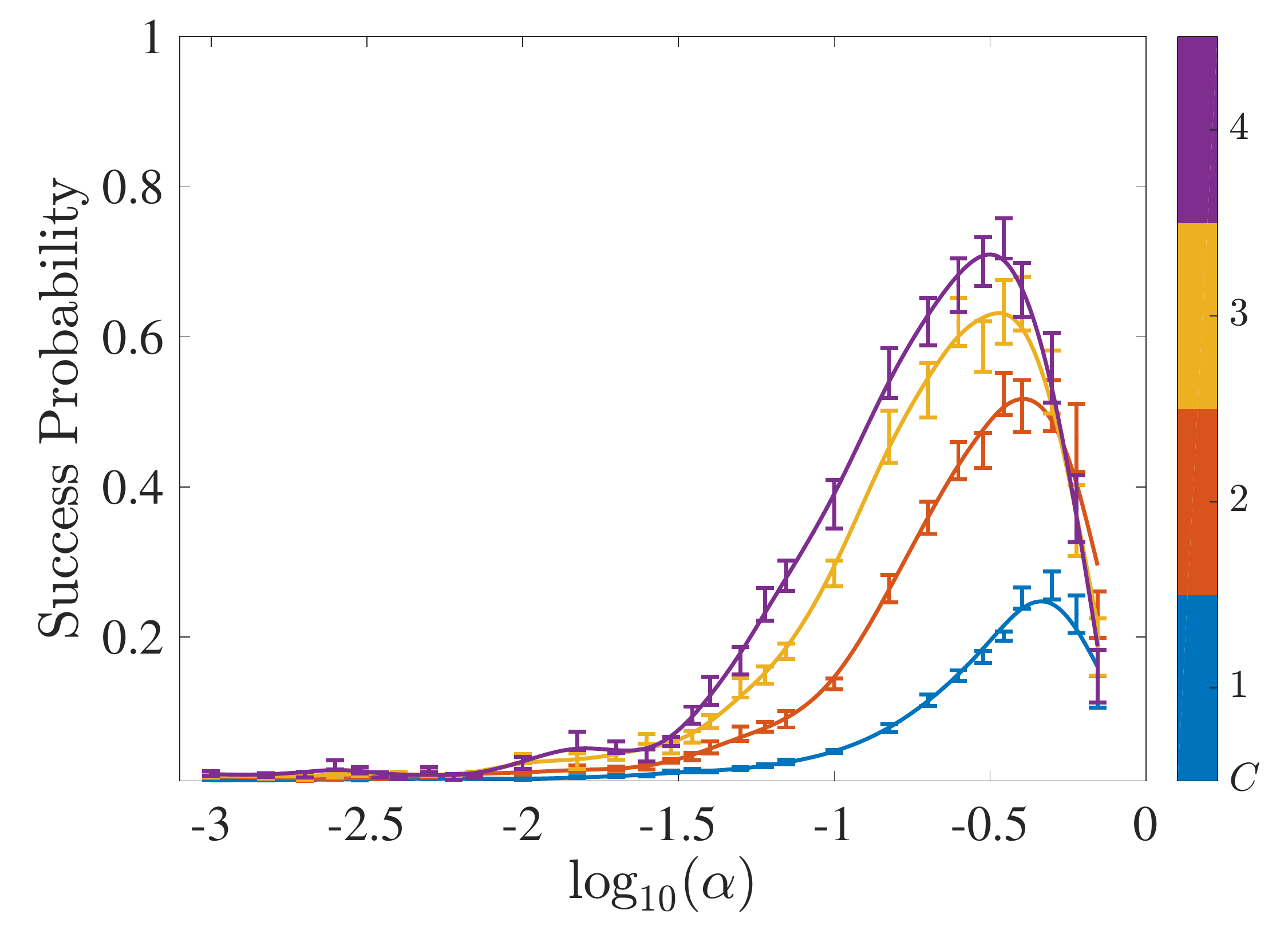}\label{fig:8x8_ideal_random_per_logical-}}
{\includegraphics[width=0.33\textwidth]{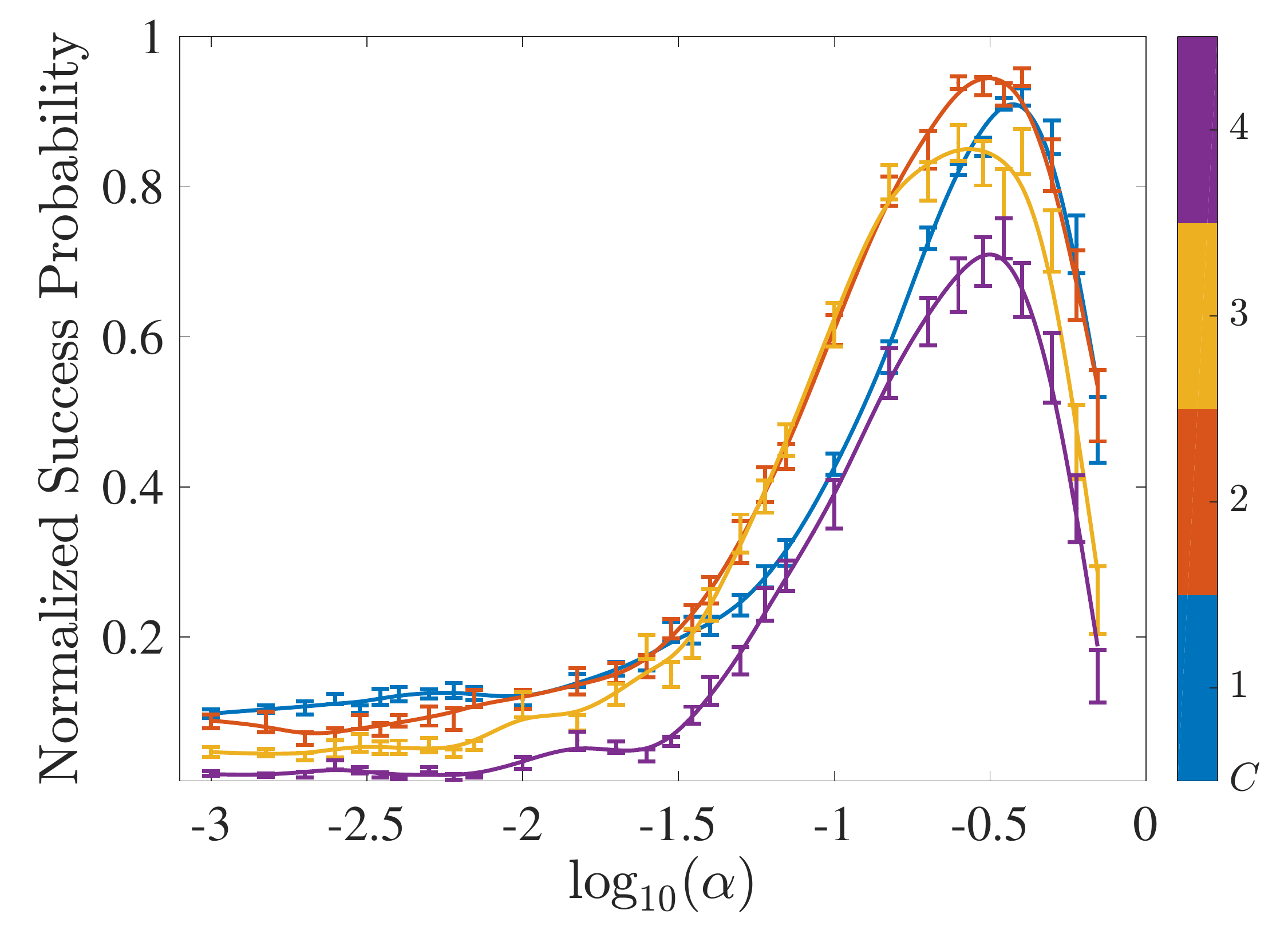}\label{fig:8x8_corrected}}
%{\includegraphics[width=0.33\textwidth]{Choi_K8_1_pGS_ME=1_MVAll_nSW=20k_beta=01_v2}\label{fig:8x8_sqa}}
{\includegraphics[width=0.33\textwidth]{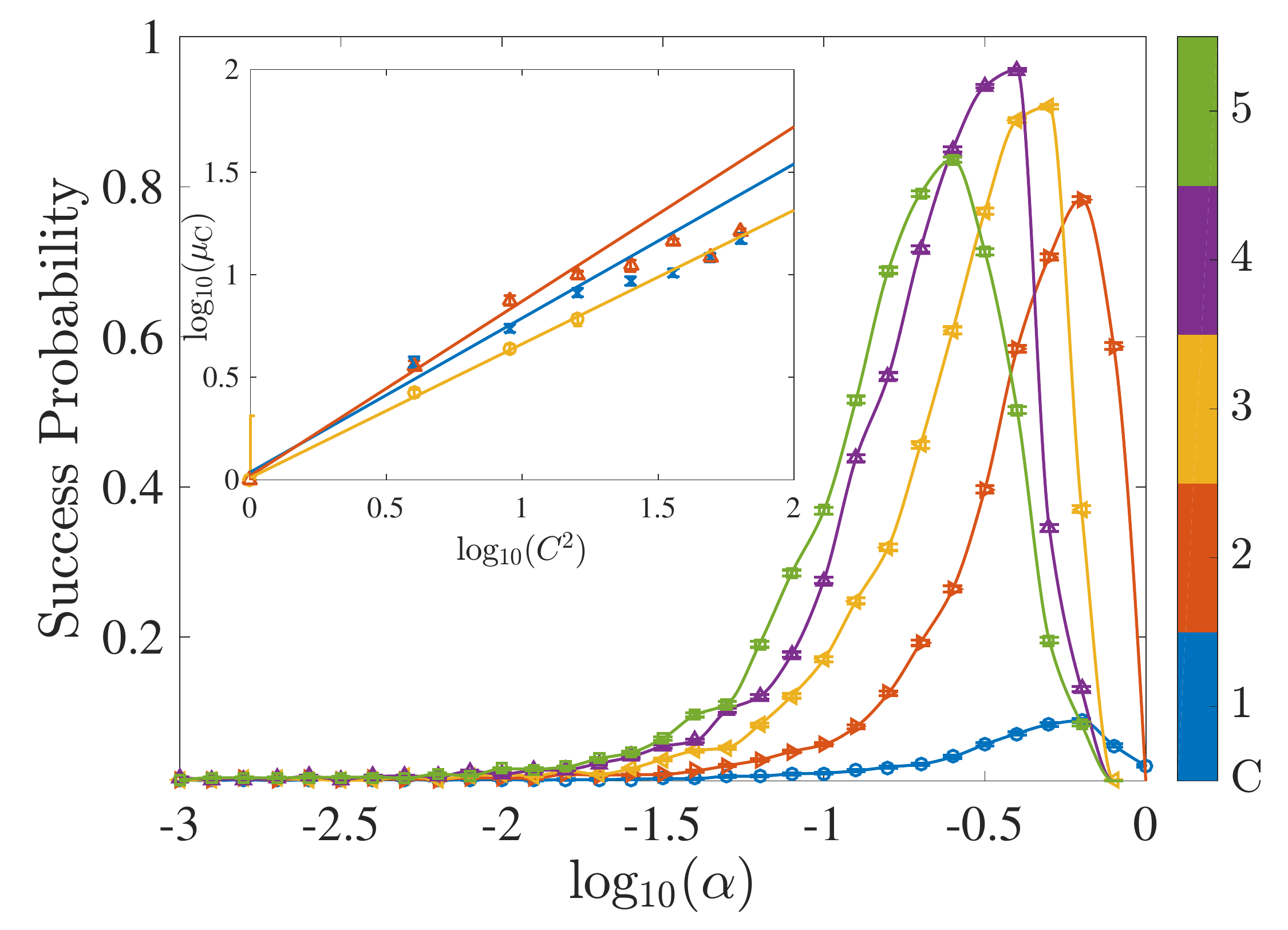}\label{fig:8x8_sqa}}
\caption{Random antiferromagnetic $K_8$: experimental and numerical results. Left: success probabilities $P_C(\alpha)$ for four nesting levels. Middle: success probabilities $P_C'(\alpha)$ adjusted for classical repetition.  Right: numerical results for SQA simulations with $20000$ sweeps, $\sigma = 0.05$ Gaussian noise on the couplings, and with the Choi embedding, showing five nesting levels. Inset: scaling of the energy boost $\mu_C$ \textit{vs} the maximal energy boost $\mu_C^{\max}$, for both the DW2 and SQA. Yellow circles: DW2 results. Blue crosses and red up-triangles: SQA for the Choi ME with $10000$ (crosses) and $20000$ (up-triangles) sweeps, and with $\sigma = 0.05$ Gaussian noise on the couplings. The flattening of $\mu_C$ for $C>4$ suggests that the energy boost becomes less effective at large $C$, but increasing the number of sweeps recovers the effectiveness.  The lines represent best fits to only the first four data points, with respective slopes $\eta/2=0.65$, $0.75$, and $0.85$.} 
\label{fig:exp-k8-nesting}
\end{center}
\end{figure*}
%%%%%%

%%%%%%%%%%%%%%%%%%%%%%%%%%%%%%%%%%%%%%%%%
\section{Results}
%%%%%%%%%%%%%%%%%%%%%%%%%%%%%%%%%%%%%%%%%
%
%%%%%%%%%%%%%%%%%%%%%%%%%%%%%%%%%%%%%%%%%
  \textbf{Free energy} --
%%%%%%%%%%%%%%%%%%%%%%%%%%%%%%%%%%%%%%%%%
 Using a mean-field analysis similar to the approach pursued in Ref.~\cite{MNAL:15} we can compute the partition function associated with the nested Hamiltonian $A(t)H_X + B(t)\tilde H_{\mathrm{P}}$ for the case with uniform antiferromagnetic couplings. This leads to the following free energy density in the low temperature and thermodynamic limits (see Appendix~\ref{sec:MeanField}):
\bea
\hspace{-.5cm} \beta F = C^2 \beta \left( \sqrt{\left[ A(t)/C \right]^2  + \left[ 2 {\gamma}  B(t) m  \right]^2 }  - {\gamma} B(t) m^2 \right)
\label{eq:freeE}
\eea
where $m$ is the mean-field magnetization. There are two key noteworthy aspects of this result. First, the driver term is rescaled as $A(t) \mapsto C^{-1} A(t)$. This shifts the crossing  between the $A$ and $B$ annealing schedules to an earlier point in the evolution and is related to the fact that QAC encodes only the problem Hamiltonian term proportional to $B(t)$.  Consequently the quantum critical point is moved to earlier in the evolution, which benefits QAC since the effective energy scale at this new point is higher~\cite{MNAL:15}.  Second, the inverse temperature is rescaled as $\beta \mapsto C^2 \beta$. This corresponds to an effective temperature reduction by $C^2$, a manifestly beneficial effect. The same conclusion, of a lower effective temperature, is reached by studying the numerically computed success probability associated with thermal distributions (see Appendix~\ref{sec:Num_Add}). We shall demonstrate that this prediction is born out by our experimental results, though it is masked to some extent by complications arising from the ME and noise. \\

%%%%%%%%%%%%%%%%%%%%%%%%%%%%%%%%%%%%%%%%%
 \textbf{NQAC results} -- 
%%%%%%%%%%%%%%%%%%%%%%%%%%%%%%%%%%%%%%%%%
The hardness of an Ising optimization problem, using a QA device, is controlled by its size $N$ as well as by an overall energy scale $\alpha$ \cite{q-sig2}. The smaller this energy scale, the higher the effective temperature and the more susceptible QA becomes to (dynamical and thermal) excitations out of the ground state and misspecification noise on the problem Hamiltonian. This provides us with an opportunity to test NQAC. Since in our experiments we were limited by the largest complete graph that can be embedded on the DW2 device, a $K_{32}$ (see Appendix~\ref{sec:two-MEs} for details), we tuned the hardness of a problem by studying the performance of NQAC as a function of $\alpha$ via $H_{\mathrm{P}} \mapsto \alpha H_{\mathrm{P}}$, with $0<\alpha\leq 1$. Note that we did not rescale $\gamma$; instead $\gamma$ was optimized for optimal post-decoding performance (see Appendix~\ref{sec:Exp_Add}). It is known that for the DW2, intrinsic coupler control noise can be taken to be Gaussian with standard deviation $\sigma\sim 0.05$ of the maximum value for the couplings \cite{King:2015zr}. Thus we may expect that, without error correction, Ising problems with $\alpha \lesssim 0.05$ are dominated by control noise.

We applied NQAC to completely antiferromagnetic ($h_i=0$ $\forall i$) Ising problems over $K_4$ ($J_{ij} = 1$ $\forall i,j$), and $K_8$ (random $J_{ij} \in [0.1,1]$ with steps of 0.1) with nesting up to $C=8$ and $C=4$, respectively.
% (the DW2 QA device used in our experiments accommodated complete graph embeddings of size up to $K_{32}$; see SI for further details). 
We denote by  $P_C(\alpha)$ the probability to obtain the logical ground state at energy scale $\alpha$ for the $C$-level nested implementation (see Appendix~\ref{sec:Exp_Meth} for data collection methods). The experimental QA data in Fig.~\ref{fig:exp-k4-nesting} (left) shows a monotonic increase of $P_C(\alpha)$ as a function of the nesting level $C$ over a wide range of energy scales $\alpha$.  As expected, $P_C(\alpha)$ drops from $P_C(1) = 1$ (solution always found) to $P_C(0) = 6/16$ (random sampling of $6$ ground states, where $4$ out of the $6$ couplings are satisfied, out of a total of $16$ states). 

%The DW2 operates at $17$mK$\approx 2.2$GHz; we have $B(t_f) \approx 20$GHz [recall that $B(t)$ is one of the annealing schedules in Eq.~\eqref{eq:encoded}], so the effective temperature at the final time $t_f$ is $k_B T \sim 0.1$ in units of the largest allowed couplings (corresponding to $\alpha = 1$ in dimensionless units), and the intrinsic coupler control noise can be taken to be Gaussian with $\sigma\sim 0.05$ of the maximum value for the couplings \cite{King:2015zr}. This coincides with the drop of $P_1(\alpha)$ (no nesting) by $\sim50\%$ when $\alpha \sim 0.1$. 
Note that $P_1(\alpha)$ (no nesting) drops by $\sim50\%$ when $\alpha \sim 0.1$, which is consistent with the aforementioned $\sigma\sim 0.05$ control noise level, while 
%On the other hand, 
$P_8(\alpha)$ exhibits a similar drop only when $\alpha \sim 0.01$. This suggests that NQAC is particularly effective in mitigating the two dominant effects that limit the performance of quantum annealers: thermal excitations and control errors. To investigate this more closely, the middle panel of Fig.~\ref{fig:exp-k4-nesting} shows that the data from the left panel can be collapsed via $P_C(\alpha) \mapsto P_C(\alpha/\mu_C)$, where $\mu_C$ is an empirical rescaling factor 
%determined to be $\mu_C \sim C^{1.37}$ 
discussed below (see also Appendix~\ref{sec:mu_C}). This implies that $P_1(\mu_C \alpha)\approx P_C(\alpha)$, and hence that the performance enhancement obtained at nesting level $C$ can be interpreted as an energy boost $\alpha \mapsto  \mu_C \alpha$ with respect to an implementation without nesting.

The existence of this energy boost is a key feature of NQAC, as anticipated above. Recall [Eq.~\eqref{eq:nesting}] that a nested graph $K_{C\times N}$ contains $C^2$ equivalent copies of the same logical coupling $J_{ij}$. Hence a level-$C$ nesting before ME can provide a maximal energy boost $\mu_C^{\max}=C^{\eta^{\max}}$, with $\eta^{\max}=2$. This simple argument agrees with the reduction of the effective temperature by $C^2$ based on the calculation of the free energy~\eqref{eq:freeE}. The right panel of Fig.~\ref{fig:exp-k4-nesting} shows $\mu_C$ as a function of $\mu_C^{\max}$, yielding $\mu_C \sim C^\eta$ with $\eta \approx {1.37}$ (purple circles).
To understand why $\eta < \eta^{\max}$, we performed simulated quantum annealing (SQA) simulations (see Appendix~\ref{sec:Num_Meth} for details).  We observe in Fig.~\ref{fig:exp-k4-nesting} (right) that without ME and control errors, the boost scaling matches $\mu_C^{\max}$ (blue stars). When including ME and control errors a performance drop results (red triangles). Both factors thus contribute to the sub-optimal energy boost observed experimentally. However, the optimal energy boost is recovered for a fully thermalized state with a sufficiently large penalty (see Appendix~\ref{sec:Num_Add}). To match the experimental DW2 results using SQA we replace the Choi ME designed for full Chimera graphs \cite{Choi2} by the heuristic ME designed for Chimera graphs with missing qubits \cite{Cai:2014nx,Boothby2015a}, and achieve a near match (yellow triangles) (see Appendix~\ref{sec:two-MEs} for more details on ME). \\

%%%%%%%%%%%%%%%%%%%%%%%%%%%%%%%%%%%%%%%%%
%\emph{NQAC \textit{vs} classical parallelism.}---% 
%\label{sec:Nest}
 \textbf{Performance of NQAC \textit{vs} classical repetition} -- 
%%%%%%%%%%%%%%%%%%%%%%%%%%%%%%%%%%%%%%%%%
%
Recall that $N^{\mathrm{phys}}_{C} = CNL$ is the total number of physical qubits used at nesting level $C$; let ${C_{\max}}$ denote the highest nesting level that can be accommodated on the QA device for a given $K_N$, i.e., $C_{\max}NL\leq N_{\mathrm{tot}} <  (C_{\max}+1)NL$, where $N_{\mathrm{tot}}$ is the total number of physical qubits ($504$ in our experiments). Then $M_C = \lfloor N^{\mathrm{phys}}_{C_{\max}}/N^{\mathrm{phys}}_{C}\rfloor$ is the number of copies that can be implemented in parallel. For NQAC at level $C$ to be useful, it must be more effective than a classical repetition scheme where $M_C$ copies of the problem are implemented in parallel. If a single implementation has success probability $P_C(\alpha)$, the probability to succeed at least once with $M_C$ statistically independent implementations is $P_C'(\alpha) = 1-[1-P_C(\alpha)]^{M_C}$. 
It turns out that the antiferromagnetic $K_4$ problem, for which a random guess succeeds with probability $6/16$, is too easy [i.e., $P_{C}'(\alpha)$ approaches $1$ too rapidly], and we therefore consider a harder problem: an antiferromagnetic $K_8$ instance with couplings randomly generated from the set $J_{ij} \in \{0.1,0.2,\dots,0.9,1\}$ (see Appendix~\ref{sec:Exp_Add} for more details and data on this and additional instances). Problems of this type turn out to have a sufficiently low success probability for our purposes, and can still be nested up to $C=4$ on the DW2 processor. 

Results for  $P_C(\alpha)$ are shown in Fig.~\ref{fig:exp-k8-nesting} (left), and again increase monotonically with $C$, as in the $K_4$ case. For each $C$, $P_C(\alpha)$ peaks at a value of $\alpha$ for which the maximum allowed strength of the energy penalties $\gamma = 1$ is optimal ($\gamma>1$ would be optimal for larger $\alpha$, as shown in Appendix~\ref{sec:Exp_Add}; the growth of the optimal penalty with problem size, and hence chain length, is a typical feature of minor-embedded problems \cite{Venturelli:2014nx}). An energy-boost interpretation of the experimental data of Fig.~\ref{fig:exp-k8-nesting} is  possible for $\alpha$ values to the left of the peak; to the right of the peak, the performance is hindered by the saturation of the energy penalties.
% (see SI, Sec.~\ref{sec:Exp_Add}).  

Figure~\ref{fig:exp-k8-nesting} (middle) compares the success probabilities $P_C'(\alpha)$ adjusted for classical repetition, where we have set ${C_{\max}}=4$, and shows that $P_2'(\alpha) > P_1'(\alpha)$, i.e., even after accounting for classical parallelism $C=2$ performs better than $C=1$. However, we also find that $P_4'(\alpha)< P_3'(\alpha) \leq P_2'(\alpha)$, so no additional gain results from increasing $C$ in our experiments. This can be attributed to the fact that even the $K_8$ problem still has a relatively large $P_1(\alpha)$.  Experimental tests on QA devices with more qubits will thus be important to test the efficacy of higher nesting levels on harder problems. 

To test the effect of increasing $C$, and also to study the effect of varying the annealing time, we present in Fig.~\ref{fig:exp-k8-nesting} (right) the performance of SQA on a random $K_8$ antiferromagnetic instance with the Choi ME.  The results are qualitatively similar to those observed on the DW2 processor with the heuristic ME [Fig.~\ref{fig:exp-k8-nesting} (left)].  Interestingly, we observe a drop in the peak performance at $C=5$ relative to the peak observed for $C=4$.  We attribute this to both a saturation of the energy penalties and a suboptimal number of sweeps. The latter is confirmed in Fig.~\ref{fig:exp-k8-nesting} (right, inset), where we observe that the scaling of $\mu_C$ with $C$ is better for the case with more sweeps, i.e., again $\mu_C\sim C^\eta$, and $\eta$ increases with the number of sweeps.\\

%%%%%%%%%%%%%%%%%%%%%%%%%%%%%%%%%%%%%%%%%
\section{DISCUSSION}
%%%%%%%%%%%%%%%%%%%%%%%%%%%%%%%%%%%%%%%%%
%
 Nested QAC offers several significant improvements over previous approaches to the problem of error correction for QA. It is a flexible method that can be used with any optimization problem, and allows the construction of a family of codes with arbitrary code distance. We have given experimental and numerical evidence that nesting is effective by performing studies with a D-Wave QA device and numerical simulations.  We have demonstrated that the protection from errors provided by NQAC can be interpreted as arising from an  increase (with nesting level $C$) in the energy scale at which the logical problem is implemented.  This represents a very useful tradeoff: the effective temperature drops as we increase the number of qubits allocated to the encoding, so that these two resources can be traded. Thus NQAC can be used to combat thermal excitations, which are the dominant source of errors in QA, and are the bottleneck for scalable QA implementations. We have also demonstrated that an appropriate nesting level can outperform classical repetition with the same number of qubits, with improvements to be expected when next-generation QA devices with larger numbers of physical qubits become available. We, therefore, believe that our results are of immediate and near-future practical use, and constitute an important step toward scalable QA.\\

%%%%%%%%%%%%%%%%%%%%%%%%%%%%%%%%%%%%%%%%%
\section*{ACKNOWLEDGEMENTS}
%%%%%%%%%%%%%%%%%%%%%%%%%%%%%%%%%%%%%%%%%
%
\noindent We thank Prof.~Hidetoshi Nishimori and Dr.~Shunji Matsuura for valuable comments, and Dr.~Aidan Roy for providing the minor embeddings used in the experiments with the D-Wave Two. Access to the D-Wave Two was made available by the USC-Lockheed Martin Quantum Computing Center.  Part of the computing resources were provided by the USC Center for High Performance Computing and Communications.  This work was supported under ARO grant number W911NF-12-1-0523, ARO MURI Grant Nos. W911NF-11-1-0268 and W911NF-15-1-0582, and NSF grant number INSPIRE-1551064.\\

\appendix

\newpage
\onecolumngrid

%%%%%%%%%% Merge with supplemental materials %%%%%%%%%%
\pagebreak
\widetext

%%%%%%%%%% Merge with supplemental materials %%%%%%%%%%

\section{Experimental Methods}
\label{sec:Exp_Meth}

We tested NQAC on the DW2 quantum annealing device at the University of Southern California's Information Sciences Institute (USC-ISI), which has been described in numerous previous publications (e.g., see \cite{q-sig2}). The largest complete graph that can be embedded on this device, featuring $504$ active qubits, is a $K_{32}$. 

\begin{figure}[h]
{\includegraphics[width=0.45\textwidth]{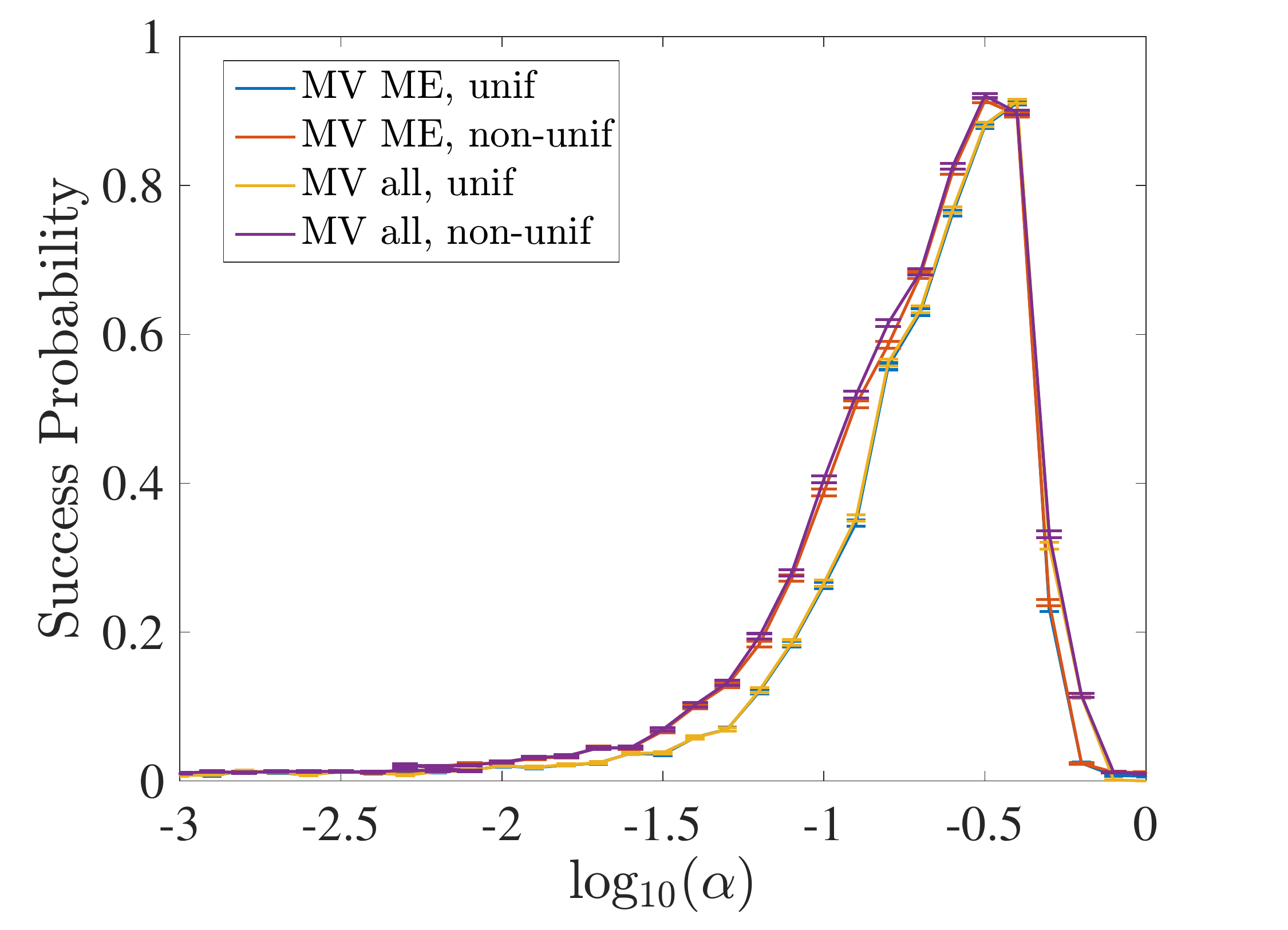}}
\caption{Effect of separately optimizing $\gamma$ for ME and penalties. The plot shows the success probability from SQA simulations, for NQAC applied to a random antiferromagnetic $K_8$ with $10,000$ sweeps, $\sigma = 0.05$ noise, Choi embedding, with $\beta=0.1$. The results obtained after separately optimizing the penalty for the nesting and for the ME are denoted ``non-unif", while the results for using a single penalty for both (the strategy used in the main text) is denoted ``unif". The former results in a small improvement. Also shown is that separate (``MV ME") or joint (``MV all") majority vote decoding of the nesting and the ME has no effect.}
\label{fig:opt-gamma-sep}
\end{figure}

We determined an experimental value of the success probability $P_C(\alpha,\gamma)$ as a function of the energy penalty strength $\gamma$. All figures show, whenever the $\gamma$ dependence is not explicitly considered, the optimal value $P_C(\alpha) = \max_\gamma P_C(\alpha,\gamma)$, with $\gamma \in \{0.05,0.1,0.2,\dots,0.9,1  \}$. We used the same penalty value for both the nesting and the ME. In principle these two values can be optimized separately for improved performance, but we did not pursue this here, since the resulting improvement is small, as shown in Fig.~\ref{fig:opt-gamma-sep}, and costly since each instance needs to be rerun at all penalty settings.

Each $P_C(\alpha,\gamma)$ is the overall success probability after $2\times 10^4$ annealing runs obtained by implementing $20$ programming cycles of $10^3$ runs each. A sufficiently large number of programming cycles is necessary to average out  intrinsic control errors (ICE) that, as explained in the main text,  prevent the physical couplings to be set with a precision better than $\sim 5\%$. To further remove possible sources of systematic noise, at each programming cycle we perform a random gauge transformation on the values of the physical qubits. A permutation of the $C\times N$ vertices is a symmetry of the nested graph but it is not a symmetry of the encoded Hamiltonian obtained after ME. This is because the $C\times N$ chains of physical qubits are physically distinguishable. In each programming cycle we also then performed a random permutation of the vertices of the nested graph, before proceeding to the ME. Error bars correspond to the standard error of the mean of the $20$ $P_C(\alpha)$ values. 

%%%%%%%%%%%%%%%%%%%%%%%%%

%%%%%%%%%%%%%%%%%%%%%
\section{Mean Field Analysis of the Partition Function}
%%%%%%%%%%%%%%%%%%%%%
%
\label{sec:MeanField}
In this section we sketch how to compute the partition function of the logical problem [Eq.~\eqref{eq:encoded} of the main text], in order to analyze the effect of nesting. Full details will be given in a subsequent publication \cite{ALMNV}. 
%Our goal is to demonstrate the conditions under which the empirically observed relation
%\beq
%P_C(\alpha) = P_1(\mu_C \alpha)\ ,
%\label{eq:scaling}
%\eeq
%for the success probabilities relating $K_{C\times N}$ and $K_N$ holds.  
In the main text we were concerned with Hamiltonians of the form 
\begin{align}
H= B(t)(H^x + H^z)
\label{eq:H/J}
\end{align}
where 
\bes
\begin{align}
H^x  &= [A(t)/B(t)] H_X = - \Gamma(t)  \sum_{i=1}^N\sum_{c_i=1}^C \sigma^x_{ic_i}   \\
H^z &= \bar{H}_P = \sum_{i,j=1}^N\sum_{c_i,c'_j=1}^C J_{(ic_i),(jc'_j)}\sigma^z_{ic_i}\sigma^z_{jc'_j} \\
& =  \frac{J}{N} \sum_{i\neq j} \sum_{c_i,c'_j=1}^C \sigma^z_{ic_i}\sigma^z_{jc'_j} - \gamma \sum_{i=1}^N \sum_{c_i\neq c'_i} \sigma^z_{ic_i}\sigma^z_{ic'_i} \ ,
\end{align}
\ees
$A(t),B(t)$ have dimensions of energy, and where $J$ and $\gamma$ are dimensionless, and have each absorbed a factor of $1/2$ to account for double counting. Note that both $H^x$ and $H^z$ are extensive (proportional to $N$). 
Throughout we use $\sigma_{ic}^z \equiv \sigma^z_{ic_i}$ ($\sigma_{ic}^x \equiv \sigma^x_{ic_i}$) to denote the Pauli $z$ ($x$) operator acting on physical qubit $c$ of encoded qubit $i$. 

We define the collective variables
\beq
S^x_i \equiv \frac{1}{C}\sum_{c_i=1}^C \sigma^x_{ic_i}\ ,\qquad S^z_i \equiv \frac{1}{C} \sum_{c_i=1}^C \sigma^z_{ic_i} \ ,\qquad S^x \equiv \frac{1}{N}\sum_{i=1}^N S^x_i\ ,\qquad S^z \equiv \frac{1}{N}\sum_{i=1}^N S^z_i \ .
\eeq
%Note that both $S^\alpha_i = O(1)$ and $S^\alpha = O(1)$. The $1/N$ normalization is needed in order to get extensivity to work out correctly below, while the $1/C$ normalization is needed to preserve the correct scaling with $C$ once we introduce the Hubbard-Stratonovich fields. 
%
We can interpret $S^{x}_i$ and $S^{z}_i$ as the mean transverse and longitudinal fields on logical qubit $i$, respectively. Then
\beq
H^x  = - \Gamma(t) C \sum_{i=1}^N S^x_i = - N C \Gamma(t) S^x \ ,
\eeq
and
\beq
\bar{H}_P = \frac{J}{N}C^2\sum_{i,j}S^z_i S^z_j - \left( \frac{J}{N}+\gamma\right)\sum_{i=1}^N\sum_{c_i, c'_i}\sigma^z_{ic_i}\sigma^z_{ic'_i} + \gamma \sum_{i=1}^N \sum_{c_i} (\sigma^z_{ic_i})^2 \ ,
\label{eq:8}
\eeq
but the last term is a constant [equal to $\gamma N C\openone$], so it can be ignored. Therefore, up to a constant we have
\begin{align}
\bar{H}_P & = JN C^2 \left((S^z)^2 - \lambda \frac{1}{N}\sum_{i=1}^N (S^z_i)^2 \right) \ ,
\label{eq:barH_P}
\end{align}
where 
\beq
\lambda = \frac{\gamma}{J}+\frac{1}{N} \geq 0 \ ,
\label{eq:lambda}
\eeq 
encodes the penalty strength; the $1/N$ correction will disappear in the thermodynamic limit. Note that $\frac{1}{N}\sum_{i=1}^N (S^z_i)^2 =O(1)$, so that $\lambda \frac{1}{N}\sum_{i=1}^N (S^z_i)^2 = O(1)$, like $(S^z)^2$, and hence $\bar{H}_P$ is extensive in $N$, as it should be. 

The form~\eqref{eq:barH_P} for $\bar{H}_P$ shows that the NQAC Hamiltonian in the fully antiferromagnetic $K_{N\times C}$ case can be interpreted as describing the collective evolution of all logical qubits. The term $\lambda \sum_{i=1}^N (S^z_i)^2$ favors all the spins of each logical qubit (where by spin we mean the qubit at $t=t_f$) being aligned, since this maximizes each summand.

\subsection{Partition Function Calculation}
%%%%%%
We are interested in the partition function 
\beq
Z = \Tr \ e^{-\beta H} = \Tr \ e^{-\beta B(t)[H^x + H^z]} = \Tr \ e^{-\theta [H^x + H^z]} \ ,
\label{eq:Z}
\eeq
where $\theta = \beta B(t)$ is the dimensionless inverse temperature.  We write the partition function explicitly as~\cite{Seoane:2012uq}
\beq
Z =  \sum_{ \{ \sigma^z \} } \bra{\{ \sigma^z \}} \exp  \left[ - \theta  \left( H^z + H^x \right) \right] \ket{\{ \sigma^z \}} = \lim_{M\to\infty} Z_M ,
\eeq
where $\sum_{ \{ \sigma^z \} }$ is a sum over all possible $2^{CN}$ spin configurations in the $z$ basis, and $\ket{\{ \sigma^z \}} = \otimes_{i=1}^N \otimes_{c=1}^C \ket{\sigma_{ic}^z}$. $Z_M$ is determined using the Trotter-Suzuki formula $e^{A+B} = \lim_{M\rightarrow \infty} \left(e^{A/M}e^{B/M}\right)^M$:
\beq
Z_M=   \sum_{ \{ \sigma^z \} } \bra{\{ \sigma^z \}} \left(   \exp \left[  -\frac{\theta}{M} H^z \right] \exp  \left[ -\frac{\theta}{M} H^x \right]  \right)^M \ket{\{ \sigma^z \}} .
\label{eq:8b}
\eeq
%
%%%%begin ignore
\ignore{
We introduce $M$ copies of the identity operator closure relations $I(\al) = \sum_{\{ \sigma^z(\al) \}} \ketbra{\{ \sigma^z(\al) \}}{\{ \sigma^z(\al) \}}$, each labeled by the Trotter time $\al$: 
\bes
\begin{align}
Z_M &=   \sum_{ \{ \sigma^z \} } \bra{\{ \sigma^z \}} \prod_{\al = 1}^{M }\left(   \exp \left[  -\frac{\theta}{M} H^z \right] I(\al) \exp  \left[ -\frac{\theta}{M} H^x \right]  \right) \ket{\{ \sigma^z \}} \\
& = \sum_{ \{ \sigma^z (1) \} } \bra{\{ \sigma^z (1)\}} \prod_{\al = 1}^{M }\left(   \exp \left[  -\frac{\theta}{M} H^z \right] \sum_{\{ \sigma^z(\al) \}} \ketbra{\{ \sigma^z(\al) \}}{\{ \sigma^z(\al) \}}\exp  \left[ -\frac{\theta}{M} H^x \right]  \right) \ket{\{ \sigma^z (1) \}} \\
&=  \prod_{\al = 1}^{M }\sum_{ \{ \sigma^z(\al) \} } \bra{\{ \sigma^z(\al) \}} \left(    \exp \left[  -\frac{\theta}{M} H^z \right] \exp  \left[ -\frac{\theta}{M} H^x \right]  \right) \ket{\{ \sigma^z(\al+1) \}}\ ,
\label{eq:8c}
\end{align}
\ees
where $\ket{\{ \sigma^z(M+1) \}} \equiv  \ket{\{ \sigma^z(1) \}}$; $M$ is known as the Trotter number. Likewise we introduce $M$ copies of the identity operator closure relations $I(\al) = \sum_{\{ \sigma^x(\al) \}} \ketbra{\{ \sigma^x(\al) \}}{\{ \sigma^x(\al) \}}$:
\bes
\begin{align}
\label{eq:15a}
Z_M= & \prod_{\al = 1}^{M }\sum_{ \{ \sigma^{x,z}(\al) \} }  \bra{\{ \sigma^z(\al) \}}  \exp \left[  -\frac{\theta}{M} H^z \right] \ket{\{ \sigma^x(\al) \}} \bra{\{ \sigma^x(\al) \}} \exp  \left[ -\frac{\theta}{M} H^x \right]  \ket{\{ \sigma^z(\al+1) \}} \\
\label{eq:15b}
= & \prod_{\al = 1}^{M}\sum_{ \{ \sigma^{x,z}(\al) \} }   \exp \left[  -\frac{\theta}{M} H^z(\al) \right]  \braket{\{ \sigma^z(\al) \}}{\{ \sigma^x(\al) \}}  \exp  \left[ -\frac{\theta}{M} H^x (\al) \right]  \braket{\{ \sigma^x(\al) \}}{\{ \sigma^z(\al+1) \}} \\
= & \prod_{\al = 1}^{M}\sum_{ \{ \sigma^{x,z}(\al) \} }   \exp \left[  -\frac{\theta}{M} \left( H^z(\al) +H^x (\al) \right) \right]  \braket{\{ \sigma^z(\al) \}}{\{ \sigma^x(\al) \}}    \braket{\{ \sigma^x(\al) \}}{\{ \sigma^z(\al+1) \}}
\label{eq:15c}
\end{align} 
\ees
The notation $\{ \sigma^{x,z}(\al) \}$ is shorthand for $\{\{ \sigma_{jc'}^x(\al),\sigma_{ic}^z(\al) \}_{c,c'=1}^C\}_{i,j=1}^N$, and 
\beq
\braket{\{ \sigma^z(\al) \}}{\{ \sigma^x(\al) \}}    \braket{\{ \sigma^x(\al) \}}{\{ \sigma^z(\al+1) \}}= \prod_{i,j=1}^{N} \prod_{c,c'=1}^{C} \langle \sigma^{z}_{ic}(\al)|\sigma^{x}_{jc'}(\al)\rangle \langle \sigma^{x}_{jc'}(\al) | \sigma^{z}_{ic}(\al+1) \rangle .
\eeq
Note that this allowed us to replace the operators $H^x$ and $H^z$ by c-numbers $H^x(\al)$ and $H^z(\al)$.

%We now consider the model from the main text, but generalize it to the $p$-body case:
%\bes
%\begin{align}
%H^z_p(\al) = NJ\left([S^z(\al)]^p - \lambda \sum_{i=1}^N [S^z_i(\al)]^p \right) \ .
%\\
%H^x(\al) = & - \Gamma  N \sum_{k=1}^K S^x_c (\al) \ .
%\end{align}
%\ees

We now consider the model from Eq.~\eqref{eq:barH_P}:
\bes
\begin{align}
H^z(\al) &= JNC^2 \left(\left[\frac{1}{N}\sum_{i=1}^N S^z_i(\al)\right]^2 - \lambda \frac{1}{N}\sum_{i=1}^N [S^z_i(\al)]^2 \right) \ ,
\\
H^x(\al) = & - N C \Gamma S^x(\al) = - C \Gamma   \sum_{i=1}^N S^x_i (\al) \ .
\end{align}
\ees
%where we chose to write $S^z(\al)$ explicity as $\frac{1}{N}\sum_{i=1}^N S^z_i(\al)$ for reasons that will become apparent below, having to do with extensivity.

We can rewrite the interaction in terms of one-body interactions by introducing auxiliary Hubbard-Stratonovich fields $m_{i \al}$ and  $m'_{i \al}$. This is done by successively using the elementary $\delta$ function identities
\begin{align}
f(a) = \int_{-\infty}^{\infty} f(m_{j \al}) \delta(m_{j \al}-a)\ dm_{j \al} , \quad \delta(m_{j \al}-a) = \frac{1}{2\pi} \int_{-\infty}^{\infty} e^{i (m_{j \al}-a) m'_{j \al}}\ dm'_{j \al} \ .
\end{align}
In this manner the $m_{i \al}$ and  $m'_{i \al}$ play the role of order parameters and Lagrange multipliers, respectively.

Consider first $\exp\left[-\frac{\theta}{M}H^z(\al)\right]$  from Eq.~\eqref{eq:15b}:
%\footnote{Note that by defining the fields $m_{i,\alpha}$ as $S^z_{i,\alpha}$ we have made a choice of the physical mean-field variables. This is important since it means that we have effectively coarse-grained the description and no longer have access to the physical qubits comprising the logical qubits. Essentially this is a $\gamma \gg J$ approximation. To see this, note that for $\gamma=0$ we simply have $C$ decoupled copies of the problem.}
\bes
\begin{align}
e^{-\frac{\theta}{M}H^z(\al)} & = e^{\frac{N}{M}\theta J C^2 \left(\lambda \frac{1}{N}\sum_{i=1}^N [S^z_i(\al)]^2 - \left[\frac{1}{N}\sum_{i=1}^N S^z_i(\al)\right]^2\right)}  \\
\label{eq:19b}
& =  \int dm_{1,\al} \delta[m_{1,\al} - S^z_1(\al)] \cdots \int dm_{N,\al} \delta[m_{N,\al} - S^z_N(\al)] e^{\frac{N}{M}\theta J C^2 \left(\lambda \frac{1}{N}\sum_{i=1}^N m_{i,\al}^2 - \left[\frac{1}{N}\sum_{i=1}^N m_{i,\al}\right]^2\right)}\\
& = \frac{1}{(2\pi)^{2N}} \int dm_{1,\al} dm'_{1,\al} \cdots \int dm_{N,\al} dm'_{N,\al} e^{i m'_{1,\al} [m_{1,\al} - S^z_1(\al)]} \cdots e^{i m'_{N,\al} [m_{N,\al} - S^z_N(\al)]} \notag \\
& \times e^{\frac{N}{M}\theta J C^2 \left(\lambda\frac{1}{N} \sum_{i=1}^N m_{i,\al}^2 - \left[\frac{1}{N}\sum_{i=1}^N m_{i,\al}\right]^2\right)}  \\
\label{eq:19d}
& \approx \int \md m_{1} \md \tilde{m}_{1}\cdots \int \md m_{N} \md \tilde{m}_{N} e^{i\frac{N}{M} \frac{1}{N}\sum_{j=1}^N \tilde{m}_{j} [m_{j} - S^z_j(\al)]} e^{\frac{N}{M}\theta J C^2\left[\lambda\frac{1}{N} \sum_{j=1}^N m_{j}^2 - \ave{m}^2\right]}
 \ ,
\end{align}
\ees
where 
\beq
\ave{m} \equiv \frac{1}{N}\sum_{j=1}^N m_j
\eeq
and we used the static approximation \cite{Bray:1980fk,PhysRevB.78.134428} in Eq.~\eqref{eq:19d}, i.e., $m_{i, \al} \mapsto m_{i}$ and $m'_{i, \al} \mapsto m'_{i}$. We also made a change of variables $m'_{i} = \frac{1}{M} \tilde{m}_i$, and absorbed a factor of $\frac{1}{(2\pi)^2 M}$ into each differential. As noted before $S^z_i$ is $O(1)$, so our expression has maintained extensivity. Namely, it was important to ensure that the Hubbard-Stratonovich fields remained in the range $[-1,1]$, since once they are introduced they lose the dependence on $N$ or $C$.

Writing $\exp  \left[ -\frac{\theta}{M} H^x (\al) \right] =  \exp  \left[ \frac{1}{M} \theta \Gamma C \sum_{j=1}^N S^x_j (\al) \right] $ and $\int \md m_{1} \md \tilde{m}_{1}\cdots \int \md m_{N} \md \tilde{m}_{N}  = \int \Pi_j \md m_{j} \md \tilde{m}_{j}$
we have
\begin{align}
e^{-\frac{\theta}{M}[H^z(\al)+H^x (\al)]} & \approx  \int \Pi_j \md m_{j} \md \tilde{m}_{j} e^{\frac{N}{M} \left[ \frac{1}{N}\sum_{j=1}^N m_j (i \tilde{m}_{j} +\theta J C^2 \lambda m_j) - \theta J C^2 \ave{m}^2 \right]}  e^{\frac{1}{M}\left[ \sum_{j=1}^N \theta \Gamma C S^x_j(\al) -i \tilde{m}_j S^z_j(\al)  \right]}\ .
\end{align}

We now reinsert $\prod_{\al = 1}^{M}\sum_{ \{ \sigma^{x,z}(\al)\} } \braket{\{ \sigma^z(\al) \}}{\{ \sigma^x(\al) \}}    \braket{\{ \sigma^x(\al) \}}{\{ \sigma^z(\al+1) \}}$ in order to compute the trace. The notation $\prod_{\al = 1}^{M}\sum_{ \{ \sigma^{x,z}(\al)\} }$ is shorthand for an $M$-fold sum $\sum_{ \{ \sigma^{x,z}(1)\}} \cdots \sum_{ \{ \sigma^{x,z}(M)\}}$ so can be moved through the integrals. 
Let $e^{g(\al)/M} \equiv e^{\frac{1}{M}\left[ \sum_{j=1}^N \theta \Gamma C S^x_j(\al) -i \tilde{m}_j S^z_j(\al)  \right]}$ and $e^{f_\al/M} \equiv e^{\frac{N}{M} \left[ \frac{1}{N}\sum_{j=1}^N m_j (i \tilde{m}_{j} +\theta J C^2 \lambda m_j) - \theta J C^2 \ave{m}^2 \right]}$, where we added the subscript $\al$ to $f$ to remind ourselves that in spite of not being explicitly $\al$-dependent, $e^{f_\al/M}$ does belong to the $\al$th term in the product over all $\al$; this is why below, when we multiply these terms, $e^{f_\al/M}$ will become $e^{f}$.  

Now, picking up the thread from Eq.~\eqref{eq:15c}, we have
\bes
\begin{align}
Z_M &=    \int \Pi_j \md m_{j} \md \tilde{m}_{j}\  \prod_{\al = 1}^{M}\sum_{ \{ \sigma^{x,z}(\al) \} }   \exp \left[  -\frac{\theta}{M} \left( H^z(\al) +H^x (\al) \right) \right]  \braket{\{ \sigma^z(\al) \}}{\{ \sigma^x(\al) \}}    \braket{\{ \sigma^x(\al) \}}{\{ \sigma^z(\al+1) \}} \\
& \approx \int \Pi_j \md m_{j} \md \tilde{m}_{j}\ \prod_{\al = 1}^{M}\sum_{ \{ \sigma^{x,z}(\al) \} } e^{f_\al/M} e^{g(\al)/M} \braket{\{ \sigma^z(\al) \}}{\{ \sigma^x(\al) \}}  \braket{\{ \sigma^x(\al) \}}{\{ \sigma^z(\al+1) \}} \\
& = \int \Pi_j \md m_{j} \md \tilde{m}_{j}\ \sum_{ \{ \sigma^{x,z}(1)\}}  e^{f/M} e^{g(1)/M} \cdots \sum_{ \{ \sigma^{x,z}(M)\}} e^{f/M} e^{g(N)/M} \prod_{\al = 1}^{M} \braket{\{ \sigma^z(\al) \}}{\{ \sigma^x(\al) \}}  \braket{\{ \sigma^x(\al) \}}{\{ \sigma^z(\al+1) \}} \ .
\end{align}
\ees
Hence, taking the $M\to\infty$ limit and replacing the subscript by $C$ to indicate that we are now tracking the $C$-dependence of the partition function:
\bes
\begin{align}
Z_C& \approx \int \Pi_j \md m_{j} \md \tilde{m}_{j}\ e^f \Tr e^g \\
& = \int \Pi_j \md m_{j} \md \tilde{m}_{j}e^{N \left[ \frac{1}{N}\sum_{j=1}^N m_j (i \tilde{m}_{j} +\theta J C^2 \lambda m_j) - \theta J C^2 \ave{m}^2 \right]} \Tr e^{ \sum_{j=1}^N \theta \Gamma C S^x_j -i \tilde{m}_j S^z_j }
\end{align}
\ees

The trace can be simplified as follows:
\bes
\begin{align}
\Tr e^{\sum_{j=1}^N \theta \Gamma C S^x_j -i \tilde{m}_j S^z_j} &= \prod_{j=1}^N \Tr e^{ \theta \Gamma C S^x_j -i \tilde{m}_j S^z_j} = \prod_{j=1}^N \prod_{c_j=1}^C \Tr e^{ \theta \Gamma \sigma^x_{j c_j} -i \tilde{m}_j \sigma^z_{j c_j}/C} \\
& = \prod_{j=1}^N \prod_{c_j=1}^C 2\cosh\left((\theta \Gamma)^2 - (\tilde{m}_j/C)^2\right)^{1/2} =  \prod_{j=1}^N \left[2\cosh\left((\theta \Gamma)^2 - (\tilde{m}_j/C)^2\right)^{1/2}\right]^C \\
&= e^{C \sum_{j=1}^N  \ln  \left[2\cosh\left((\theta \Gamma)^2 - (\tilde{m}_j/C)^2\right)^{1/2}\right]}\ .
\end{align}
\ees
}
%%%end ignore

After a lengthy calculation \cite{ALMNV}  
we find
%Combining it all, we have:
%\bes 
\begin{align}
\label{eqt:Zc}
%Z_C & \approx \int \Pi_j \md m_{j} \md \tilde{m}_{j}e^{N \left[ \frac{1}{N}\sum_{j=1}^N m_j (i \tilde{m}_{j} +\theta J C^2 \lambda m_j) - \theta J C^2 \ave{m}^2 \right]} e^{C\sum_{j=1}^N \ln  \left[2\cosh\left((\theta \Gamma)^2 - (\tilde{m}_j/C)^2\right)^{1/2}\right]} \\
%& = 
Z \approx \int \Pi_j \md m_{j} \md \tilde{m}_{j}e^{N \left[ \frac{1}{N}\sum_{j=1}^N \left\{ C \ln  \left[2\cosh\left((\theta \Gamma)^2 - (\tilde{m}_j/C)^2\right)^{1/2}\right] + m_j (i \tilde{m}_{j} +\theta J C^2 \lambda m_j)\right\} - \theta J C^2 \ave{m}^2 \right]} \ ,
\end{align}
%\ees
where  
%\beq
$\ave{m} \equiv \frac{1}{N}\sum_{j=1}^N m_j$,
%\eeq
and where $m_j$ is the Hubbard-Stratonovich field that represents $S^z_j(\al)$ after the static approximation (i.e., dropping the $\al$ dependence) \cite{Bray:1980fk,PhysRevB.78.134428}. The second Hubbard-Stratonovich field $\tilde{m}_j$ acts as a Lagrange multiplier.

%%%%begin ignore
\ignore{
\subsection{Saddle point analysis}

The saddle point condition for ${m}_j$ from Eq.~\eqref{eqt:Zc} is:
\bes
\begin{align}
0 &= N\frac{\partial}{\partial {m}_k}\left[ \frac{1}{N}\sum_{j=1}^N \left\{ C \ln  \left[2\cosh\left((\theta \Gamma)^2 - (\tilde{m}_j/C)^2\right)^{1/2}\right] + i \tilde{m}_{j}m_j  +\theta J C^2 \lambda m_j^2\right\} - \theta J C^2 \ave{m}^2 \right] \\
& = i \tilde{m}_{k} +2 \theta J C^2 \lambda m_k  - 2 \theta J C^2 \ave{m} \ ,
\end{align}
\ees
i.e.,
\beq
\tilde{m}_j = 2i \theta J C^2 \left( \lambda m_j-\ave{m}  \right)  \ .
\label{eq:tildem_k}
\eeq
The saddle point condition for $\tilde{m}_j$ is: 
\bes
\begin{align}
0 &= N\frac{\partial}{\partial \tilde{m}_k}\left[ \frac{1}{N}\sum_{j=1}^N \left\{ C \ln  \left[2\cosh\left((\theta \Gamma)^2 - (\tilde{m}_j/C)^2\right)^{1/2}\right] + i \tilde{m}_{j}m_j  +\theta J C^2 \lambda m_j^2\right\} - \theta J C^2 \ave{m}^2 \right] \\
& = -\frac{1}{C}\frac{\tilde{m}_k}{\left((\theta \Gamma)^2 - (\tilde{m}_k/C)^2\right)^{1/2}} \tanh\left((\theta \Gamma)^2 - (\tilde{m}_k/C)^2\right)^{1/2}  + i m_k  \ ,
\end{align}
\ees
i.e.,
\beq
m_k = -i \frac{\tilde{m}_k}{\sqrt{(\theta \Gamma C)^2 - \tilde{m}_k^2}} \tanh\sqrt{(\theta \Gamma)^2 - (\tilde{m}_k/C)^2} \ .
\label{eq:mk}
\eeq
Substituting $\tilde{m}_k$ from Eq.~\eqref{eq:tildem_k} into Eq.~\eqref{eq:mk}, we have
\beq
m_j =  \frac{2 J C\left( \lambda m_j-\ave{m} \right) }{\sqrt{\Gamma^2 + \left[2 J C \left(\lambda m_j - \ave{m}\right)\right]^2}} \tanh\left[\theta\sqrt{\Gamma^2 +\left[2 J C (\lambda m_j - \ave{m})\right]^2}\right] \ .
\label{eq:m_j}
\eeq
Recall that $\theta = \beta B(t)$ and $\Gamma = A(t)/B(t)$. Let 
\bes
\begin{align}
\label{eq:x_j-def}
x_j &\equiv \lambda m_j - \ave{m}\\
\label{eq:a-def}
a & \equiv 2 J C 
%\\
%= 2\beta B(t) J C\\
%\label{eq:b-def}
%b & \equiv \theta \Gamma = \beta A(t) \ .
\end{align}
\ees
Multiplying Eq.~\eqref{eq:m_j} by $\lambda$ and subtracting $\ave{m}$ we obtain
\beq
x_j = \frac{a\lambda  x_j}{\sqrt{\Gamma^2 + (a x_j)^2}} \tanh\left[\theta\sqrt{\Gamma^2+(a x_j)^2}\right] - \ave{m} \ .
\label{eq:x_j}
\eeq
This is the self-consistency equation for the magnetization.

We focus on solutions of the consistency equation with $\ave{m} = 0$.  Note that if $\Gamma >0$ and $x_j = 0$ $\forall j$ and Eq.~\eqref{eq:x_j} is trivially satisfied. Assuming $x_j\neq 0$ we can solve Eq.~\eqref{eq:x_j} in general by letting $y_j \equiv \sqrt{\Gamma^2+(a x_j)^2}$ and still assuming $\ave{m}=0$. We then have
\beq
y_j =  a\lambda  \tanh(\theta y_j) \ .
\label{eq:y_j}
\eeq
This is again similar to the standard mean field model equation for the magnetization of the fully connected classical Ising model. The derivative of the RHS is $\theta a\lambda/\cosh^2(\theta y_j)$, so there is again a unique paramagnetic-like solution if $\theta a\lambda < 1$, this time resulting in $y_j=\sqrt{\Gamma^2+(a x_j)^2}=0$. This is possible only if both $\Gamma=0$ and $x_j=0$, but we assumed $x_j\neq 0$, so this case is ruled out.

If $\theta a\lambda > 1$ an additional pair of solutions is found at $y_j = \pm y \neq 0$, with $y$ a monotonically increasing function of $a\lambda$ and $\theta$. Thus, since $x_j=\lambda m_j $ when $\ave{m}=0$:
$m_j = \pm \frac{1}{a\lambda }\sqrt{y^2-\Gamma^2}$, provided $y>\Gamma$. When $y\leq \Gamma$ the solution remain $m_j=0$. Thus, summarizing in terms of the original variables:
\bes
\begin{align}
2\beta B(t_f)  C \left({\gamma}+\frac{J}{N}\right) &< 1 \ \textrm{or} \ \beta\to 0 \ \textrm{or} \ \beta A(t) \geq y\ : \quad m_j = 0 \quad \textrm{(paramagnetic phase)}\\
2\beta B(t)  C \left({\gamma}+\frac{J}{N}\right) &> 1 \ : \quad m_j = \pm \frac{1}{2\beta B(t)  C \left({\gamma}+\frac{J}{N}\right) }\sqrt{y^2-[\beta A(t)]^2} \quad \textrm{provided  $\beta A(t) < y$ (ordered phase)}\ .
\end{align}
\ees
Thus, at fixed inverse temperature $\beta$, increasing $C$ or $\gamma$ linearly reduces the arrival time of the ordered phase.
}
%%%%end ignore

\subsection{Free energy}
In the large $\beta$ (low temperature) limit, the partition function is dominated by the global minimum.  This minimum is given by $\ave{m} = 0$, which corresponds to either a paramagnetic phase (all $m_j=0$) or a symmetric phase ($m_j=\pm m$ in equal numbers). It can be shown that the system undergoes a second order QPT, with the critical point moving to the left as $C$ and $\gamma$ grow \cite{ALMNV}.  Using 
%Eq.~\eqref{eq:tildem_k} we have 
a saddle point analysis of the partition function we can show that $\tilde{m}_j = \pm 2i \theta C^2 J \lambda m $, and hence, the dominant contribution to the partition function is given by:
\bes
\begin{align}
Z_C & \approx %\nu_g 
e^{N \left\{  C \ln  \left[2\cosh\left((\theta \Gamma)^2 + (2 \theta J C  \lambda m)^2\right)^{1/2}\right] - \theta J C^2 \lambda m^2  \right\}}\\
& =  %\nu_g 
e^{N \left\{  C \ln  \left[2\cosh\left(\left[\beta A(t)\right]^2 + \left[2 \beta B(t)  C  (\gamma + \frac{J}{N}) m\right]^2\right)^{1/2}\right] - \beta B(t)  C^2 \left(\gamma + \frac{J}{N}\right) m^2  \right\}} \ .
\end{align}
\ees
%where $\nu_g = {N \choose N/2}$ is the ground state degeneracy.
%Thus the free energy density $F$ in $Z = e^{-\beta N F}$ in the large $\beta$ limit is
%\beq
%F =   \frac{1}{\beta} C \ln  \left[2\cosh\left(\left[\beta A(t)\right]^2 + \left[2 \beta B(t)  C  \left(\gamma + \frac{J}{N}\right) m\right]^2\right)^{1/2}\right] +  B(t)  C^2 \left(\gamma + \frac{J}{N}\right) m^2 - \frac{1}{\beta N} \log {N \choose N/2} \ .
%\label{eq:Free}
%\eeq
For $B(t)>0$ and in the low temperature limit ($\theta \gg 1$) we can approximate $2\cosh(\theta |x|)$ as $e^{\theta |x|}$, 
\bes
\begin{align}
Z_C & \approx  e^{N \theta \left\{  \left( (C\Gamma)^2 + (2   J \lambda C^2 m)^2\right)^{1/2} -  J  \lambda C^2 m^2  \right\}} \\
&=  e^{N \beta \left\{  \left( \left[C A(t)\right]^2 + \left[ 2B(t)   (\gamma  +  \frac{J}{N}) C^2 m\right]^2\right)^{1/2} -  B(t)\left(\gamma  +  \frac{J}{N}\right) C^2 m^2  \right\}} =  e^{-\beta N F}\ ,
\label{eq:Z_C}
\end{align}
\ees
where in the second line we reintroduced the physical inverse temperature $\beta$ [recall Eq.~\eqref{eq:Z}]. 
%We can conclude that the introduction of $C$ has the effect of rescaling as follows:
%\bes
%\begin{align}
%A(t) &\mapsto A(t) C\\
%B(t)  & \mapsto   B(t)  C^2 \ .
%\label{eq:29}
%\end{align}
%\ees
%This moves the crossing point between the $A$ and $B$ annealing schedules to the left, which is associated with the minimum gap point arriving earlier in the evolution.
Factoring out $C^2$ and taking the large $N$ limit then directly yields the free energy density expression~\eqref{eq:freeE} given in the main text.

%%%%%%%%%%%%%%%%%%%%%%%%%

\section{Additional Numerical Data}
\label{sec:Num_Add}
\begin{figure*}[ht]
\begin{center}
\subfigure[]{\includegraphics[width=0.45\textwidth]{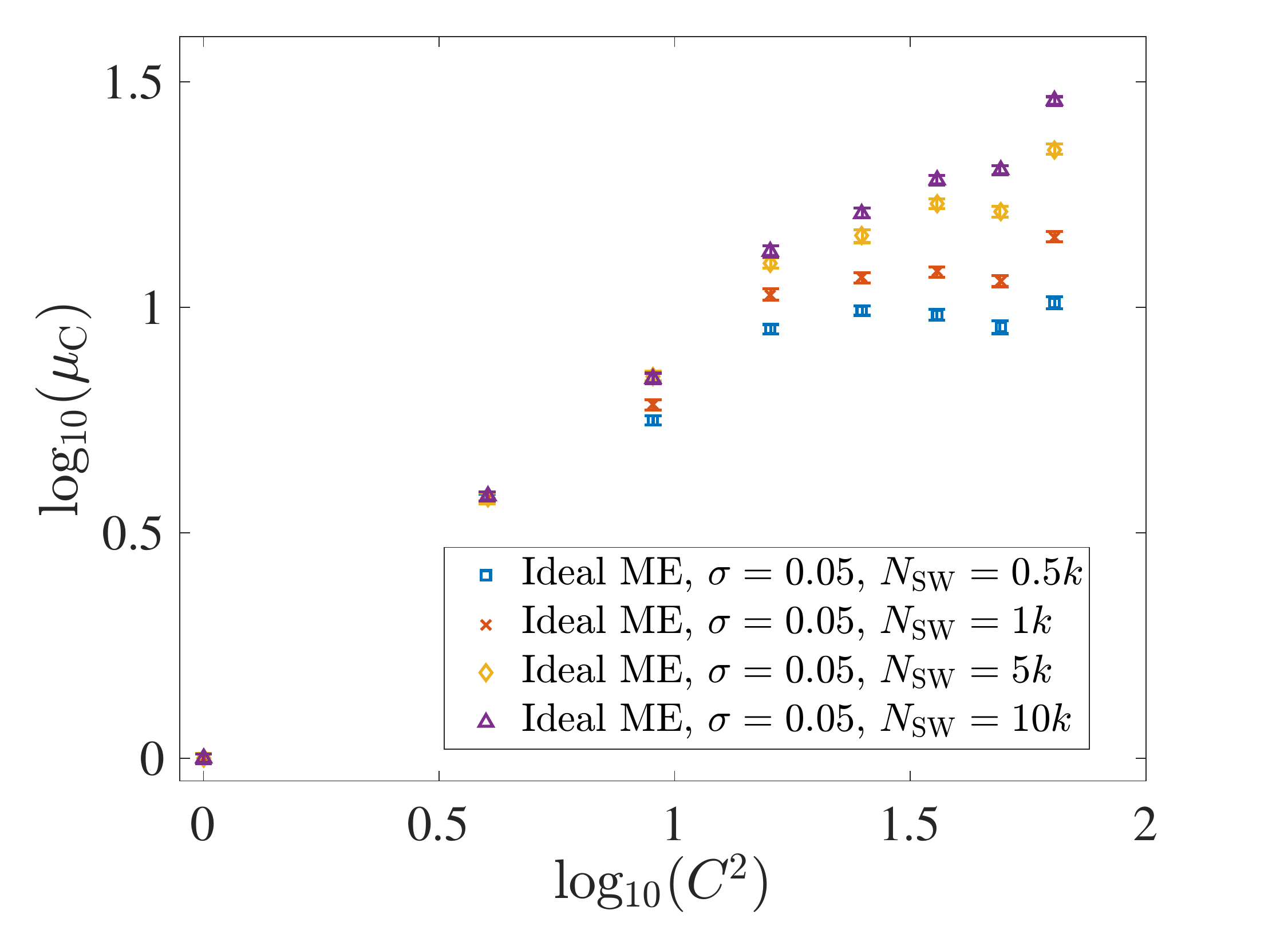}\label{fig:EBoost_SQA2}}
\subfigure[]{\includegraphics[width=0.45\textwidth]{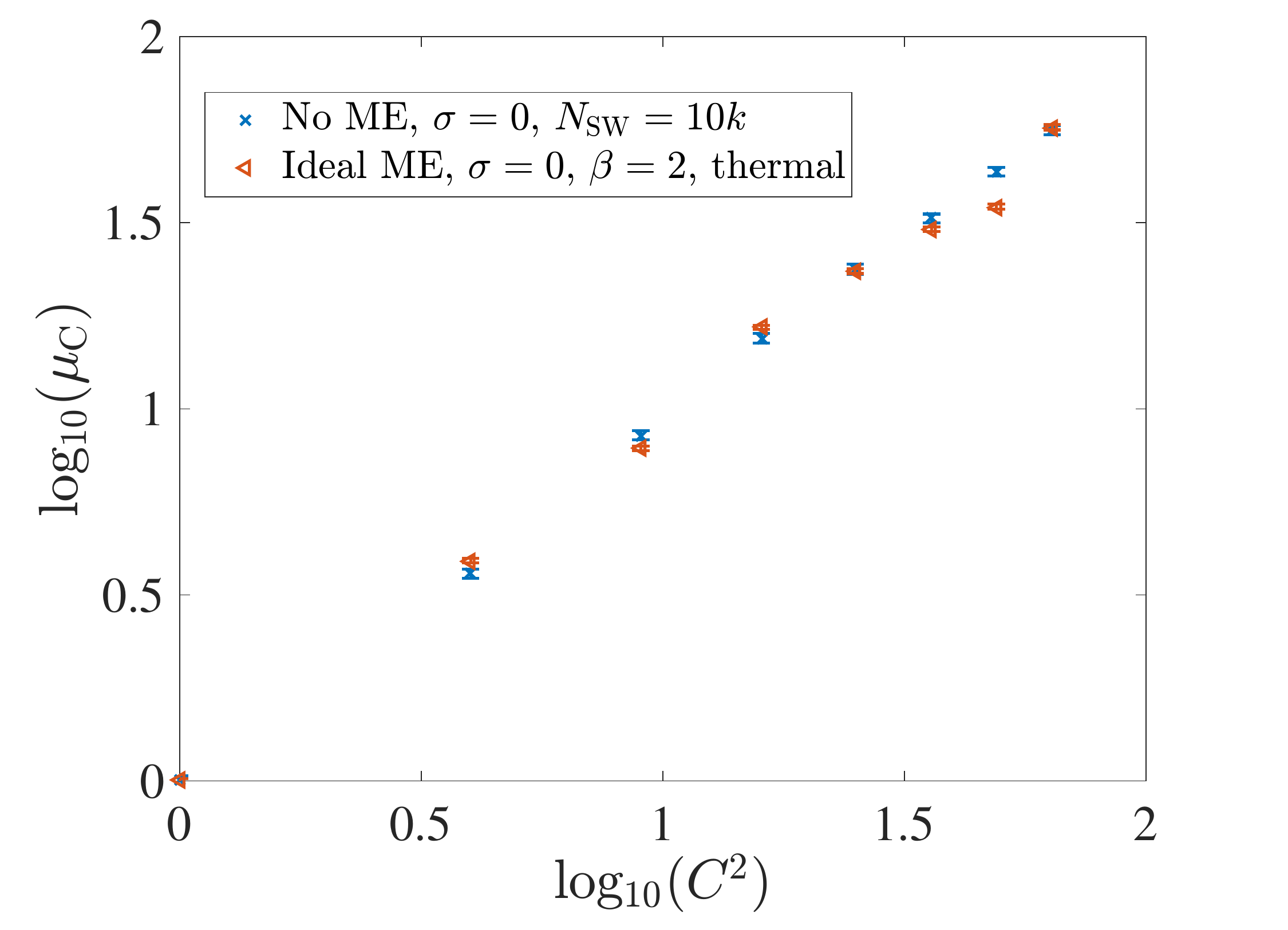}\label{fig:EBoost_SQA3}}
\caption{Saturation removal for NQAC applied to antiferromagnetic $K_4$: (a) SQA results.  As we increase the number of sweeps, the flattening of $\mu_C$ is slowly removed.  (b) Parallel tempering (infinite sweep number) results. A thermal distribution on the ME fully recovers the no-ME scaling.} 
\label{fig:Mu2}
\end{center}
\end{figure*}

Figure~\ref{fig:EBoost_SQA2} shows that the saturation of $\mu_C$ at large $C$ is removed when the number of sweeps is increased. The thermal state, where the system has fully thermalized, can be understood as the limit of an infinite number of sweeps. Figure~\ref{fig:EBoost_SQA3} shows that the saturation is fully removed for the thermal state (generated using parallel tempering), and nesting is then equivalent to an energy (or temperature) boost close to the ideal result $\mu_C^{\max} = C^2$. This suggests that for a sufficiently large sweep number, performance can be brought to near the ideal result.

\begin{figure*}[ht]
\begin{center}
\subfigure[]{\includegraphics[width=0.45\textwidth]{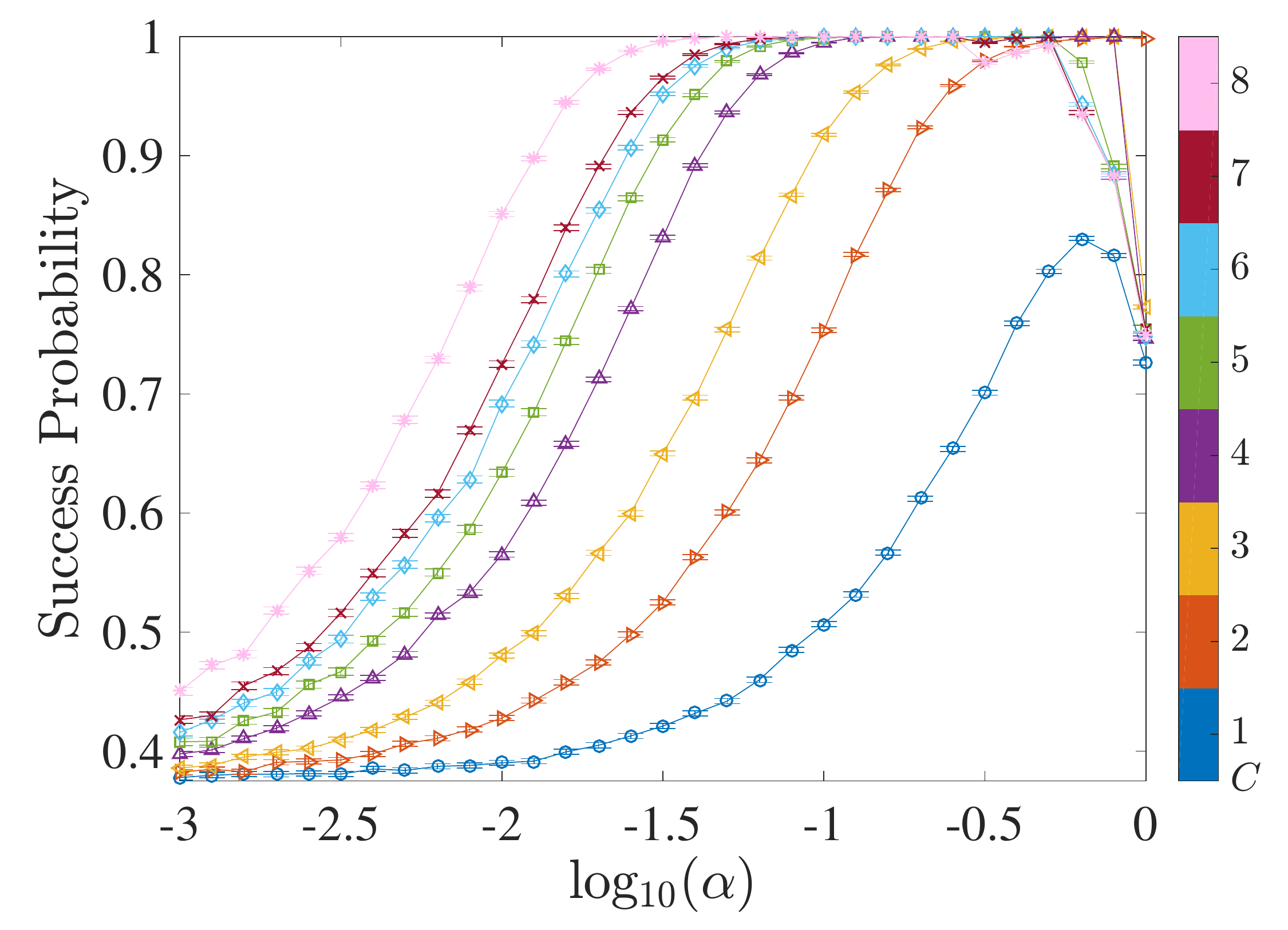}\label{fig:EBoost_PT1}}
\subfigure[]{\includegraphics[width=0.45\textwidth]{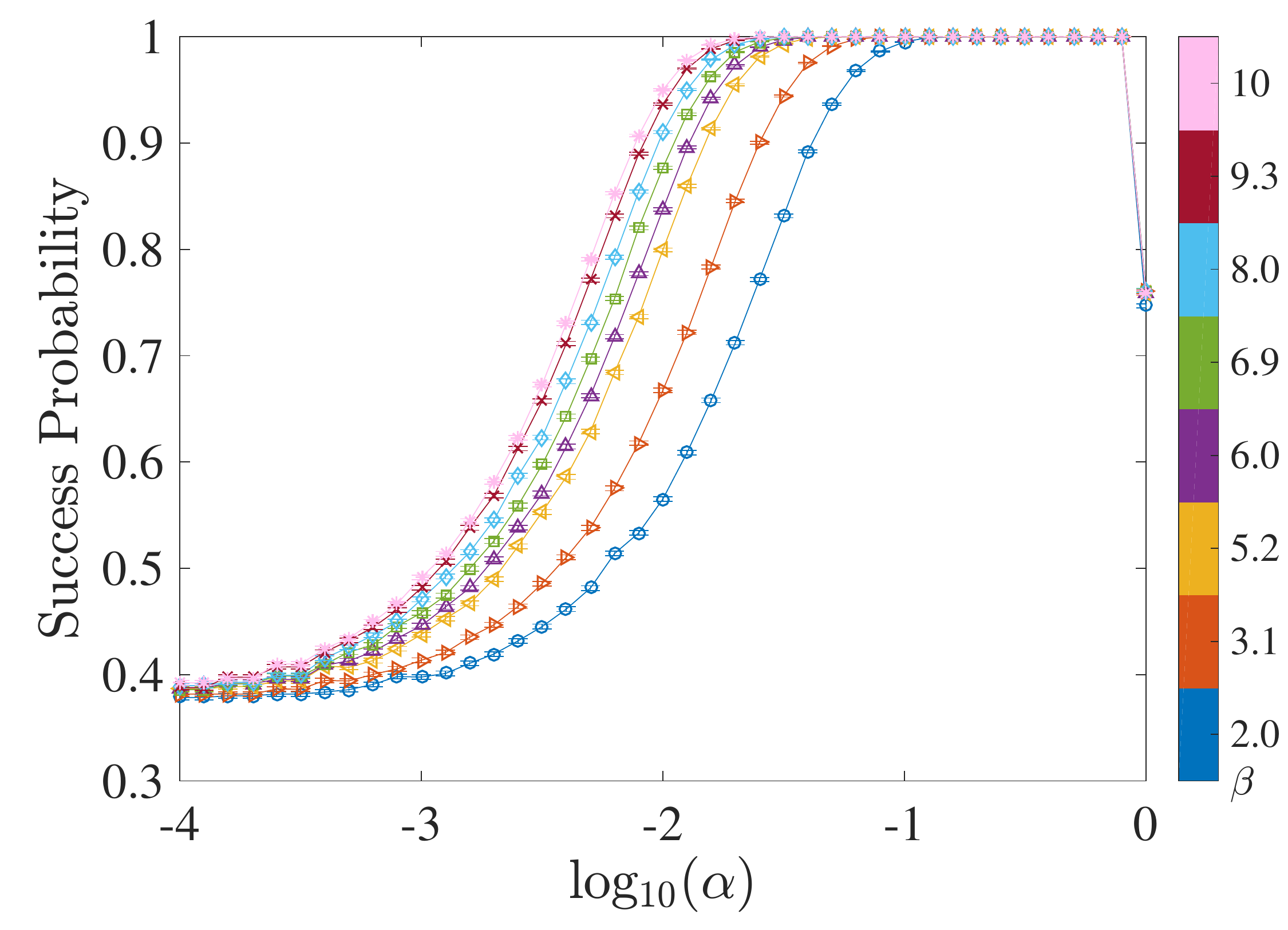}\label{fig:EBoost_PT2}}
%\subfigure[]{\includegraphics[width=0.45\textwidth]{PT_Choi_pS_K4_0_std=0AlphaC}\label{fig:EBoost_PT3}}
\subfigure[]{\includegraphics[width=0.45\textwidth]{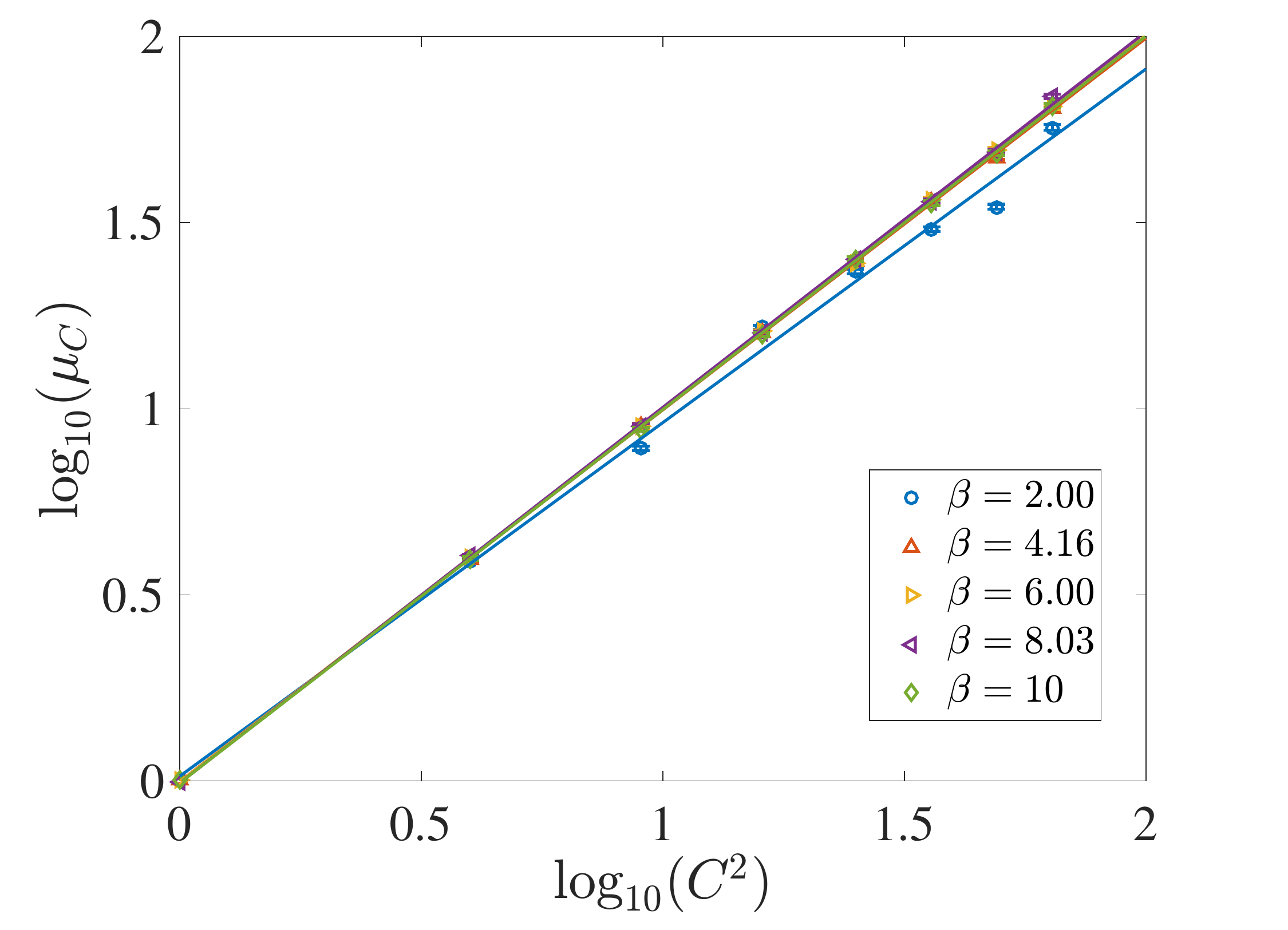}\label{fig:EBoost_PT3}}
\caption{Parallel tempering (PT) results for antiferromagnetic $K_4$ with no noise on the couplers. PT was used to generate a thermal state with respect to the ME, which was then decoded using majority voting. (a) Shows success probabilities for different nesting levels $C$ at $\beta=2$. (b)  Shows success probabilities for different inverse temperatures at $C=4$.  (c) Scaling of $\mu_C$ for the thermal state.  The solid lines represent the best linear fit to all the data points.  All the best-fit lines have slopes greater than $0.95$, so we find that the optimal scaling of $\mu_C \sim C^2$ is recovered at all (sufficiently large) inverse temperatures tested. This illustrates that for a sufficiently cold equilibrated system ME does not result in a suboptimal energy boost.} 
\label{fig:Mu3}
\end{center}
\end{figure*}

Figure~\ref{fig:Mu3} gives further evidence that nesting can be interpreted as an effective reduction of temperature by studying the success probability associated with the thermal distribution on the ME.  We used parallel tempering (PT) to sample from the thermal state associated with the ME of the different NQAC cases shown in Fig.~\ref{fig:exp-k4-nesting}, and decoded using majority voting.  We find that the thermal state at different temperatures but fixed $C$, exhibits the same qualitative behavior as the thermal state at fixed temperature but different $C$ [see Fig.~\ref{fig:EBoost_PT1} \emph{vs.} Fig.~\ref{fig:EBoost_PT2}].  Therefore, the performance improvement associated with reducing the temperature can also be reproduced by increasing $C$. This enforces that the energy boost can also be interpreted as decrease of the effective temperature of the device. We also find that the thermal state exhibits an energy boost scaling of $\mu_C \sim C^2$ [see Fig.~\ref{fig:EBoost_PT3}].

%%%%%%%%%%%%%%%%%%%%%%%%%

\section{Choi and Heuristic Embeddings}
\label{sec:two-MEs}
\begin{figure*}[ht]
\begin{center}
\subfigure[\ $K_{32}$: Choi ME.]{\includegraphics[width=0.49\textwidth]{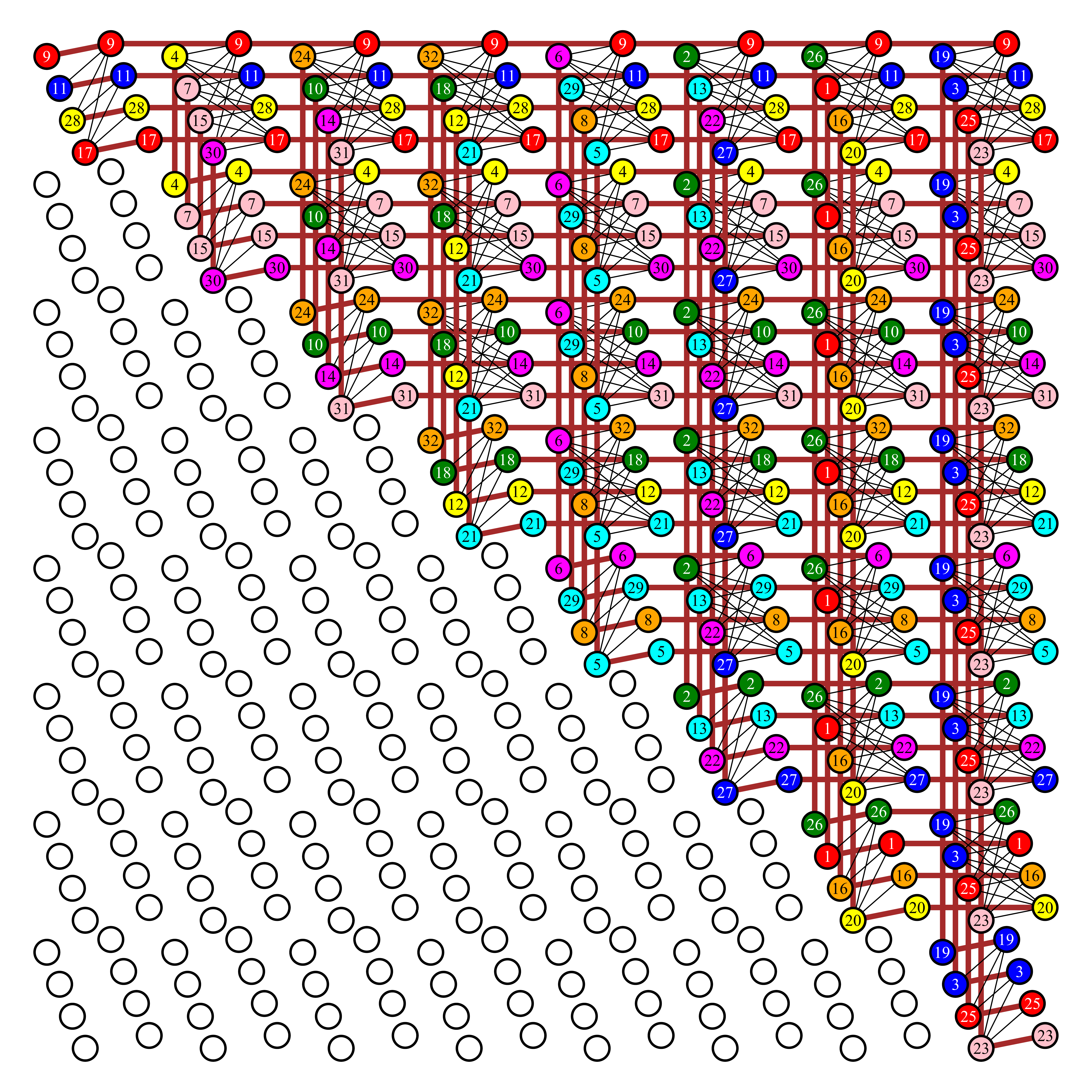}\label{fig:Choi}}
\subfigure[\ $K_{32}$: heuristic ME. ]{\includegraphics[width=0.49\textwidth]{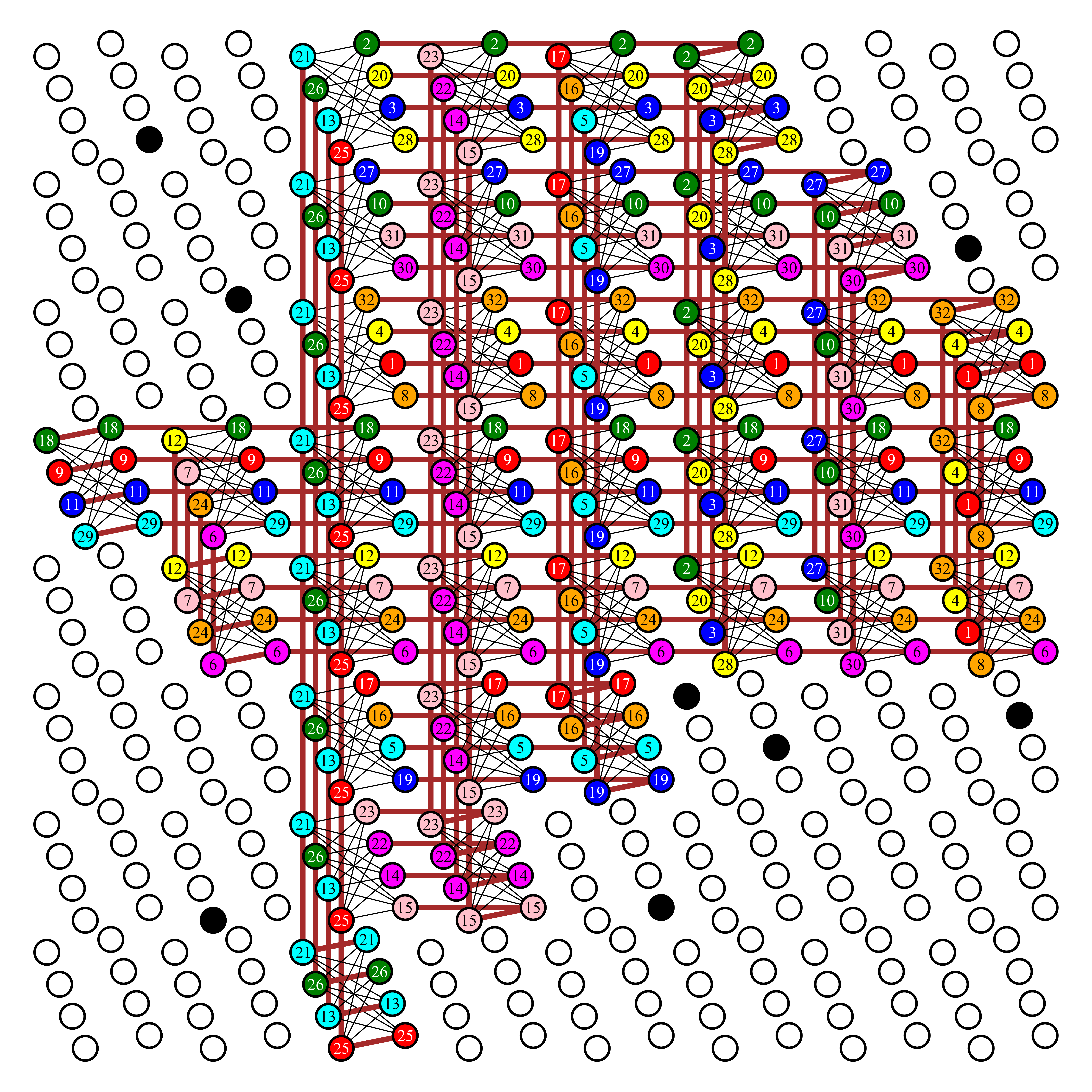}\label{fig:Heu}}
\caption{MEs of a $K_{32}$. We used these, e.g., to minor-embed a $C=8$ nesting of a $K_4$, or a $C=4$ nesting of a $K_8$. (a) The Choi embedding implemented on a perfect Chimera graph. (b) A heuristic ME for the actual DW2 device used in this work, whose Chimera graph contains $8$ unusable qubits (black circles). Different colors (and labels) denote chains representing minor-embedded logical qubits. Black (thin) lines are logical couplings, while brown (thick) lines represent energy penalties (ferromagnetic couplings).} 
\label{fig:Choi-vs-Heu}
\end{center}
\end{figure*}

The ``Chimera" hardware connectivity graph of the D-Wave devices allows for a ME of complete graphs as described in Refs.~\cite{Choi1,Choi2}. In the main text we called this the ``Choi minor embedding". The Choi ME of a $K_{32}$ graph is shown in Fig.~\ref{fig:Choi}. The Choi technique requires a perfect Chimera graph, without missing vertices. In actual devices, however, imperfections in fabrication or the calibration process lead to the presence of unusable qubits (e.g., due to trapped flux). These qubits, along with their couplings are then permanently disabled and cannot be used in the QA process. Efficient heuristic algorithms have been developed to search for MEs for the resulting induced Chimera subgraphs \cite{klymko_adiabatic_2012,Cai:2014nx,Boothby2015a}.  Figure~\ref{fig:Heu} shows the ME of a $K_{32}$ obtained when the heuristic algorithm developed in Ref.~\cite{Boothby2015a} is applied to the actual hardware graph of the DW2 ``Vesuvius" chip installed at USC-ISI. Note how the ME avoids the unusable qubits, depicted as black circles in Fig.~\ref{fig:Heu}.

The MEs shown in Fig.~\ref{fig:Choi-vs-Heu} are the actual ``Choi" and ``heuristic" MEs used in our experiments and simulations. As discussed in the main text, SQA simulations demonstrate that the choice of the ME has a significant impact on the performance of NQAC. In particular, it turns out that the Choi ME outperforms the heuristic ME. Since the two MEs use the same amount of physical resources (a logical qubit is represented by chains of equal lengths), it is unclear why the Choi embedding should perform better than the heuristic embedding,
%. It is possible that embeddings with a more symmetric structure (like the Choi embedding) generically have an advantage, but 
and further investigations are needed in order to clarify this point. In the present work we limit ourselves to stressing the importance of the embedding choice when assessing the performance of minor-embedded problems in QA.

%%%%%%%%%%%%%%%%%%%%%%%%%

\section{Additional Experimental Data}
\label{sec:Exp_Add}
\begin{figure*}[ht]
\begin{center}
\subfigure[\ $P_C(\alpha)$ for the hard $K_8$. ]{\includegraphics[width=0.45\textwidth]{8x8_ideal_random_per_logical-}}
\subfigure[\ Adjusted $P_C'(\alpha)$ for the hard $K_8$. ]{\includegraphics[width=0.45\textwidth]{8x8_ideal_random_per_logical-corrected-}}\\
\subfigure[\ $P_C(\alpha)$ for the easy $K_8$. ]{\includegraphics[width=0.45\textwidth]{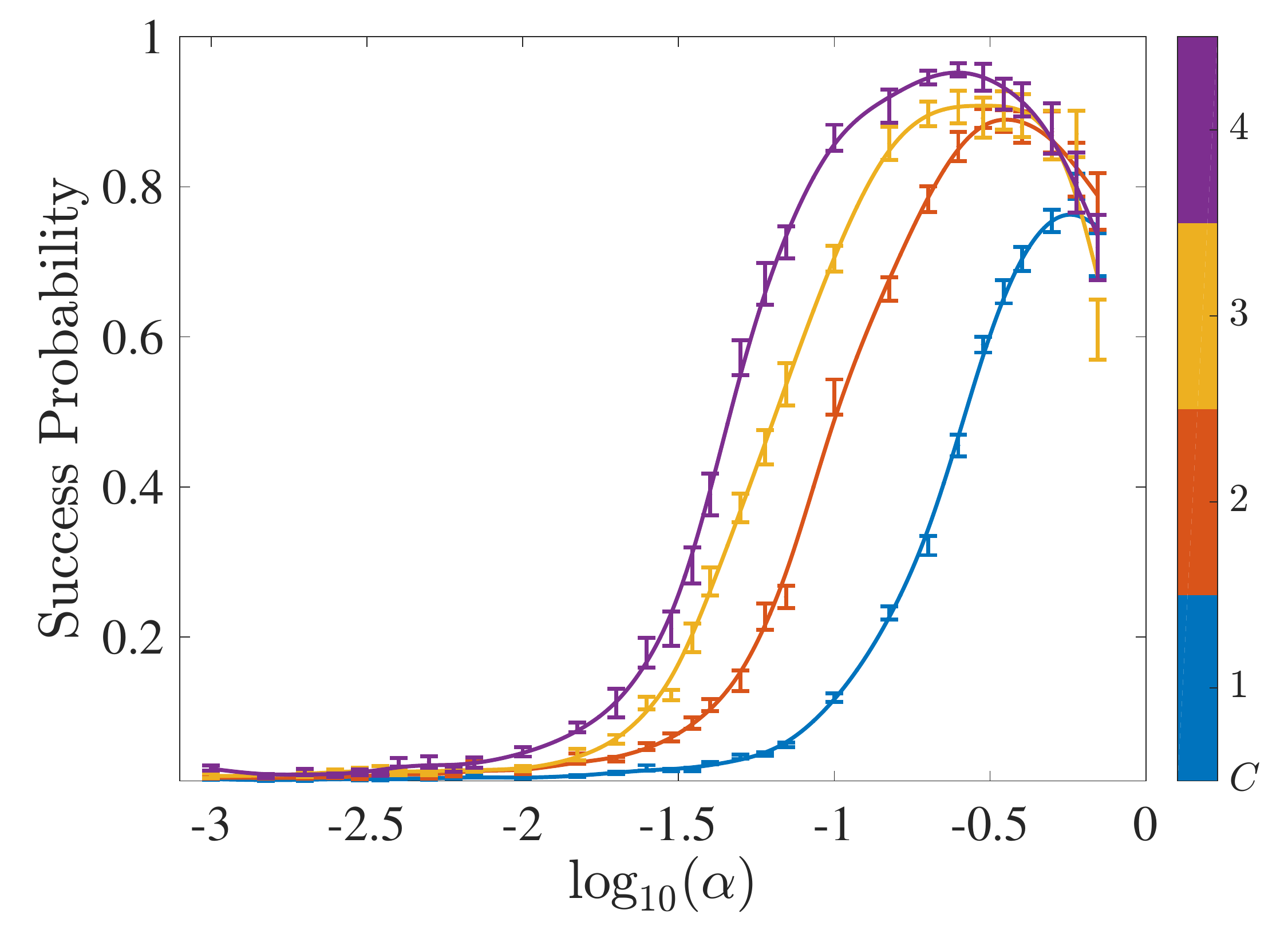}}
\subfigure[\ Adjusted $P_C'(\alpha)$ for the easy $K_8$. ]{\includegraphics[width=0.45\textwidth]{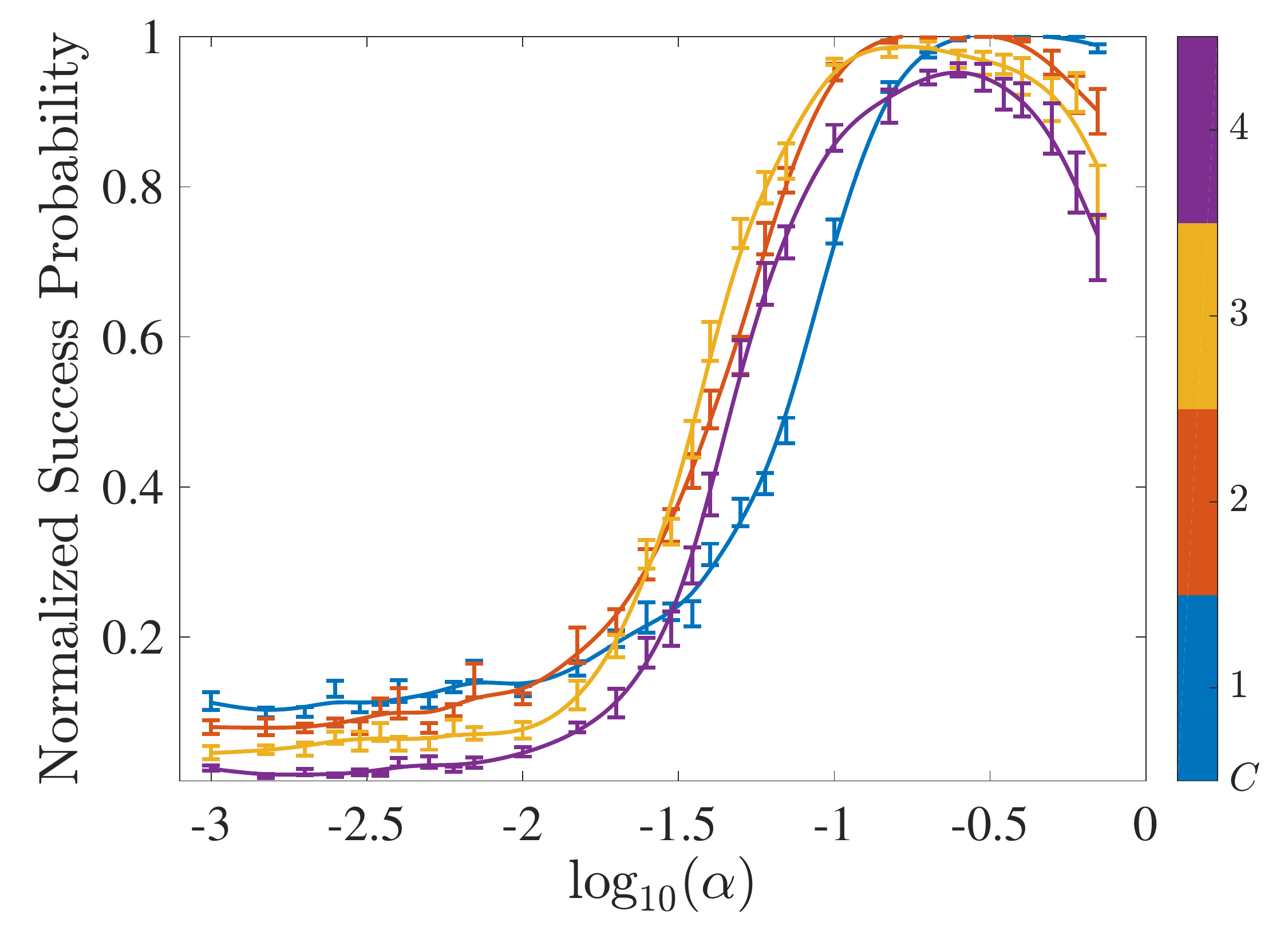}}
\caption{Random antiferromagnetic $K_8$ nesting: experimental results for a harder and easier instance. } 
\label{fig:exp-k8-nesting-full}
\end{center}
\end{figure*}

\begin{figure*}[ht]
\begin{center}
\subfigure[\ $P_C(\alpha)$ for the hard $K_{10}$. ]{\includegraphics[width=0.45\textwidth]{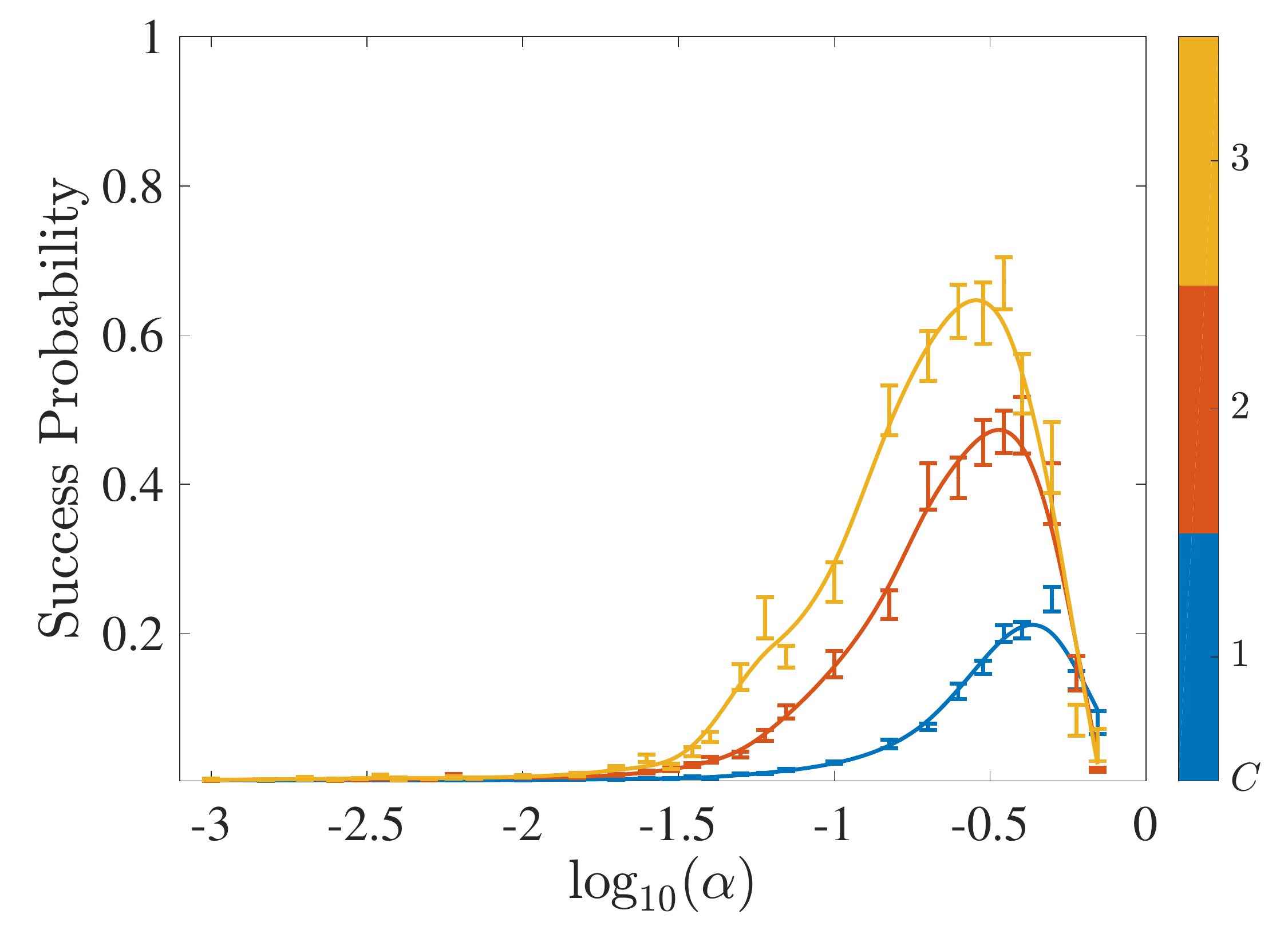}}
\subfigure[\ Adjusted $P_C'(\alpha)$ for the hard $K_{10}$. ]{\includegraphics[width=0.45\textwidth]{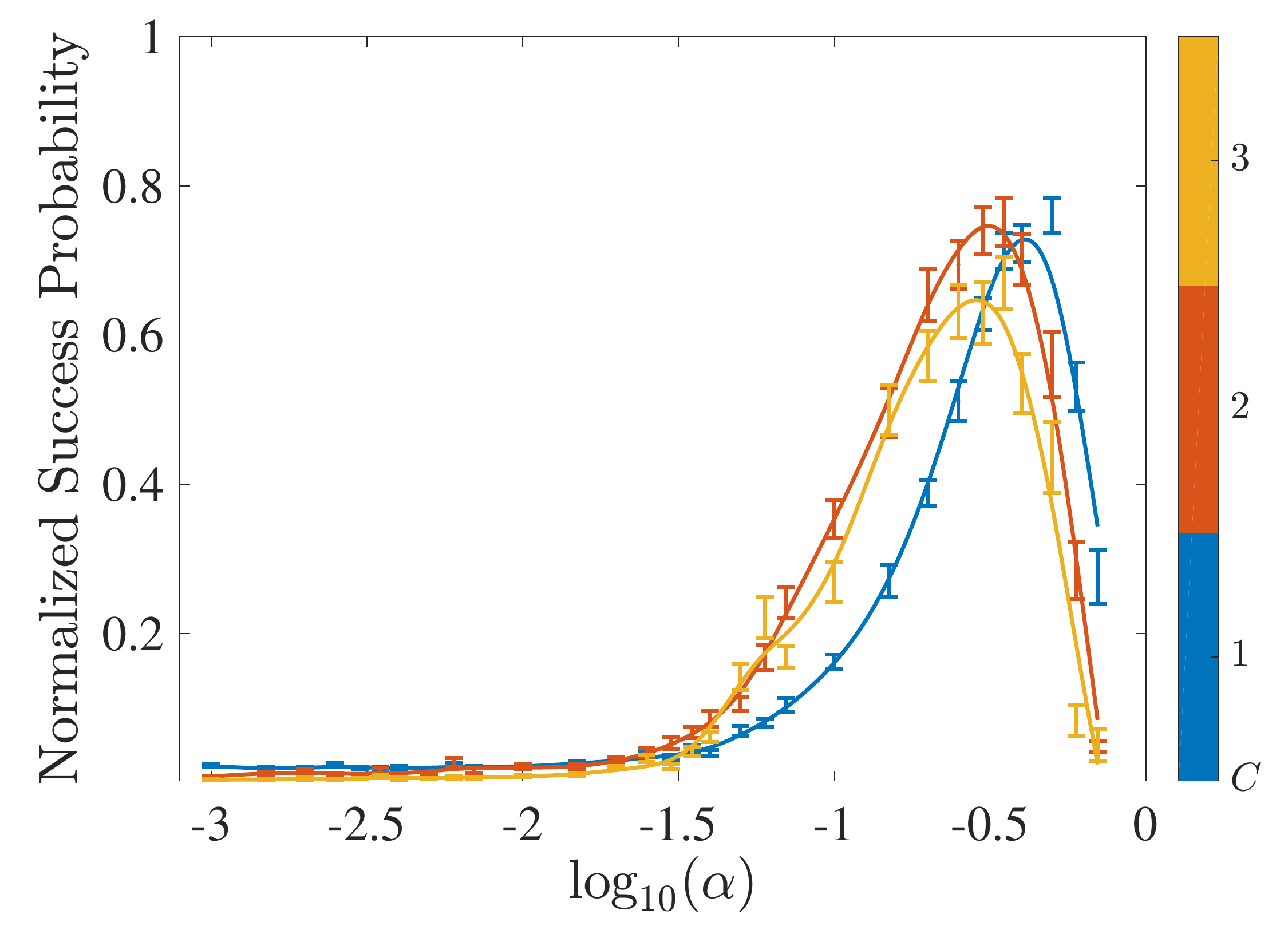}}
\subfigure[\ $P_C(\alpha)$ for the easy $K_{10}$. ]{\includegraphics[width=0.45\textwidth]{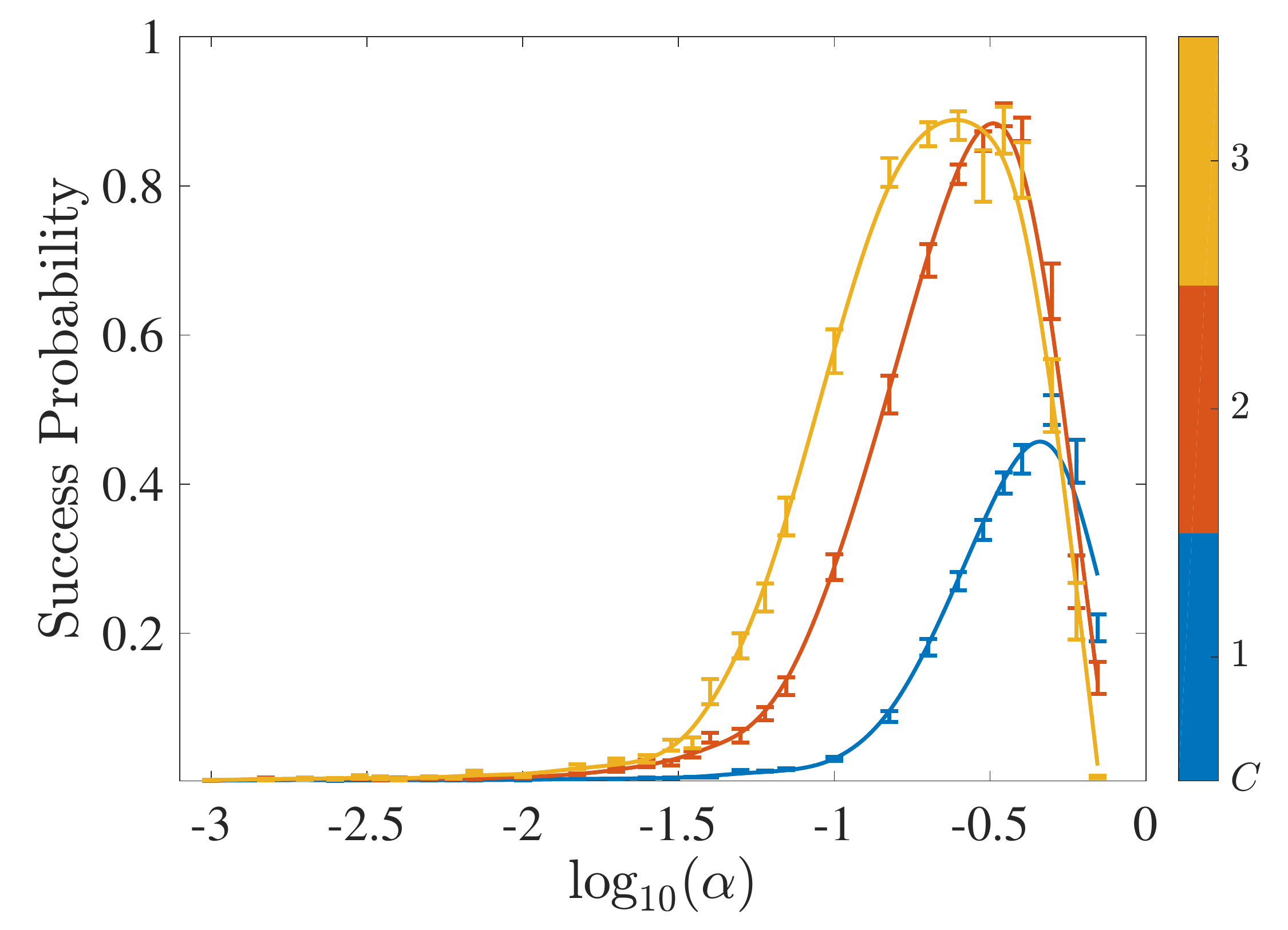}}
\subfigure[\ Adjusted $P_C'(\alpha)$ for the easy $K_{10}$. ]{\includegraphics[width=0.45\textwidth]{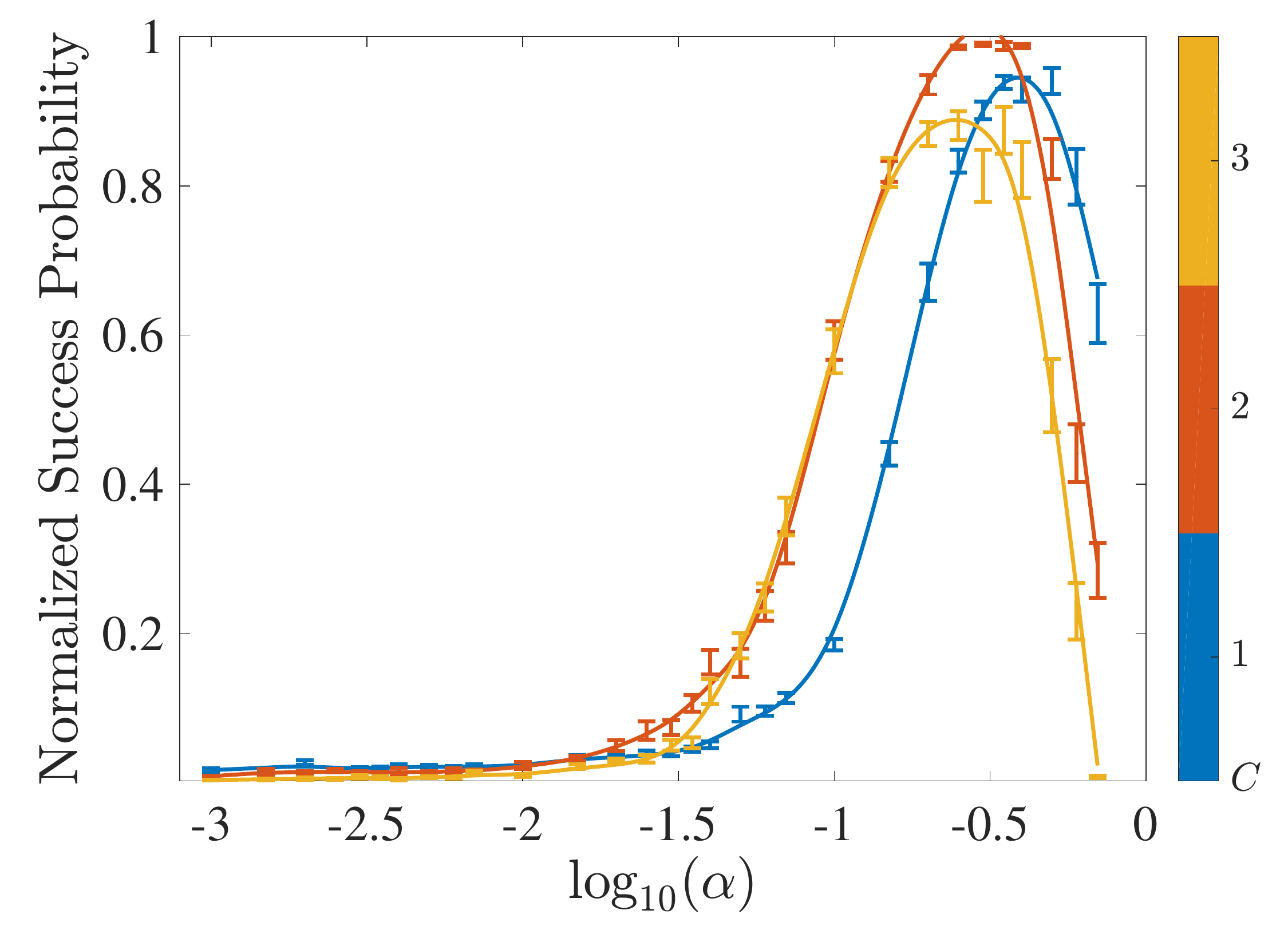}}\\
\caption{Random antiferromagnetic $K_{10}$ nesting: experimental results for a harder and easier instance.} 
\label{fig:exp-k10-nesting-full}
\end{center}
\end{figure*}

In this section we present additional experimental data for $K_N$'s with couplings randomly generated from the set $J_{ij} \in \{0.1,0.2,\dots,0.9,1\}$.  For large $N$, $K_N$ generated in this manner have a finite temperature spin glass phase transition~\cite{Kirkpatrick:1978dn}. This property renders simulated annealing inefficient in finding the ground state of such problems~\cite{2014Katzgraber}.  The main text reports data for a random $K_8$ instance that is referred to here as ``harder-$K_8$":
\beq
K_8^{\rm{h}} = 
\left(
\begin{array}{cccccccc}
 0 &  0.4    & 0.7 & 0.5 & 0.3 & 0.5 & 0.2 & 0.5  \\
 0 &  0 & 0.3 & 0.8 & 0.8  & 0.3 & 0.5 & 0.7  \\
 0 &  0 & 0  & 0.5 & 0.9 & 0.9 & 0.3 & 0.9  \\
 0 &  0 & 0 & 0 & 1 & 0.8 & 0.8  & 0.7  \\
 0 &  0 & 0 &0  &0  & 0.9  & 0.3 & 0.6  \\
 0 &  0 & 0 & 0 & 0 & 0 &  0.9 & 0.4  \\
 0 & 0  & 0 & 0 & 0 & 0 & 0 &  0.5 \\
 0 & 0  & 0 &0 & 0 &0  & 0 & 0  
\end{array}
\right)\ .
\eeq
Figure~\ref{fig:exp-k8-nesting-full} includes similar data for another random $K_8$ instance that turned out to have a higher success probability, so we refer to it as ``easier-$K_8$":
\beq
K_8^{\rm{e}} = 
\left(
\begin{array}{cccccccc}
 0 &   0.8  & 0.7 & 0.8 & 0.9 & 0.4 & 0.2 & 0.9  \\
 0 &  0 &0.7  &0.8  & 0.3 & 0.7 & 1& 0.3  \\
 0 &  0 & 0  & 0.7 & 0.6 & 0.1 & 0.5 & 0.6  \\
 0 &  0 & 0 & 0 & 0.1 & 0.8 & 0.1 & 0.5  \\
 0 &  0 & 0 &0  &0  & 0.5 & 0.8 & 0.2  \\
 0 &  0 & 0 & 0 & 0 & 0 & 0.6  & 0.7   \\
 0 & 0  & 0 & 0 & 0 & 0 & 0 &  1 \\
 0 & 0  & 0 &0 & 0 &0  & 0 & 0  
\end{array}
\right)\ .
\eeq
Figure~\ref{fig:exp-k10-nesting-full} displays results for NQAC applied to an ``easier-$K_{10}$":
\beq
K_{10}^{\rm{e}} = 
\left(
\begin{array}{cccccccccc}
          0  &   0.2  &    0.7  &    0.8  &    0.5  &    0.3  &    0.8  &    0.9  &    0.4  &    0.1  \\
         0  &        0  &   0.1  &    0.1  &    0.4  &    0.7  &    0.3  &    0.3  &    0.9  &    0.1  \\
         0  &        0  &        0  &   0.3  &    0.8  &    0.7  &    0.6  &    0.9  &    0.6  &    0.6  \\
         0  &        0  &        0  &        0  &   0.8  &    0.2  &    0.7  &    0.3  &    0.6  &    0.8   \\
         0  &        0  &        0  &        0  &        0  &   0.2  &    0.9  &    1 &  1 &   1  \\
         0  &        0  &        0  &        0  &        0  &        0  &   1  &    0.4  &    0.3  &    0.2  \\
         0  &        0  &        0  &        0  &        0  &        0  &        0  &   0.2  &    0.8  &    0.6  \\
         0  &        0  &        0  &        0  &        0  &        0  &        0  &        0  &   0.8  &    0.5  \\
         0  &        0  &        0  &        0  &        0  &        0  &        0  &        0  &        0  &   0.1  \\
         0  &        0  &        0  &        0  &        0  &        0  &        0  &        0  &        0  &        0
\end{array}
\right)\ ,
\eeq
and a ``harder-$K_{10}$":
\beq
K_{10}^{\rm{h}} = 
\left(
\begin{array}{cccccccccc}
         0  &   0.6  &   0.9  &   0.8  &   0.5  &   1  &   0.4  &   0.2  &   0.1  &   0.5 \\
         0  &        0  &   0.8  &   0.9  &   0.1  &   0.6  &   0.2  &   0.7  &   0.7  &   0.9 \\
         0  &        0  &        0  &   0.8  &   0.6  &   0.3  &   0.8  &   0.2  &   0.6  &   0.6 \\
         0  &        0  &        0  &        0  &   0.1  &   0.3  &   0.8  &   0.4  &   0.6  &   0.5 \\
         0  &        0  &        0  &        0  &        0  &   0.7  &   0.6  &   0.4  &   0.3  &   0.1 \\
         0  &        0  &        0  &        0  &        0  &        0  &   0.1  &   1  &   0.9  &   0.6 \\
         0  &        0  &        0  &        0  &        0  &        0  &        0  &   0.9  &   0.9  &   0.9 \\
         0  &        0  &        0  &        0  &        0  &        0  &        0  &        0  &   0.1  &   1.0 \\
         0  &        0  &        0  &        0  &        0  &        0  &        0  &        0  &        0  &   0.3 \\
         0  &        0  &        0  &        0  &        0  &        0  &        0  &        0  &        0  &        0 
\end{array}
\right)\ .
\eeq
In all cases we display results up to nesting level $C = 3$.

\begin{figure*}[ht]
\begin{center}
\subfigure[\ Hard $K_8$. ]{\includegraphics[width=0.45\textwidth]{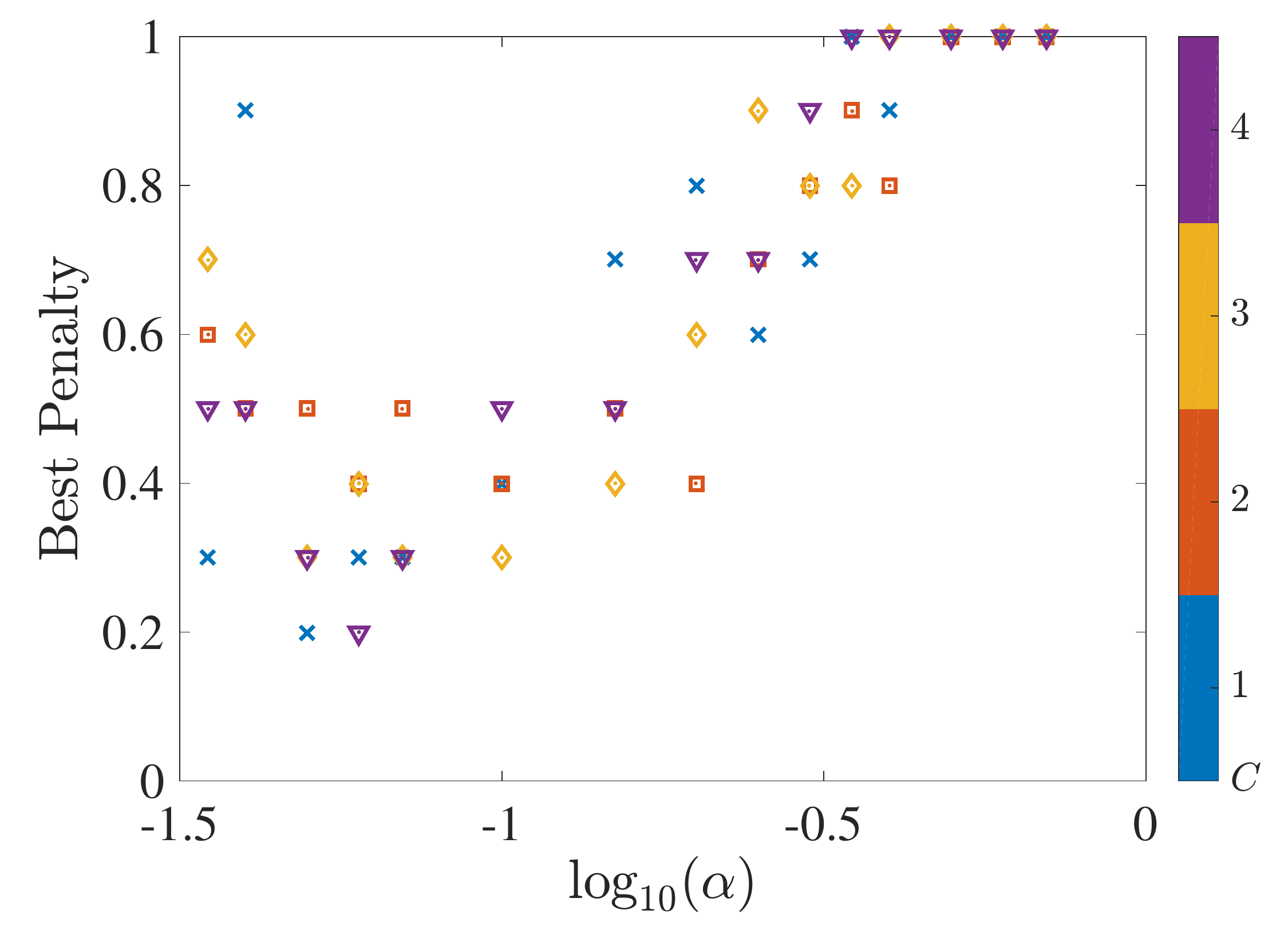}}
\subfigure[\ Easy $K_8$. ]{\includegraphics[width=0.45\textwidth]{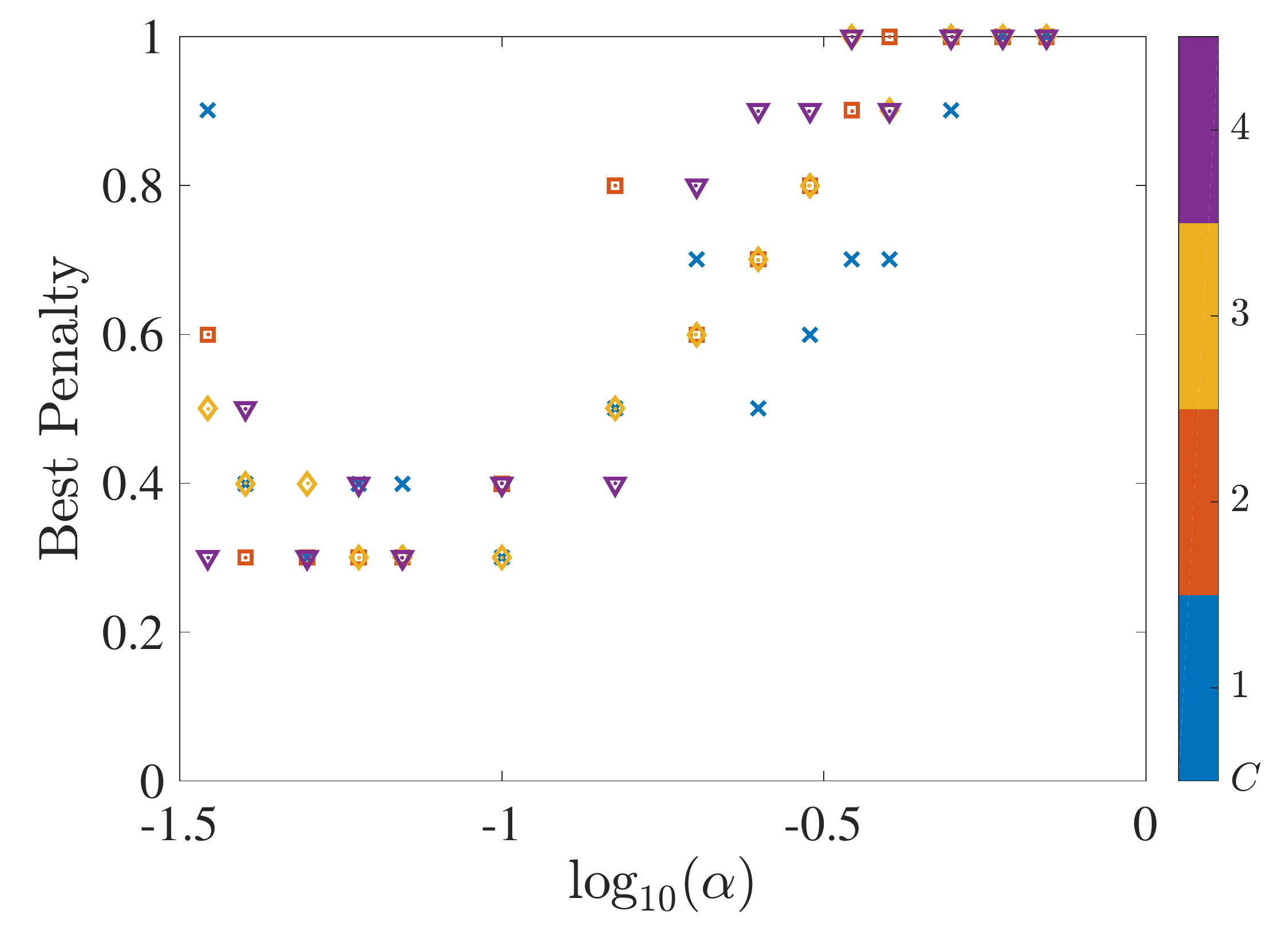}}
\subfigure[\ Hard $K_{10}$. ]{\includegraphics[width=0.45\textwidth]{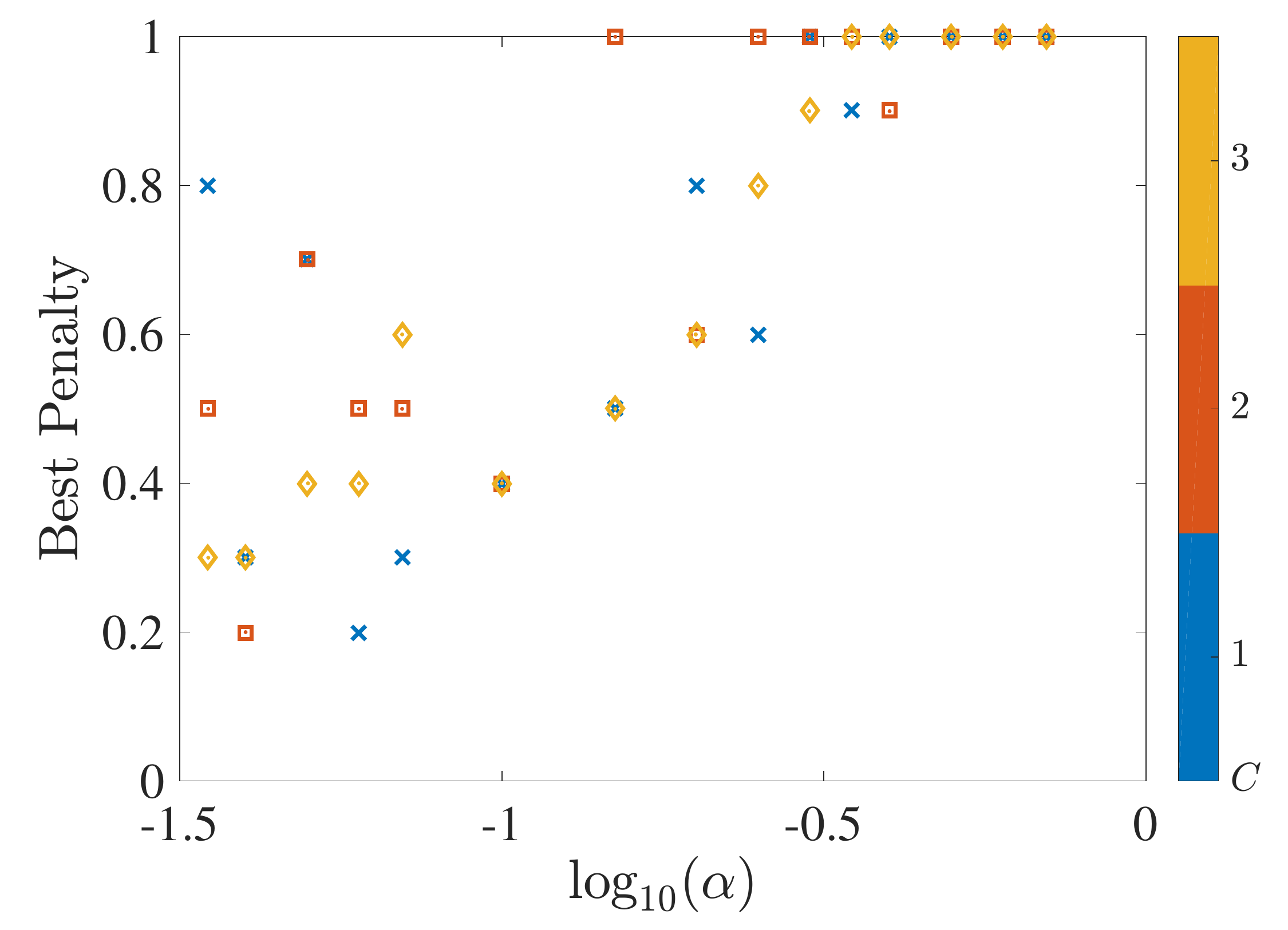}}
\subfigure[\ Easy $K_{10}$. ]{\includegraphics[width=0.45\textwidth]{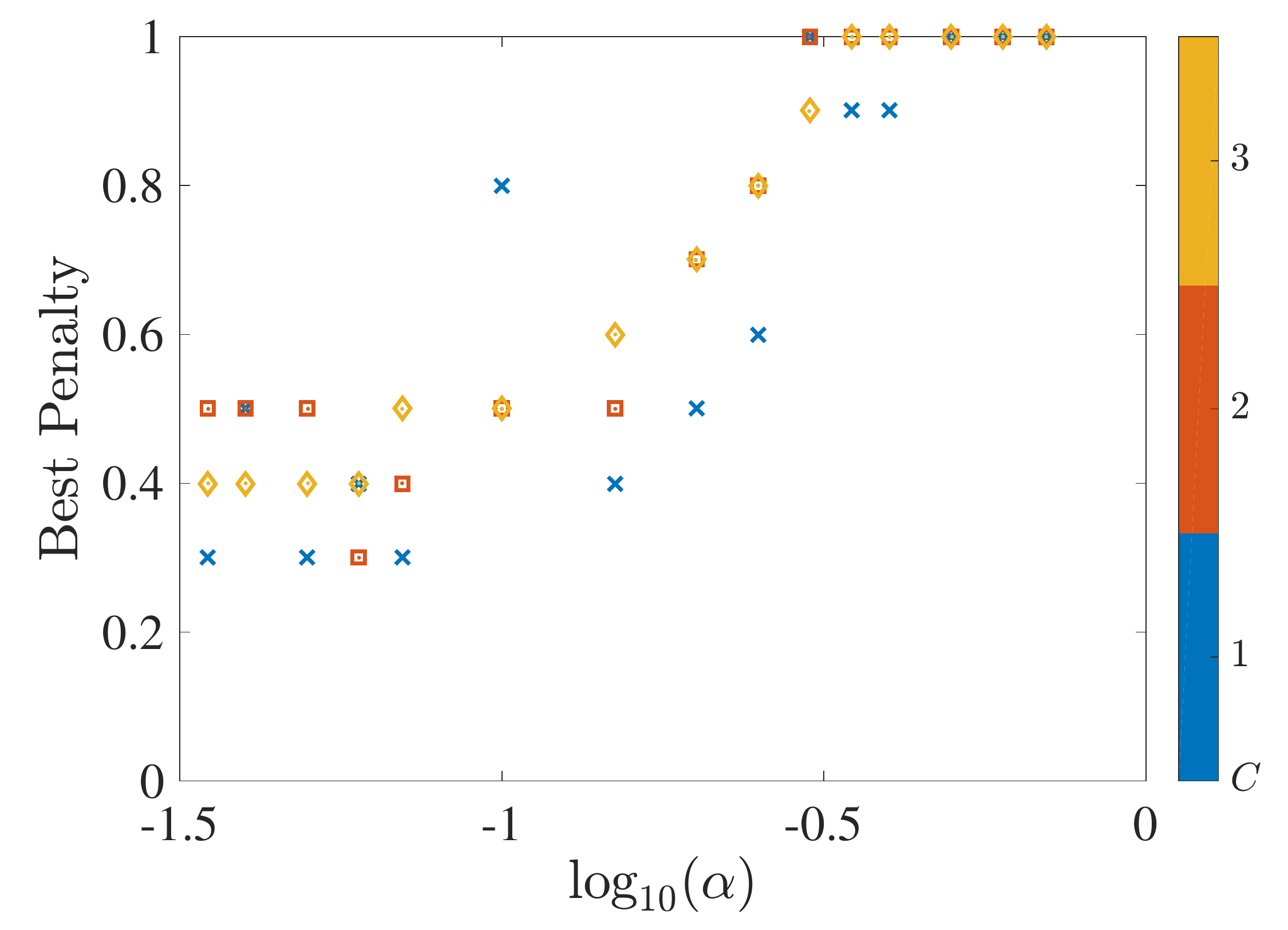}}
\caption{Optimal penalties $\gamma$. } 
\label{fig:exp-penalties}
\end{center}
\end{figure*}

Figure~\ref{fig:exp-penalties} shows the optimal penalty strength as a function of the energy scale $\alpha$ for the four instances considered. A saturation of the optimal penalty is visible at the maximal possible value $|\gamma| = 1$ for $\alpha$ close to $1$, implying that the true optimal penalty values are $>1$ in this range.

\begin{figure*}[ht]
\begin{center}
\subfigure[\, Data collapse for hard $K_8$]{\includegraphics[width=0.45\textwidth]{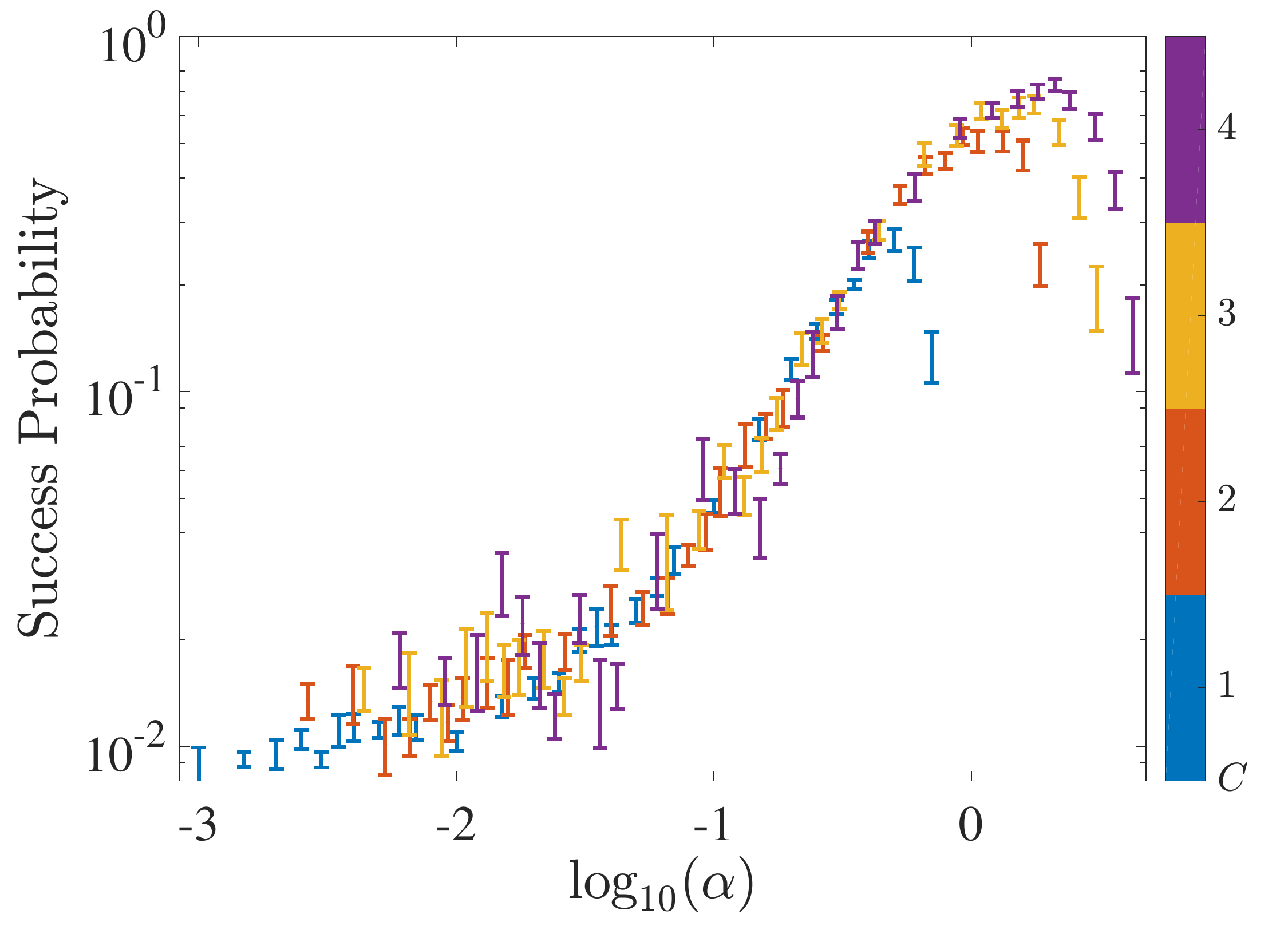}\label{fig:overlap_K8_1}}
\subfigure[\, Hard $K_8$ scaling of $\mu_c$]{\includegraphics[width=0.45\textwidth]{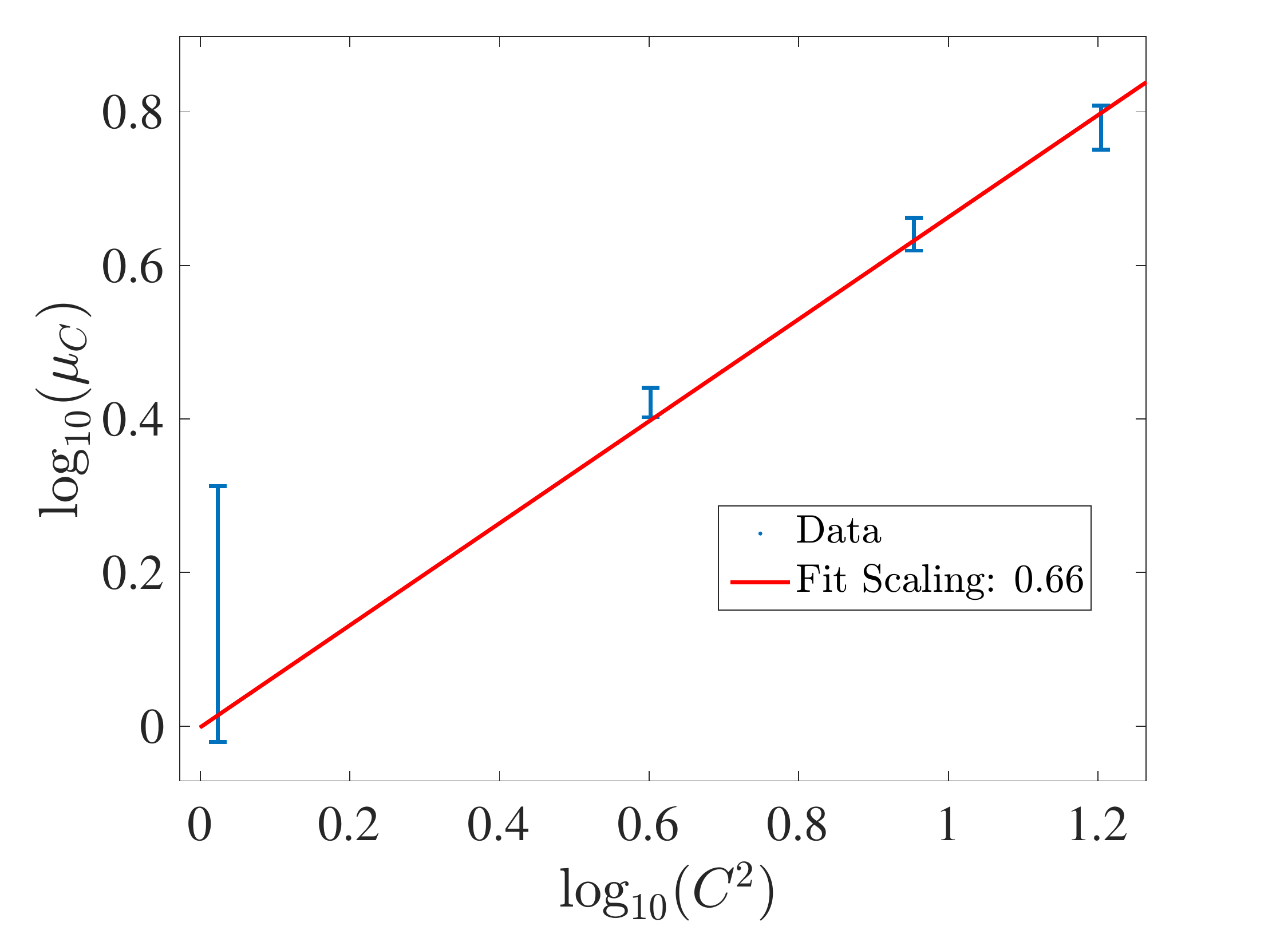}\label{fig:scaling_K8_1}}
\subfigure[\, Data collapse for easy $K_8$]{\includegraphics[width=0.45\textwidth]{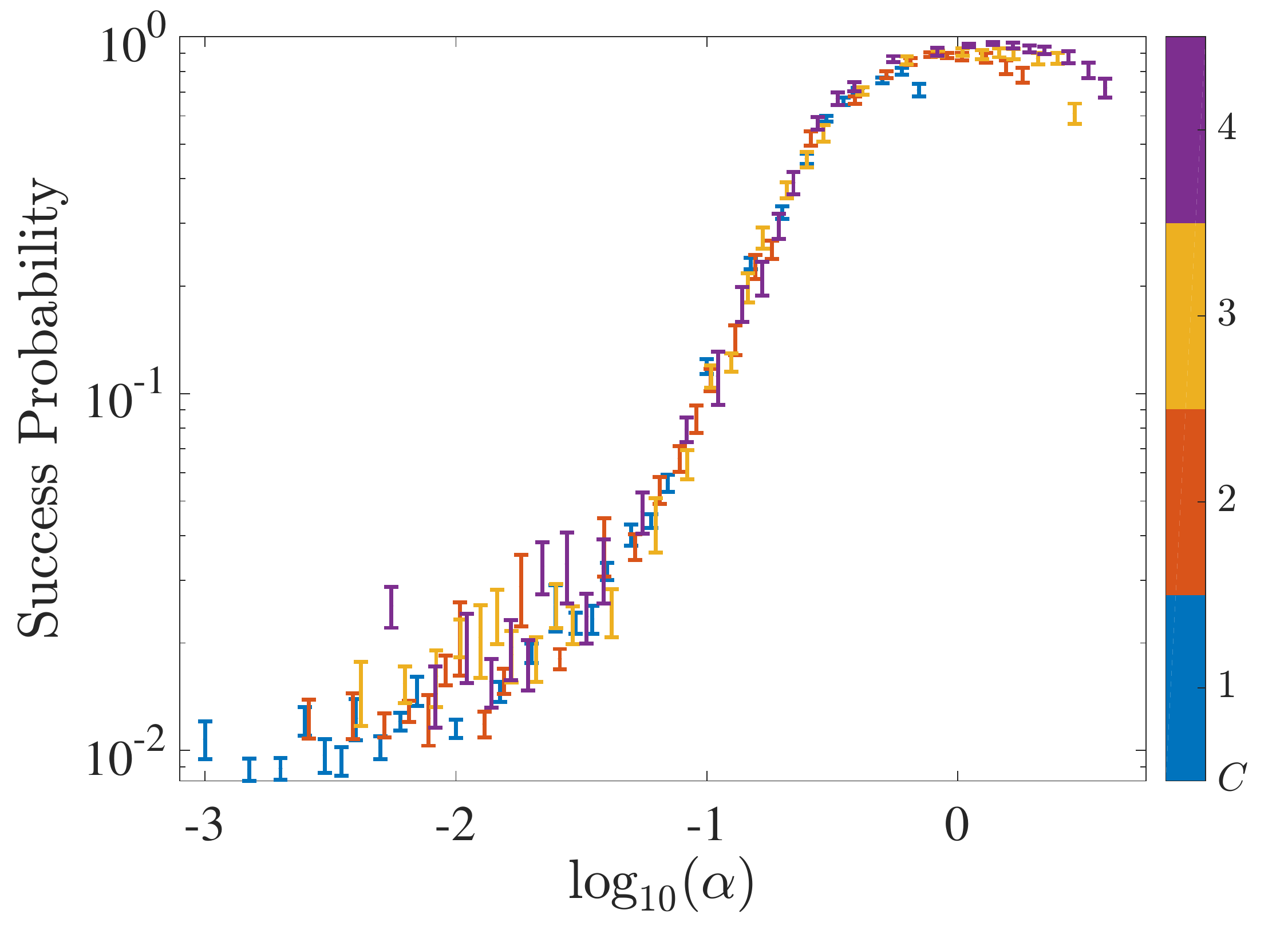}\label{fig:overlap_K8_2}}
\subfigure[\, Easy $K_8$ scaling of $\mu_c$]{\includegraphics[width=0.45\textwidth]{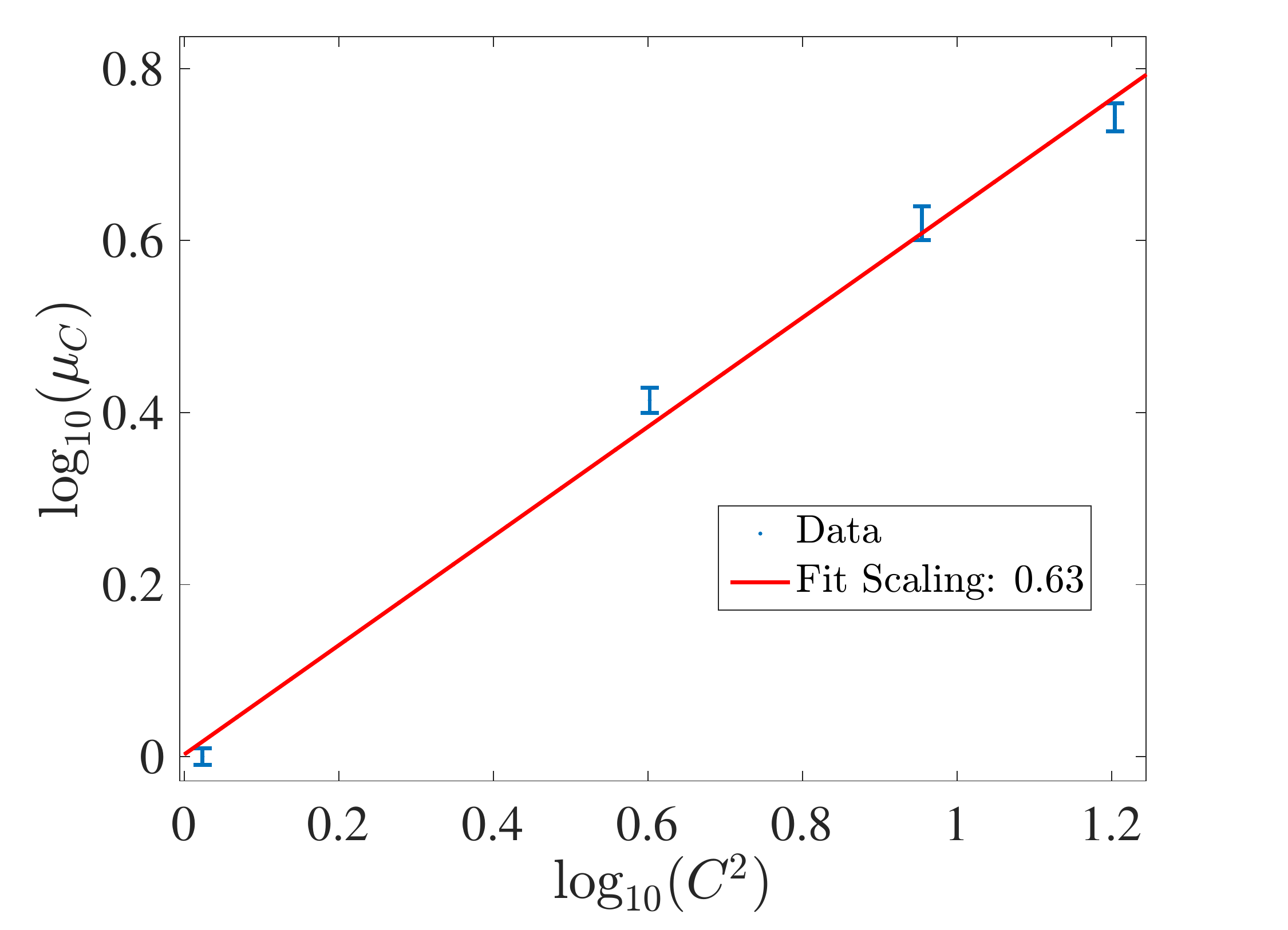}\label{fig:scaling_K8_2}}
\caption{Data collapse (left) and scaling of $\mu_C$ (right) for fully antiferromagnetic $K_8$. } 
\label{fig:extra1}
\end{center}
\end{figure*}

\begin{figure*}[ht]
\begin{center}
\subfigure[\, Data collapse for hard $K_{10}$]{\includegraphics[width=0.45\textwidth]{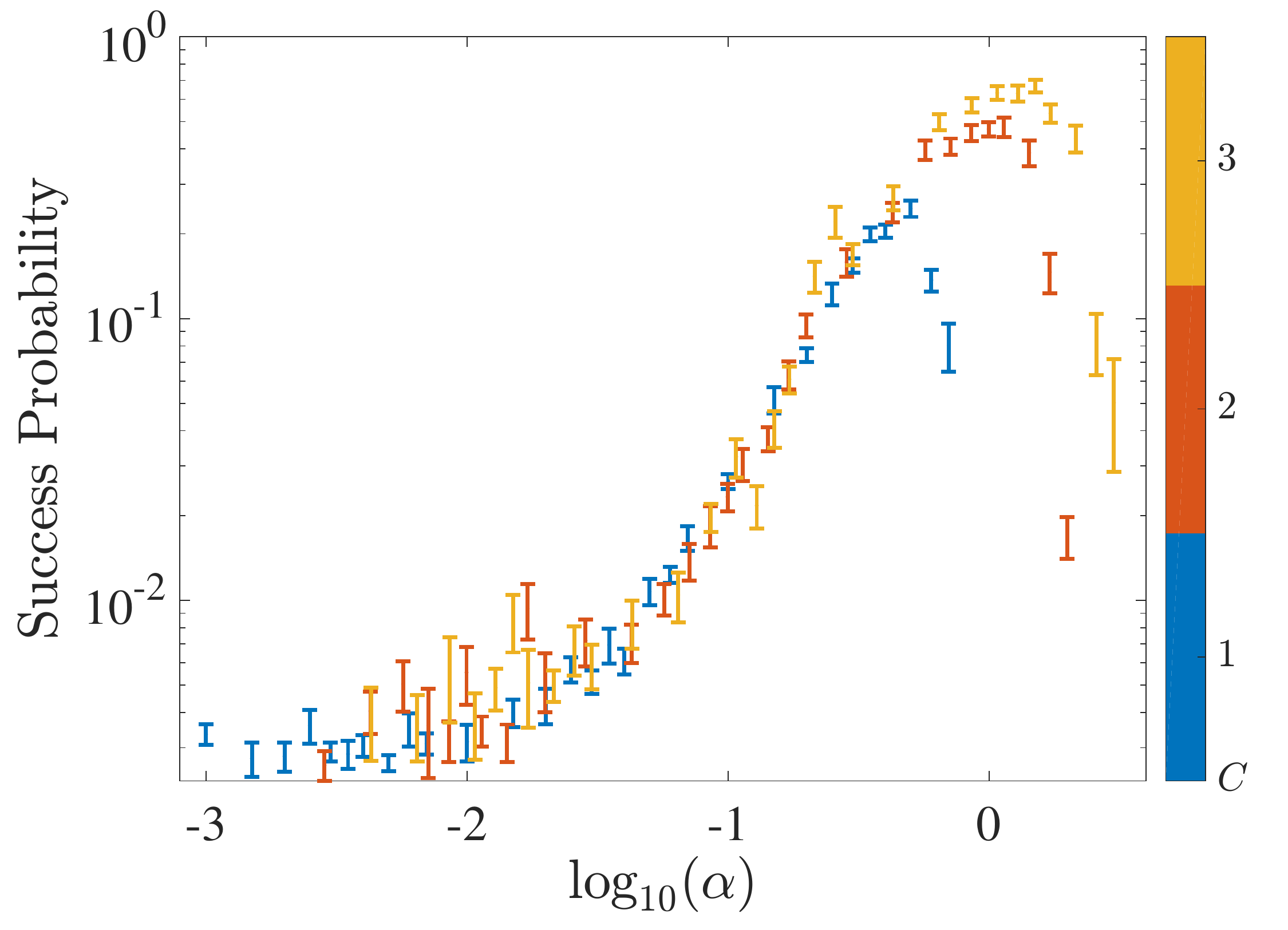}\label{fig:overlap_K10_1}}
\subfigure[\, Hard $K_{10}$ scaling of $\mu_c$]{\includegraphics[width=0.45\textwidth]{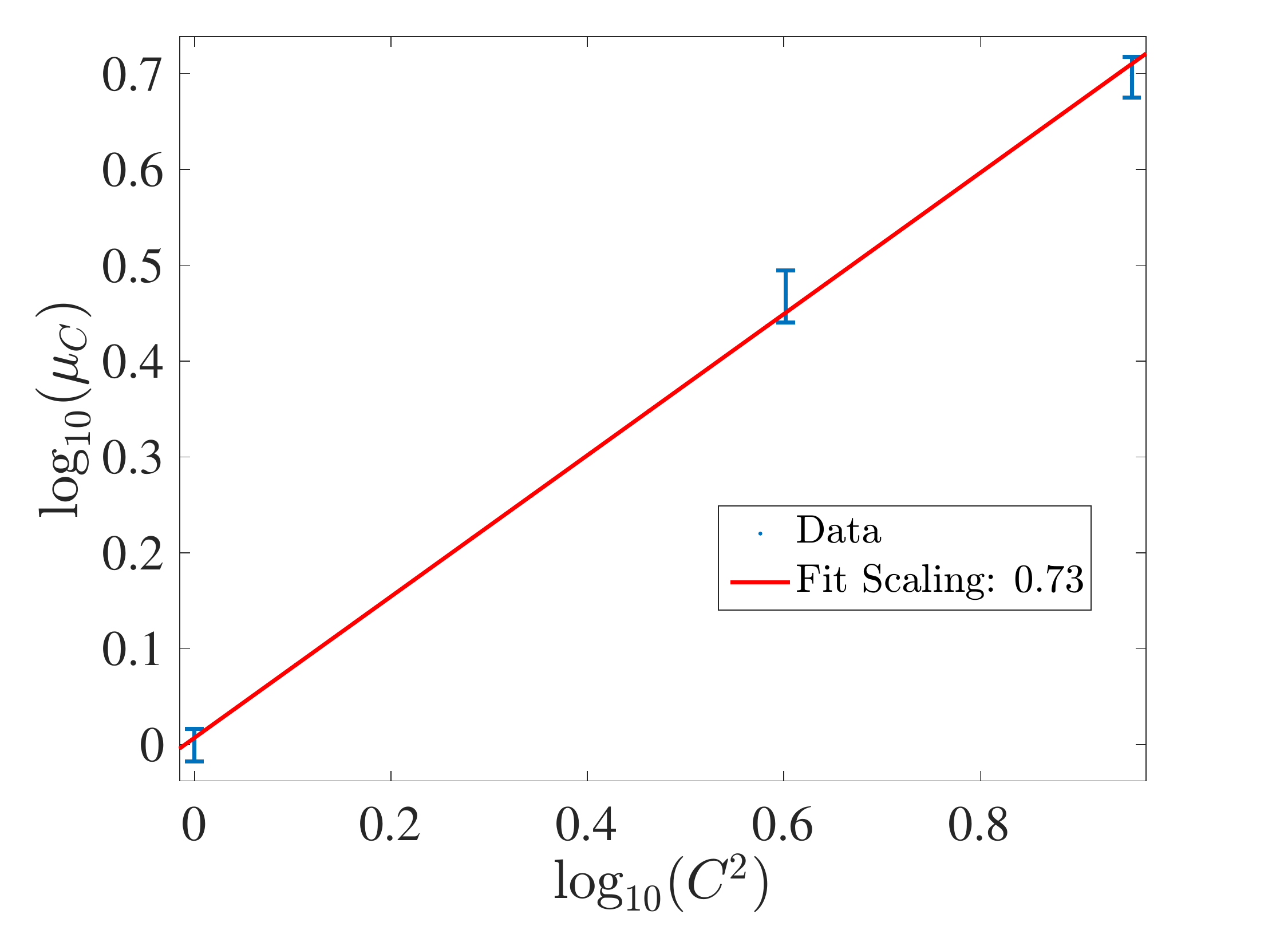}\label{fig:scaling_K10_1}}
\subfigure[\, Data collapse for easy $K_{10}$]{\includegraphics[width=0.45\textwidth]{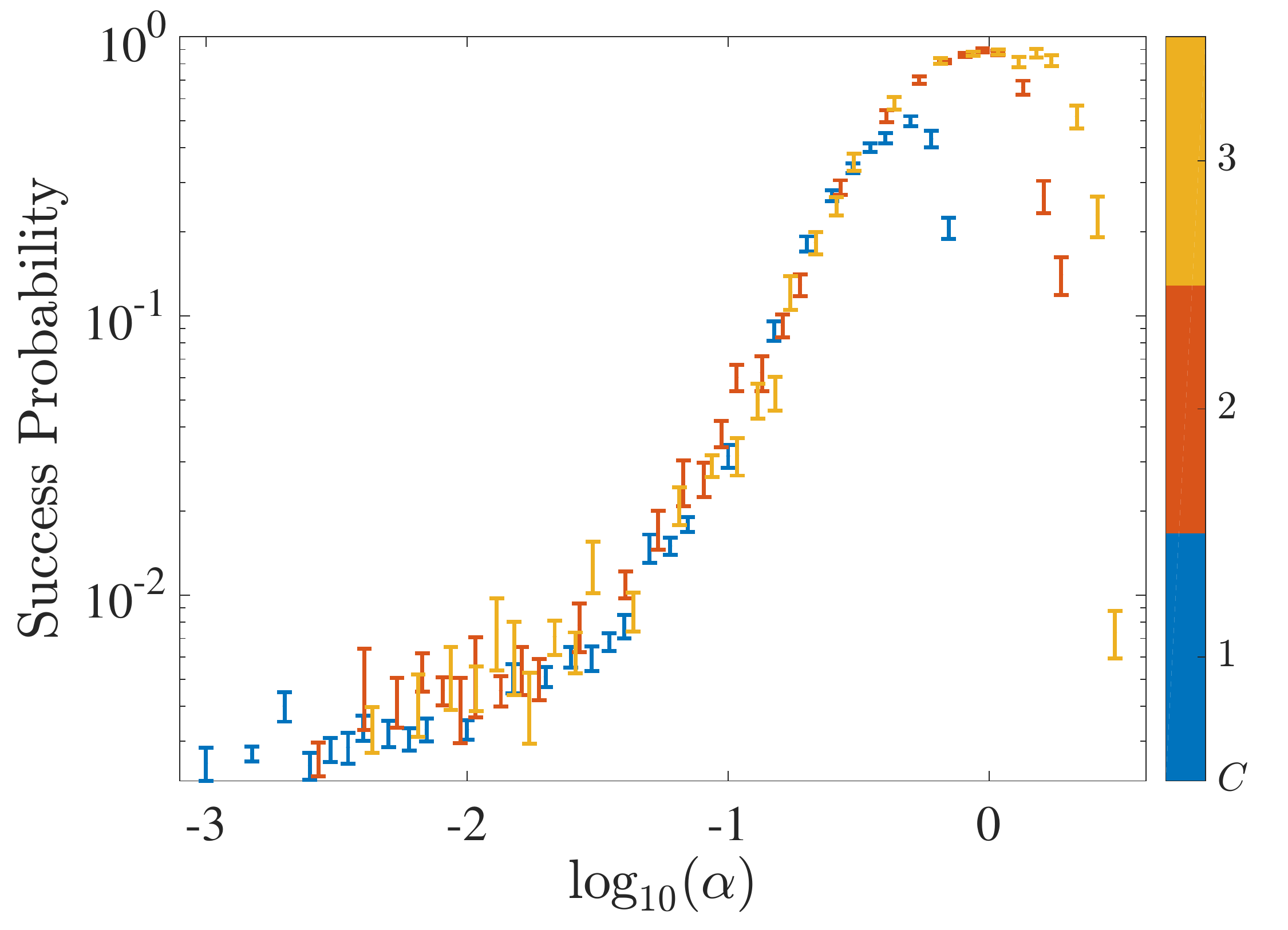}\label{fig:overlap_K10_2}}
\subfigure[\, Easy $K_{10}$ scaling of $\mu_c$]{\includegraphics[width=0.45\textwidth]{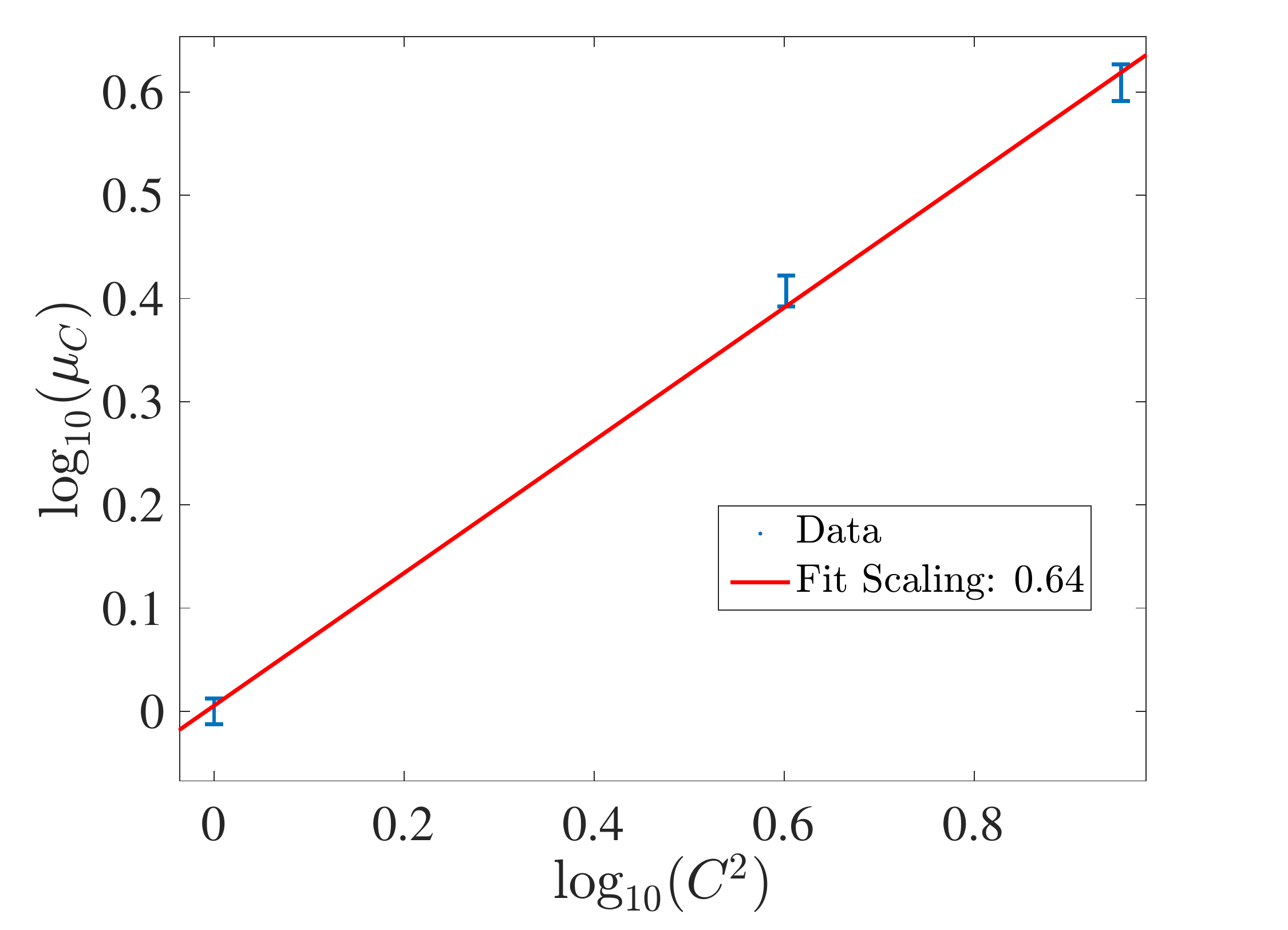}\label{fig:scaling_K10_2}}
\caption{Data collapse (left) and scaling of $\mu_C$ (right) for fully antiferromagnetic $K_{10}$. } 
\label{fig:extra2}
\end{center}
\end{figure*}

Figure~\ref{fig:extra1} shows that the antiferromagnetic harder-$K_8$ problem considered in the main text, as well as the easier-$K_8$ problem, also admit a data collapse (left), to the left of the peak. Recall that the peak is due to having reached the maximum penalty value, as illustrated in Fig.~\ref{fig:exp-penalties}.  The associated scaling of the energy boost $\mu_C$ is shown in the right column, yielding $\mu_C \sim C^{1.32}$ (harder-$K_8$) and $\mu_C \sim C^{1.26}$ (easier-$K_8$).
Figure~\ref{fig:extra2} shows the same for harder-$K_{10}$ and easier-$K_{10}$ problems. There we find $\mu_C \sim C^{1.34}$ for both problems.

%%%%%%%%%%%%%%%%%%%%%%%%%

\section{Determination of $\mu_C$}
\label{sec:mu_C}

To determine the values of $\mu_C$ and estimate error bars, we proceeded as follows. First, we used smoothing splines to determine a continuous  interpolation $P^{\mathrm{mid}}_C(\alpha)$ of the  discrete data points $P_C(\alpha)$. In the same way we also determined the higher and lower interpolating curves $P^{\mathrm{high}}_C(\alpha)$ and $P^{\mathrm{low}}_C(\alpha)$ for the data points $P_C(\alpha)+\delta P_C(\alpha)$ and $P_C(\alpha)-\delta P_C(\alpha)$ respectively, where $\delta P_C(\alpha)$ denotes the standard error of $P_C(\alpha)$. A reference value $\alpha^{\mathrm{mid}}_C$ was then determined such that $P_C^{\mathrm{mid}}(\alpha^{\mathrm{mid}}_C) = P_0$, where we used the smooth interpolation of the experimental data. The energy boost was then determined as $\mu_C =  \alpha^{\mathrm{mid}}_1/\alpha^{\mathrm{mid}}_C$.  $P_0$ is an arbitrarily chosen reference value where the different $P_C(\alpha)$ curves are overlapped. This reference  serves as a base point for computing $\mu_C$. As shown in the main text for the $K_4$, the overlap of the $P_C$ data over the entire $\alpha$ range means that the specific choice of $P_0$ is arbitrary.

We similarly determined  $\mu^{\mathrm{high}}_C =  \alpha^{\mathrm{high}}_1/\alpha^{\mathrm{high}}_C$ and $\mu^{\mathrm{low}}_C =  \alpha^{\mathrm{low}}_1/\alpha^{\mathrm{low}}_C$ using the corresponding interpolating curves. The error bars shown in the figures were then centered at  $\mu_C$, with lower and upper error bars being $\mu^{\mathrm{high}}_C$ and $\mu^{\mathrm{low}}_C$, respectively.

\section{Numerical Methods}
\label{sec:Num_Meth}
We reported results based on quantum Monte Carlo techniques in the main text. Here we briefly review this technique. Simulated Quantum Annealing (SQA) is a quantum Monte Carlo based algorithm whereby Monte Carlo dynamics are used to sample from the instantaneous Gibbs state associated with the Hamiltonian $H(t)$ of the system.  The state at the end of the quantum Monte Carlo simulation of the quantum Hamiltonian $H(t)$ is used as the initial state for the next Monte Carlo simulation with Hamiltonian $H(t+\Delta t)$.  This is repeated until $H(t_f)$ is reached.  SQA was originally proposed as an optimization algorithm \cite{sqa1,Santoro}, but it has since gained traction as a computationally efficient classical description for $T>0$ quantum annealers \cite{q108,speedup,Albash:2014if,Hen:2015rt}.  An important caveat is that SQA does not capture the unitary dynamics of the quantum system, but it is hoped that the sampling of the instantaneous Gibbs state captures thermal processes in the quantum annealer, which may be the dominant dynamics if the evolution is sufficiently slow.  Although there is strong evidence that SQA does not completely capture the final-time output of the D-Wave processors \cite{Albash:2014if,Boixo:2014yu}, at present it is the only viable means to simulate large ($\gtrsim 15$ qubits) open QA systems.  We used discrete-time quantum Monte Carlo in our simulations with the number of Trotter slices fixed to $64$.  Spin updates were performed via Wolff-cluster updates \cite{PhysRevLett.62.361} along the Trotter direction only.

\begin{thebibliography}{73}
\expandafter\ifx\csname natexlab\endcsname\relax\def\natexlab#1{#1}\fi
\expandafter\ifx\csname url\endcsname\relax
  \def\url#1{\texttt{#1}}\fi
\expandafter\ifx\csname urlprefix\endcsname\relax\def\urlprefix{URL }\fi

\bibitem[{Kadowaki \& Nishimori(1998)}]{kadowaki_quantum_1998}
Kadowaki, T. \& Nishimori, H.
\newblock Quantum annealing in the transverse \uppercase{I}sing model.
\newblock \emph{Phys. Rev. E} \textbf{58}, 5355 (1998).

\bibitem[{Brooke \emph{et~al.}(1999)Brooke, Bitko, F., Rosenbaum \&
  Aeppli}]{Brooke1999}
Brooke, J., Bitko, D., F., T., Rosenbaum \& Aeppli, G.
\newblock Quantum Annealing of a Disordered Magnet.
\newblock \emph{Science} \textbf{284}, 779--781 (1999).
\newblock
  \urlprefix\url{http://www.sciencemag.org/content/284/5415/779.abstract}.

\bibitem[{Brooke \emph{et~al.}(2001)Brooke, Rosenbaum \&
  Aeppli}]{brooke_tunable_2001}
Brooke, J., Rosenbaum, T.~F. \& Aeppli, G.
\newblock Tunable quantum tunnelling of magnetic domain walls.
\newblock \emph{Nature} \textbf{413}, 610--613 (2001).

\bibitem[{Farhi \emph{et~al.}(2001)}]{farhi_quantum_2001}
Farhi, E. \emph{et~al.}
\newblock A {Quantum} {Adiabatic} {Evolution} {Algorithm} {Applied} to {Random}
  {Instances} of an {NP}-{Complete} {Problem}.
\newblock \emph{Science} \textbf{292}, 472--475 (2001).
\newblock \urlprefix\url{http://www.sciencemag.org/content/292/5516/472}.

\bibitem[{Morita \& Nishimori(2008)}]{morita:125210}
Morita, S. \& Nishimori, H.
\newblock Mathematical foundation of quantum annealing.
\newblock \emph{J. Math. Phys.} \textbf{49}, 125210--47 (2008).

\bibitem[{Das \& Chakrabarti(2008)}]{RevModPhys.80.1061}
Das, A. \& Chakrabarti, B.~K.
\newblock \textit{Colloquium}: Quantum annealing and analog quantum
  computation.
\newblock \emph{Rev. Mod. Phys.} \textbf{80}, 1061--1081 (2008).

\bibitem[{{S. Suzuki and A. Das (guest eds.)}(2015)}]{EPJ-ST:2015}
{S. Suzuki and A. Das (guest eds.)}.
\newblock {Discussion and Debate - Quantum Annealing: The Fastest Route to
  Quantum Computation?}
\newblock \emph{Eur. Phys. J. Spec. Top.} \textbf{224}, 1 (2015).
\newblock
  \urlprefix\url{http://epjst.epj.org/articles/epjst/abs/2015/01/contents/contents.html}.

\bibitem[{Farhi \emph{et~al.}(2000)Farhi, Goldstone, Gutmann \&
  Sipser}]{farhi_quantum_2000}
Farhi, E., Goldstone, J., Gutmann, S. \& Sipser, M.
\newblock Quantum {Computation} by {Adiabatic} {Evolution}.
\newblock \emph{arXiv:quant-ph/0001106}  (2000).
\newblock \urlprefix\url{http://arxiv.org/abs/quant-ph/0001106}.

\bibitem[{Aharonov \emph{et~al.}(2007)}]{aharonov_adiabatic_2007}
Aharonov, D. \emph{et~al.}
\newblock Adiabatic Quantum Computation is Equivalent to Standard Quantum
  Computation.
\newblock \emph{SIAM J. Comput.} \textbf{37}, 166--194 (2007).

\bibitem[{Mizel \emph{et~al.}(2007)Mizel, Lidar \&
  Mitchell}]{PhysRevLett.99.070502}
Mizel, A., Lidar, D.~A. \& Mitchell, M.
\newblock Simple Proof of Equivalence between Adiabatic Quantum Computation and
  the Circuit Model.
\newblock \emph{Phys. Rev. Lett.} \textbf{99}, 070502 (2007).
\newblock
  \urlprefix\url{http://link.aps.org/doi/10.1103/PhysRevLett.99.070502}.

\bibitem[{Gosset \emph{et~al.}(2015)Gosset, Terhal \&
  Vershynina}]{Gosset:2014rp}
Gosset, D., Terhal, B.~M. \& Vershynina, A.
\newblock Universal Adiabatic Quantum Computation via the Space-Time
  Circuit-to-Hamiltonian Construction.
\newblock \emph{Phys. Rev. Lett.} \textbf{114}, 140501-- (2015).
\newblock
  \urlprefix\url{http://link.aps.org/doi/10.1103/PhysRevLett.114.140501}.

\bibitem[{Lloyd \& Terhal(2015)}]{Lloyd:2015fk}
Lloyd, S. \& Terhal, B.
\newblock Adiabatic and Hamiltonian computing on a 2D lattice with simple
  2-qubit interactions.
\newblock \emph{arXiv:1509.01278}  (2015).
\newblock \urlprefix\url{http://arXiv.org/abs/1509.01278}.

\bibitem[{Childs \emph{et~al.}(2001)Childs, Farhi \&
  Preskill}]{childs_robustness_2001}
Childs, A.~M., Farhi, E. \& Preskill, J.
\newblock Robustness of adiabatic quantum computation.
\newblock \emph{Phys. Rev. A} \textbf{65}, 012322 (2001).

\bibitem[{Sarandy \& Lidar(2005)}]{PhysRevLett.95.250503}
Sarandy, M.~S. \& Lidar, D.~A.
\newblock Adiabatic Quantum Computation in Open Systems.
\newblock \emph{Phys. Rev. Lett.} \textbf{95}, 250503-- (2005).
\newblock
  \urlprefix\url{http://link.aps.org/doi/10.1103/PhysRevLett.95.250503}.

\bibitem[{Amin \emph{et~al.}(2008)Amin, Love \& Truncik}]{TAQC}
Amin, M. H.~S., Love, P.~J. \& Truncik, C. J.~S.
\newblock Thermally Assisted Adiabatic Quantum Computation.
\newblock \emph{Phys. Rev. Lett.} \textbf{100}, 060503 (2008).

\bibitem[{Lloyd(2008)}]{Lloyd:2008zr}
Lloyd, S.
\newblock Robustness of Adiabatic Quantum Computing.
\newblock \emph{arXiv:0805.2757}  (2008).
\newblock \urlprefix\url{http://arXiv.org/abs/0805.2757}.

\bibitem[{Amin \emph{et~al.}(2009)Amin, Averin \&
  Nesteroff}]{amin_decoherence_2009}
Amin, M. H.~S., Averin, D.~V. \& Nesteroff, J.~A.
\newblock Decoherence in adiabatic quantum computation.
\newblock \emph{Phys. Rev. A} \textbf{79}, 022107 (2009).
\newblock \urlprefix\url{http://link.aps.org/doi/10.1103/PhysRevA.79.022107}.

\bibitem[{Albash \& Lidar(2015)}]{Albash:2015nx}
Albash, T. \& Lidar, D.~A.
\newblock Decoherence in adiabatic quantum computation.
\newblock \emph{Phys. Rev. A} \textbf{91}, 062320-- (2015).
\newblock \urlprefix\url{http://link.aps.org/doi/10.1103/PhysRevA.91.062320}.

\bibitem[{Lidar \& Brun(2013)}]{Lidar-Brun:book}
Lidar, D. \& Brun, T. (eds.).
\newblock \emph{Quantum Error Correction} (Cambridge University Press,
  {Cambridge, UK}, 2013).
\newblock \urlprefix\url{http://www.cambridge.org/9780521897877}.

\bibitem[{Jordan \emph{et~al.}(2006)Jordan, Farhi \& Shor}]{jordan2006error}
Jordan, S.~P., Farhi, E. \& Shor, P.~W.
\newblock Error-correcting codes for adiabatic quantum computation.
\newblock \emph{{Phys. Rev. A}} \textbf{74}, 052322 (2006).
\newblock \urlprefix\url{http://link.aps.org/doi/10.1103/PhysRevA.74.052322}.

\bibitem[{Lidar(2008)}]{PhysRevLett.100.160506}
Lidar, D.~A.
\newblock Towards Fault Tolerant Adiabatic Quantum Computation.
\newblock \emph{{Phys.~Rev.~Lett.}} \textbf{100}, 160506 (2008).
\newblock
  \urlprefix\url{http://link.aps.org/doi/10.1103/PhysRevLett.100.160506}.

\bibitem[{Quiroz \& Lidar(2012)}]{PhysRevA.86.042333}
Quiroz, G. \& Lidar, D.~A.
\newblock High-fidelity adiabatic quantum computation via dynamical decoupling.
\newblock \emph{Phys. Rev. A} \textbf{86}, 042333 (2012).

\bibitem[{Young \emph{et~al.}(2013{\natexlab{a}})Young, Sarovar \&
  Blume-Kohout}]{Young:13}
Young, K.~C., Sarovar, M. \& Blume-Kohout, R.
\newblock Error Suppression and Error Correction in Adiabatic Quantum
  Computation: Techniques and Challenges.
\newblock \emph{Phys. Rev. X} \textbf{3}, 041013-- (2013{\natexlab{a}}).
\newblock \urlprefix\url{http://link.aps.org/doi/10.1103/PhysRevX.3.041013}.

\bibitem[{Sarovar \& Young(2013)}]{Sarovar:2013kx}
Sarovar, M. \& Young, K.~C.
\newblock Error suppression and error correction in adiabatic quantum
  computation: non-equilibrium dynamics.
\newblock \emph{New J. of Phys.} \textbf{15}, 125032 (2013).
\newblock \urlprefix\url{http://stacks.iop.org/1367-2630/15/i=12/a=125032}.

\bibitem[{Young \emph{et~al.}(2013{\natexlab{b}})Young, Blume-Kohout \&
  Lidar}]{Young:2013fk}
Young, K.~C., Blume-Kohout, R. \& Lidar, D.~A.
\newblock Adiabatic quantum optimization with the wrong Hamiltonian.
\newblock \emph{Phys. Rev. A} \textbf{88}, 062314-- (2013{\natexlab{b}}).
\newblock \urlprefix\url{http://link.aps.org/doi/10.1103/PhysRevA.88.062314}.

\bibitem[{Pudenz \emph{et~al.}(2014)Pudenz, Albash \& Lidar}]{PAL:13}
Pudenz, K.~L., Albash, T. \& Lidar, D.~A.
\newblock Error-corrected quantum annealing with hundreds of qubits.
\newblock \emph{Nat. Commun.} \textbf{5}, 3243 (2014).
\newblock \urlprefix\url{dx.doi.org/10.1038/ncomms4243}.

\bibitem[{Ganti \emph{et~al.}(2014)Ganti, Onunkwo \& Young}]{Ganti:13}
Ganti, A., Onunkwo, U. \& Young, K.
\newblock Family of [[6k,2k,2]] codes for practical, scalable adiabatic quantum
  computation.
\newblock \emph{Phys. Rev. A} \textbf{89}, 042313-- (2014).
\newblock \urlprefix\url{http://link.aps.org/doi/10.1103/PhysRevA.89.042313}.

\bibitem[{Bookatz \emph{et~al.}(2015)Bookatz, Farhi \& Zhou}]{Bookatz:2014uq}
Bookatz, A.~D., Farhi, E. \& Zhou, L.
\newblock Error suppression in Hamiltonian-based quantum computation using
  energy penalties.
\newblock \emph{Physical Review A} \textbf{92}, 022317-- (2015).
\newblock \urlprefix\url{http://link.aps.org/doi/10.1103/PhysRevA.92.022317}.

\bibitem[{Mizel(2014)}]{Mizel:2014sp}
Mizel, A.
\newblock Fault-tolerant, Universal Adiabatic Quantum Computation.
\newblock \emph{arXiv:1403.7694}  (2014).
\newblock \urlprefix\url{http://arXiv.org/abs/1403.7694}.

\bibitem[{Pudenz \emph{et~al.}(2015)Pudenz, Albash \& Lidar}]{PAL:14}
Pudenz, K.~L., Albash, T. \& Lidar, D.~A.
\newblock {Quantum annealing correction for random Ising problems}.
\newblock \emph{{Phys. Rev. A}} \textbf{91}, 042302 (2015).
\newblock \urlprefix\url{http://link.aps.org/doi/10.1103/PhysRevA.91.042302}.

\bibitem[{Vinci \emph{et~al.}(2015)Vinci, Albash, Paz-Silva, Hen \&
  Lidar}]{Vinci:2015jt}
Vinci, W., Albash, T., Paz-Silva, G., Hen, I. \& Lidar, D.~A.
\newblock Quantum annealing correction with minor embedding.
\newblock \emph{{Phys. Rev. A}} \textbf{92}, 042310-- (2015).
\newblock \urlprefix\url{http://link.aps.org/doi/10.1103/PhysRevA.92.042310}.

\bibitem[{Mishra \emph{et~al.}(2015)Mishra, Albash \& Lidar}]{Mishra:2015ye}
Mishra, A., Albash, T. \& Lidar, D.
\newblock Performance of two different quantum annealing correction codes.
\newblock \emph{arXiv:1508.02785}  (2015).
\newblock \urlprefix\url{http://arXiv.org/abs/1508.02785}.

\bibitem[{Matsuura \emph{et~al.}(2015)Matsuura, Nishimori, Albash \&
  Lidar}]{MNAL:15}
Matsuura, S., Nishimori, H., Albash, T. \& Lidar, D.~A.
\newblock Mean Field Analysis of Quantum Annealing Correction.
\newblock \emph{arXiv:1510.07709}  (2015).
\newblock \urlprefix\url{http://arXiv.org/abs/1510.07709}.

\bibitem[{Aliferis \emph{et~al.}(2006)Aliferis, Gottesman \&
  Preskill}]{Aliferis:05}
Aliferis, P., Gottesman, D. \& Preskill, J.
\newblock Quantum accuracy threshold for concatenated distance-3 codes.
\newblock \emph{Quantum Inf. Comput.} \textbf{6}, 97 (2006).
\newblock \urlprefix\url{http://www.rintonpress.com/xqic6/qic-6-2/097-165.pdf}.

\bibitem[{Johnson \emph{et~al.}(2011)}]{Dwave}
Johnson, M.~W. \emph{et~al.}
\newblock Quantum annealing with manufactured spins.
\newblock \emph{Nature} \textbf{473}, 194--198 (2011).

\bibitem[{Johnson \emph{et~al.}(2010)}]{Johnson:2010ys}
Johnson, M.~W. \emph{et~al.}
\newblock A scalable control system for a superconducting adiabatic quantum
  optimization processor.
\newblock \emph{Superconductor Science and Technology} \textbf{23}, 065004
  (2010).
\newblock \urlprefix\url{http://stacks.iop.org/0953-2048/23/i=6/a=065004}.

\bibitem[{Berkley \emph{et~al.}(2010)}]{Berkley:2010zr}
Berkley, A.~J. \emph{et~al.}
\newblock A scalable readout system for a superconducting adiabatic quantum
  optimization system.
\newblock \emph{Superconductor Science and Technology} \textbf{23}, 105014
  (2010).
\newblock \urlprefix\url{http://stacks.iop.org/0953-2048/23/i=10/a=105014}.

\bibitem[{Harris \emph{et~al.}(2010)}]{Harris:2010kx}
Harris, R. \emph{et~al.}
\newblock Experimental investigation of an eight-qubit unit cell in a
  superconducting optimization processor.
\newblock \emph{Phys. Rev. B} \textbf{82}, 024511 (2010).

\bibitem[{Boixo \emph{et~al.}(2014{\natexlab{a}})}]{q108}
Boixo, S. \emph{et~al.}
\newblock Evidence for quantum annealing with more than one hundred qubits.
\newblock \emph{Nat. Phys.} \textbf{10}, 218--224 (2014{\natexlab{a}}).

\bibitem[{Shin \emph{et~al.}(2014)Shin, Smith, Smolin \& Vazirani}]{SSSV}
Shin, S.~W., Smith, G., Smolin, J.~A. \& Vazirani, U.
\newblock How ``Quantum" is the {D-Wave} Machine?
\newblock \emph{arXiv:1401.7087}  (2014).
\newblock \urlprefix\url{http://arXiv.org/abs/1401.7087}.

\bibitem[{Albash \emph{et~al.}(2015{\natexlab{a}})Albash, R{\o}nnow, Troyer \&
  Lidar}]{Albash:2014if}
Albash, T., R{\o}nnow, T.~F., Troyer, M. \& Lidar, D.~A.
\newblock Reexamining classical and quantum models for the D-Wave One
  processor.
\newblock \emph{Eur. Phys. J. Spec. Top.} \textbf{224}, 111--129
  (2015{\natexlab{a}}).
\newblock \urlprefix\url{dx.doi.org/10.1140/epjst/e2015-02346-0}.

\bibitem[{Albash \emph{et~al.}(2015{\natexlab{b}})Albash, Vinci, Mishra,
  Warburton \& Lidar}]{q-sig2}
Albash, T., Vinci, W., Mishra, A., Warburton, P.~A. \& Lidar, D.~A.
\newblock Consistency tests of classical and quantum models for a quantum
  annealer.
\newblock \emph{Phys. Rev. A} \textbf{91}, 042314-- (2015{\natexlab{b}}).
\newblock \urlprefix\url{http://link.aps.org/doi/10.1103/PhysRevA.91.042314}.

\bibitem[{Crowley \emph{et~al.}(2014)Crowley, Duri\'c, Vinci, Warburton \&
  Green}]{Crowley:2014qp}
Crowley, P. J.~D., Duri\'c, T., Vinci, W., Warburton, P.~A. \& Green, A.~G.
\newblock Quantum and classical dynamics in adiabatic computation.
\newblock \emph{Phys. Rev. A} \textbf{90}, 042317-- (2014).
\newblock \urlprefix\url{http://link.aps.org/doi/10.1103/PhysRevA.90.042317}.

\bibitem[{Martin-Mayor \& Hen(2015)}]{Martin-Mayor:2015dq}
Martin-Mayor, V. \& Hen, I.
\newblock Unraveling Quantum Annealers using Classical Hardness.
\newblock \emph{arXiv:1502.02494}  (2015).
\newblock \urlprefix\url{http://arXiv.org/abs/1502.02494}.

\bibitem[{King \emph{et~al.}(2015)King, Lanting \& Harris}]{King:2015zr}
King, A.~D., Lanting, T. \& Harris, R.
\newblock Performance of a quantum annealer on range-limited constraint
  satisfaction problems.
\newblock \emph{arXiv:1502.02098}  (2015).
\newblock \urlprefix\url{http://arXiv.org/abs/1502.02098}.

\bibitem[{Vinci \emph{et~al.}(2014)}]{vinci2014hearing}
Vinci, W. \emph{et~al.}
\newblock Hearing the shape of the Ising model with a programmable
  superconducting-flux annealer.
\newblock \emph{Scientific reports} \textbf{4} (2014).

\bibitem[{Bunyk \emph{et~al.}(Aug. 2014)}]{Bunyk:2014hb}
Bunyk, P.~I. \emph{et~al.}
\newblock Architectural Considerations in the Design of a Superconducting
  Quantum Annealing Processor.
\newblock \emph{IEEE Transactions on Applied Superconductivity} \textbf{24},
  1--10 (Aug. 2014).

\bibitem[{Venturelli \emph{et~al.}(2015)}]{Venturelli:2014nx}
Venturelli, D. \emph{et~al.}
\newblock Quantum Optimization of Fully Connected Spin Glasses.
\newblock \emph{Phys. Rev. X} \textbf{5}, 031040-- (2015).
\newblock \urlprefix\url{http://link.aps.org/doi/10.1103/PhysRevX.5.031040}.

\bibitem[{Kato(1950)}]{Kato:50}
Kato, T.
\newblock On the adiabatic theorem of Quantum Mechanics.
\newblock \emph{J. Phys. Soc. Jap.} \textbf{5}, 435 (1950).

\bibitem[{Jansen \emph{et~al.}(2007)Jansen, Ruskai \& Seiler}]{Jansen:07}
Jansen, S., Ruskai, M.-B. \& Seiler, R.
\newblock Bounds for the adiabatic approximation with applications to quantum
  computation.
\newblock \emph{J. Math. Phys.} \textbf{48}, -- (2007).
\newblock
  \urlprefix\url{http://scitation.aip.org/content/aip/journal/jmp/48/10/10.1063/1.2798382}.

\bibitem[{Lidar \emph{et~al.}(2009)Lidar, Rezakhani \& Hamma}]{lidar:102106}
Lidar, D.~A., Rezakhani, A.~T. \& Hamma, A.
\newblock Adiabatic approximation with exponential accuracy for many-body
  systems and quantum computation.
\newblock \emph{J. Math. Phys.} \textbf{50}, -- (2009).
\newblock
  \urlprefix\url{http://scitation.aip.org/content/aip/journal/jmp/50/10/10.1063/1.3236685}.

\bibitem[{Wiebe \& Babcock(2012)}]{Wiebe:12}
Wiebe, N. \& Babcock, N.~S.
\newblock Improved error-scaling for adiabatic quantum evolutions.
\newblock \emph{New J. Phys.} \textbf{14}, 013024 (2012).

\bibitem[{Ge \emph{et~al.}(2015)Ge, Moln{\'a}r \& Cirac}]{Ge:2015wo}
Ge, Y., Moln{\'a}r, A. \& Cirac, J.~I.
\newblock Rapid adiabatic preparation of injective PEPS and Gibbs states.
\newblock \emph{arXiv:1508.00570}  (2015).
\newblock \urlprefix\url{http://arXiv.org/abs/1508.00570}.

\bibitem[{Avron \emph{et~al.}(2012)Avron, Fraas, Graf \& Grech}]{Avron:2012tv}
Avron, J.~E., Fraas, M., Graf, G.~M. \& Grech, P.
\newblock Adiabatic Theorems for Generators of Contracting Evolutions.
\newblock \emph{Comm. Math. Phys.} \textbf{314}, 163--191 (2012).
\newblock \urlprefix\url{dx.doi.org/10.1007/s00220-012-1504-1}.

\bibitem[{Venuti \emph{et~al.}(2015)Venuti, Albash, Lidar \&
  Zanardi}]{Venuti:2015kq}
Venuti, L.~C., Albash, T., Lidar, D.~A. \& Zanardi, P.
\newblock Adiabaticity in open quantum systems.
\newblock \emph{arXiv:1508.05558}  (2015).
\newblock \urlprefix\url{http://arXiv.org/abs/1508.05558}.

\bibitem[{Choi(2011)}]{Choi2}
Choi, V.
\newblock {Minor-embedding in adiabatic quantum computation: II.
  Minor-universal graph design}.
\newblock \emph{Quant. Inf. Proc.} \textbf{10}, 343--353 (2011).
\newblock \urlprefix\url{dx.doi.org/10.1007/s11128-010-0200-3}.

\bibitem[{Cai \emph{et~al.}(2014)Cai, Macready \& Roy}]{Cai:2014nx}
Cai, J., Macready, W.~G. \& Roy, A.
\newblock A practical heuristic for finding graph minors.
\newblock \emph{arXiv:1406.2741}  (2014).
\newblock \urlprefix\url{http://arXiv.org/abs/1406.2741}.

\bibitem[{Boothby \emph{et~al.}(2015)Boothby, King \& Roy}]{Boothby2015a}
Boothby, T., King, A.~D. \& Roy, A.
\newblock Fast clique minor generation in Chimera qubit connectivity graphs.
\newblock \emph{arXiv:1507.04774}  (2015).
\newblock \urlprefix\url{http://arXiv.org/abs/1507.04774}.

\bibitem[{Kaminsky \emph{et~al.}(2004)Kaminsky, Lloyd \&
  Orlando}]{Kaminsky-Lloyd}
Kaminsky, W.~M., Lloyd, S. \& Orlando, T.~P.
\newblock \emph{Quantum Computing and Quantum Bits in Mesoscopic Systems},
  chap.~25, 229--236 (Springer, New York, 2004).

\bibitem[{Klymko \emph{et~al.}(2014)Klymko, Sullivan \&
  Humble}]{klymko_adiabatic_2012}
Klymko, C., Sullivan, B.~D. \& Humble, T.~S.
\newblock Adiabatic quantum programming: minor embedding with hard faults.
\newblock \emph{Quantum Information Processing} \textbf{13}, 709--729 (2014).
\newblock \urlprefix\url{http://dx.doi.org/10.1007/s11128-013-0683-9}.

\bibitem[{R{\o}nnow \emph{et~al.}(2014)}]{speedup}
R{\o}nnow, T.~F. \emph{et~al.}
\newblock Defining and detecting quantum speedup.
\newblock \emph{Science} \textbf{345}, 420--424 (2014).

\bibitem[{Hen \emph{et~al.}(2015)}]{Hen:2015rt}
Hen, I. \emph{et~al.}
\newblock Probing for quantum speedup in spin-glass problems with planted
  solutions.
\newblock \emph{{Phys. Rev. A}} \textbf{92}, 042325-- (2015).
\newblock \urlprefix\url{http://link.aps.org/doi/10.1103/PhysRevA.92.042325}.

\bibitem[{Albash \emph{et~al.}(????)Albash, Lidar, Matsuura, Nishimori \&
  Vinci}]{ALMNV}
Albash, T., Lidar, D., Matsuura, S., Nishimori, H. \& Vinci, W.
\newblock {In preparation}.

\bibitem[{Seoane \& Nishimori(2012)}]{Seoane:2012uq}
Seoane, B. \& Nishimori, H.
\newblock Many-body transverse interactions in the quantum annealing of the p
  -spin ferromagnet.
\newblock \emph{J. Phys. A} \textbf{45}, 435301 (2012).
\newblock \urlprefix\url{http://stacks.iop.org/1751-8121/45/i=43/a=435301}.

\bibitem[{Bray \& Moore(1980)}]{Bray:1980fk}
Bray, A.~J. \& Moore, M.~A.
\newblock Replica theory of quantum spin glasses.
\newblock \emph{Journal of Physics C: Solid State Physics} \textbf{13}, L655
  (1980).
\newblock \urlprefix\url{http://stacks.iop.org/0022-3719/13/i=24/a=005}.

\bibitem[{Krzakala \emph{et~al.}(2008)Krzakala, Rosso, Semerjian \&
  Zamponi}]{PhysRevB.78.134428}
Krzakala, F., Rosso, A., Semerjian, G. \& Zamponi, F.
\newblock Path-integral representation for quantum spin models: Application to
  the quantum cavity method and Monte Carlo simulations.
\newblock \emph{Phys. Rev. B} \textbf{78}, 134428 (2008).
\newblock \urlprefix\url{http://link.aps.org/doi/10.1103/PhysRevB.78.134428}.

\bibitem[{Choi(2008)}]{Choi1}
Choi, V.
\newblock {Minor-embedding in adiabatic quantum computation: I. The parameter
  setting problem}.
\newblock \emph{Quant. Inf. Proc.} \textbf{7}, 193--209 (2008).
\newblock \urlprefix\url{dx.doi.org/10.1007/s11128-008-0082-9}.

\bibitem[{Kirkpatrick \& Sherrington(1978)}]{Kirkpatrick:1978dn}
Kirkpatrick, S. \& Sherrington, D.
\newblock Infinite-ranged models of spin-glasses.
\newblock \emph{Physical Review B} \textbf{17}, 4384--4403 (1978).
\newblock \urlprefix\url{http://link.aps.org/doi/10.1103/PhysRevB.17.4384}.

\bibitem[{Katzgraber \emph{et~al.}(2014)Katzgraber, Hamze \&
  Andrist}]{2014Katzgraber}
Katzgraber, H.~G., Hamze, F. \& Andrist, R.~S.
\newblock Glassy Chimeras Could Be Blind to Quantum Speedup: Designing Better
  Benchmarks for Quantum Annealing Machines.
\newblock \emph{Phys. Rev. X} \textbf{4}, 021008-- (2014).
\newblock \urlprefix\url{http://link.aps.org/doi/10.1103/PhysRevX.4.021008}.

\bibitem[{Marto\ifmmode~\check{n}\else \v{n}\fi{}\'ak
  \emph{et~al.}(2002)Marto\ifmmode~\check{n}\else \v{n}\fi{}\'ak, Santoro \&
  Tosatti}]{sqa1}
Marto\ifmmode~\check{n}\else \v{n}\fi{}\'ak, R., Santoro, G.~E. \& Tosatti, E.
\newblock Quantum annealing by the path-integral {M}onte {C}arlo method: The
  two-dimensional random {I}sing model.
\newblock \emph{Phys. Rev. B} \textbf{66}, 094203 (2002).

\bibitem[{Santoro \emph{et~al.}(2002)Santoro, Marto\v{n}\'{a}k, Tosatti \&
  Car}]{Santoro}
Santoro, G.~E., Marto\v{n}\'{a}k, R., Tosatti, E. \& Car, R.
\newblock Theory of Quantum Annealing of an {I}sing Spin Glass.
\newblock \emph{Science} \textbf{295}, 2427--2430 (2002).

\bibitem[{Boixo \emph{et~al.}(2014{\natexlab{b}})}]{Boixo:2014yu}
Boixo, S. \emph{et~al.}
\newblock Computational Role of Collective Tunneling in a Quantum Annealer.
\newblock \emph{arXiv:1411.4036}  (2014{\natexlab{b}}).
\newblock \urlprefix\url{http://arXiv.org/abs/1411.4036}.

\bibitem[{Wolff(1989)}]{PhysRevLett.62.361}
Wolff, U.
\newblock Collective {M}onte {C}arlo Updating for Spin Systems.
\newblock \emph{Phys. Rev. Lett.} \textbf{62}, 361--364 (1989).
\newblock \urlprefix\url{http://link.aps.org/doi/10.1103/PhysRevLett.62.361}.

\end{thebibliography}
\end{document}